\documentclass[letter,11pt]{article}

\usepackage{ltexpprt}
\usepackage{setspace}
\usepackage{subcaption}
\usepackage{enumerate}
\usepackage{amsmath}
\usepackage{amsfonts}
\usepackage{graphicx}
\usepackage{amssymb}
\usepackage{multirow}
\usepackage{geometry}
\usepackage{soul}
\usepackage{colortbl}
\usepackage{wrapfig}
\usepackage{algorithm}
\usepackage{algorithmicx}
\usepackage{algpseudocode}
\usepackage{times}
\usepackage{hyperref,url}

\usepackage{thmtools}
\usepackage{thm-restate}

\algtext*{EndWhile}% Remove "end while" text
\algtext*{EndIf}% Remove "end if" text
\algtext*{EndFor}% Remove "end for" text

\newcommand{\eps}{\varepsilon}
\newcommand{\R}{\mathbb{R}}
\newcommand{\opt}{\mathsf{opt}}
\newcommand{\tO}{\widetilde{O}}

\newcommand{\TIMOTHY}[1]{{\color{red} (( Timothy: #1 ))}}

\newcommand{\ignore}[1]{}

\usepackage{wrapfig}
\usepackage{bm}

\title{\LARGE Dynamic Geometric Set Cover, Revisited\thanks{The research of Timothy M. Chan and Qizheng He was supported in part by NSF Grant CCF-1814026. The research of Subhash Suri and Jie Xue was supported in part by NSF grant CCF-1814172.}}
\author{Timothy M. Chan\footnote{Department of Computer Science, University of Illinois at Urbana-Champaign, USA.} \and Qizheng He\footnotemark[2] \and  Subhash Suri\footnote{Department of Computer Science, University of California at Santa Barbara, USA.} \and Jie Xue\footnote{New York University Shanghai, China.}}
\date{}

\newenvironment{myproof}{\begin{proof}}{\hspace*{\fill}\end{proof}}

\begin{document}

\pagenumbering{gobble}

\maketitle

\begin{abstract}
Geometric set cover is a classical problem in computational geometry, which has been extensively studied in the past.
In the dynamic version of the problem, points and ranges may be inserted and deleted, and our goal is to efficiently maintain a set cover solution (satisfying certain quality requirement) for the dynamic problem instance.
In this paper, we give a plethora of new dynamic geometric set cover data structures in 1D and 2D, which significantly improve and extend the previous results.
Our results include the following:
\begin{itemize}
    \item The first data structure for $(1+\varepsilon)$-approximate dynamic interval set cover with polylogarithmic  amortized update time. Specifically, we achieve an update time of $O(\log^3 n/\varepsilon)$, improving the $O(n^\delta/\varepsilon)$ bound of Agarwal et al. [SoCG'20], where $\delta>0$ denotes an arbitrarily small constant.
    \item A data structure for $O(1)$-approximate dynamic unit-square set cover with $2^{O(\sqrt{\log n})}$ amortized update time, substantially improving the $O(n^{1/2+\delta})$ update time of Agarwal et al. [SoCG'20].
    \item A data structure for $O(1)$-approximate dynamic square set cover with $O(n^{1/2+\delta})$ randomized amortized update time,
    improving the $O(n^{2/3+\delta})$ update time of Chan and He [SoCG'21].
    \item A data structure for $O(1)$-approximate dynamic 2D halfplane set cover
    with $O(n^{17/23+\delta})$ randomized amortized update time.
    The previous solution for halfplane set cover by Chan and He [SoCG'21] is slower and can only report the size of the approximate solution.
    \item The first sublinear results for the \textit{weighted} version of dynamic geometric set cover.
    Specifically, we give a data structure for $(3+o(1))$-approximate dynamic weighted interval set cover with $2^{O(\sqrt{\log n \log\log n})}$ amortized update time and a data structure for $O(1)$-approximate dynamic weighted unit-square set cover with $O(n^\delta)$ amortized update time.
\end{itemize}

%, where $\delta > 0$ denotes an arbitrarily small constant.

%\Subhash{In these bounds, shall we replace $O(n^{1/2+\delta})$ with $O(n^{\frac{1}{2}+\delta})$ etc? Otherwise, it is ambiguous whether we mean $(1/2) + \delta$ or $1/(2 + \delta)$.}
%\TIMOTHY{doesn't look ambiguous to me}
%\Subhash{Ok, then we keep it as is.}
%for a constant $\alpha>0$,
%and 2D unit-disk set cover
%2^{O(\sqrt{\log n \log\log n}+\sqrt{\log n\log(1/\varepsilon)})}$

%(for any constant $\delta>0$)
%\TIMOTHY{for arbitrarily small constants, would $\delta$ look better than $\alpha$?  (I see $\delta$ is used for other things...)}
%\Qizheng{I agree with that, but in the current writing $\delta$ is used in some of the error analysis parts. Maybe we want to change the variable names of those? @Jie}
%\Jie{It's fine actually. The sections where I used $\delta$ notations do not have a $\delta$ exponent in the time bounds.}

%\TIMOTHY{taking out 2D unit-disk result because of bug...}

\end{abstract}

%we revisit the dynamic geometric set cover problem.
%Furthermore, we give a $(3+\varepsilon)$-approximation data structure with for dynamic weighted interval set cover with $2^{O(\sqrt{\log n \log\log n})}/\varepsilon$ amortized update time.

%\newpage

\pagenumbering{arabic}

\section{Introduction} \label{sec-intro}
\emph{Geometric set cover} is a classical problem in computational geometry, with a long history and 
applications \cite{agarwal2012near,agarwal2014near,bronnimann1995almost,bus2018practical,ChanG14,chan2012weighted,ChanH20,ChanH15,clarkson2007improved,ErlebachL10,MustafaRR15,mustafa2009ptas,varadarajan2010weighted}.  A typical formulation involves a set of points $S$ in $\R^d$ 
and a family ${\cal R}$ of subsets of $S$, often called ranges, defined by a simple class of geometric 
objects. For instance, the sets may be defined by intervals of $\R^1$ in one dimension or balls, 
hypercubes or halfspaces in higher dimensions. The goal is to find a smallest subfamily of ${\cal R}$ 
covering all the points of $S$. In the \emph{weighted} set cover problem, each range is associated 
with a non-negative weight, and the goal is to find a minimum weight set cover.  In general, 
these problems are NP-complete even for the simplest of geometric families such as unit disks
or unit squares in two dimensions, but they often allow efficient approximation algorithms with 
better (worst-case) performance than the general (combinatorial) set cover.

Recently, an exciting line of research was launched by Agarwal et al.~\cite{agarwal2020dynamic} on \emph{dynamic} geometric 
set covering with the introduction of \emph{sublinear} time data structures for \emph{fully dynamic} 
maintenance of approximate set covers for intervals in one dimension and unit squares in two dimensions.
These sublinear bounds are in sharp contrast with the $\Omega(f)$ update time bottleneck faced
by the general (combinatorial) set cover problem in dynamic setting~\cite{AbboudA0PS19,BhattacharyaHNW21,GuptaK0P17}, where $f$ is the number of sets
containing an element, because inserting an element at a minimum requires updating all the
sets that contain it. The \emph{implicit} form of sets in geometric set covering---an interval or a disk,
for instance, takes only $O(1)$ pieces of information to add or delete---provides a natural yet challenging 
problem setting in which to explore the possibility of truly sublinear (possibly polylogarithmic)
updates of both the elements and the sets.
Indeed, following the work of Agarwal et al.~\cite{agarwal2020dynamic}, Chan and He~\cite{chan2021dynamic} pushed the envelope further
and managed to achieve sublinear update time for arbitrary axis-aligned squares, and
if only the size of an approximate solution is needed,
for disks in the plane and halfspaces in three dimensions as well.

In spite of these recent developments, the state of the art for dynamic geometric set covering is 
far from satisfactory even for the simplest of the set systems: covering points on the line by
intervals or covering points in the plane by axis-aligned squares. For instance, the best update
bound for the former is $O(n^\delta / \eps)$ for
a $(1 + \eps)$ approximation,
and for the latter is $O(n^{2/3 + \delta })$ for an $O(1)$ approximation, where $\delta>0$ is an arbitrarily small constant. More importantly, none of these schemes are able to handle the case
of \emph{weighted} set covers. In this paper, we make substantial progress on these fronts.

\subsection{Results.}

We present a large collection of new results, as summarized in Table~\ref{table1},
which substantially improve \emph{all} the main results of  Agarwal et al.~\cite{agarwal2020dynamic} on
unweighted intervals in 1D and unweighted unit squares in 2D, as well as the main result
of Chan and He~\cite{chan2021dynamic} on unweighted arbitrary squares in 2D. Throughout the paper, 
all the update bounds are amortized, and $\delta>0$ denotes an arbitrarily small constant; constant factors hidden in $O$ notation may depend on $\delta$.
In particular, our results include the following:

\begin{enumerate}
\item 
For unweighted intervals in 1D, we obtain the first dynamic data structure with polylogarithmic update time and constant approximation factor.  We achieve $1+\eps$ approximation with $O(\log^3 n/\eps)$ update time,
which improves Agarwal et al.'s previous update bound of $O(n^\delta/\eps)$.
(The dynamic hitting set data structure for 1D intervals in ~\cite{agarwal2020dynamic} does have 
polylogarithmic update time but not the set cover data structure.)
\item 
For unweighted unit squares in 2D, we obtain the first dynamic data structure
with $n^{o(1)}$ update time and constant approximation factor.
(All squares are axis-aligned throughout the paper.)
The precise update bound is $2^{O(\sqrt{\log n})}$, which significantly improves
Agarwal et al.'s previous update bound of $O(n^{1/2+\delta})$.
\item 
For unweighted arbitrary squares in 2D, we obtain a dynamic data structure
with $O(n^{1/2+\delta})$ update time (with Monte Carlo randomization) and constant approximation factor. 
This improves Chan and He's previous (randomized) update bound of $O(n^{2/3+\delta})$.
\item 
For unweighted halfplanes %and unit disks 
in 2D, we obtain the first
dynamic data structure with sublinear update time and constant approximation
factor that can efficiently
report an approximate solution (in time linear in the solution size).  
The (randomized) update bound is $O(n^{17/23+\delta})=o(n^{0.74})$.
Although Chan and He's previous solution~\cite{chan2021dynamic} can more generally handle halfspaces in 3D, % and arbitrary disks in 2D, 
it has a larger (randomized) update bound of $O(n^{12/13+\delta})$ and can only output the size of an approximate solution.
(Specializing Chan and He's solution to halfplanes %and unit disks 
in 2D can
lower the update time a bit, but it would still be worse than the new bound.)
%Note that by a known transformation~\cite{ChanLP11,PachT13},
%the 2D quadrant case reduces to the 2D halfplane case, 
%and the 2D unit square case is equivalent to the 2D quadrant case
%(for $O(1)$ approximation); thus, 2D halfplanes are more general than 2D unit squares. 
%% true only for one type of quadrants...
\end{enumerate}
Note that although for the static problem, PTASs were known for unweighted arbitrary squares 
and disks in 2D~\cite{mustafa2009ptas} (and exact polynomial-time algorithms were known for halfplanes in 2D~\cite{Har-PeledL12}), the running times of these static algorithms are superquadratic.  Thus, for any of the 2D problems above, constant approximation factor is the best one could hope for under the current state of the art if the goal is sublinear update time.

A second significant contribution of our paper is to extend the dynamic set cover data structures to
\emph{weighted} instances, thus providing the \emph{first} nontrivial results for dynamic weighted geometric set cover. 
(Although there were previous results on weighted independent 
set for 1D intervals and other ranges by Henzinger, Neumann, and Wiese~\cite{Henzinger0W20} and Bhore et al.~\cite{BhoreCIK20}, no results on dynamic weighted geometric set cover were known  even in 1D\@.
This is in spite of the considerable work on static weighted geometric set cover  \cite{chan2012weighted,ErlebachL10,Har-PeledL12,MustafaRR15,varadarajan2010weighted}.) In particular, we present the following results:

\begin{enumerate}
\item[5.] 
For weighted intervals in 1D, we obtain a dynamic data structure
with $n^{o(1)}$ update time and constant approximation factor.
The update bound is $2^{O(\sqrt{\log n\log\log n})}$ and the approximation
factor is $3+o(1)$.
\item[6.]
For weighted unit squares in 2D, we also obtain a dynamic data structure
with $O(n^\delta)$ update time and constant approximation factor (where the constant depends on $\delta$ and weights are assumed to be  polynomially bounded integers).
Even when compared to Agarwal et al.'s unweighted result~\cite{agarwal2020dynamic}, our result
is a substantial improvement, besides being more general.  
\end{enumerate}

For the cases of (unweighted or weighted) unit squares in 2D and
unweighted halfplanes %and unit disks 
in 2D, the same results hold for
the \emph{hitting set} problem---given a set of points and a set of ranges, find the smallest (or minimum weight) subset of points that hit all the given ranges---because hitting set is equivalent to 
set cover for these types of ranges by duality.

\begin{table}\centering
\begin{tabular}{|l|l|l|l|}
\hline
Ranges & Approx.\ & Previous update time & New update time  \\\hline
Unweighted 1D intervals & $1+\eps$ & $n^\delta$ \cite{agarwal2020dynamic} & $\log^3n$ %& (Sec.~\ref{sec-shortint}, \ref{sec-unweightedint}) 
\\
Unweighted 2D unit squares & $O(1)$ & $n^{1/2+\delta}$ \cite{agarwal2020dynamic} & $2^{O(\sqrt{\log n})}$ %& (Sec.~\ref{??},~\ref{??})
\\
Unweighted 2D arbitrary squares & $O(1)$ & $n^{2/3+\delta}$ \cite{chan2021dynamic} & $n^{1/2+\delta}$ %& (Sec.~\ref{??},~\ref{??})
\\
Unweighted 2D halfplanes & $O(1)$ & $n^{12/13+\delta}$ ($\ast$) \cite{chan2021dynamic} & $n^{17/23+\delta}$ %& (Sec.~\ref{??},~\ref{??})
\\
%unweighted 2D unit disks & $O(1)$ & $n^{12/13+\delta}$ ($\ast$) \cite{chan2021dynamic} & $n^{2/3+\delta}$ & (Sec.~\ref{??},~\ref{??})\\
Weighted 1D intervals & $3+\eps$ & none & $2^{O(\sqrt{\log n\log\log n})}$ %& (Sec.~\ref{??},~\ref{??})
\\
Weighted 2D unit squares & $O(1)$ & none & $n^{\delta}$ %& (Sec.~\ref{??},~\ref{??})
\\\hline
\end{tabular}
\caption{Summary of data structures for approximate dynamic geometric set cover. Here, $\delta>0$ denotes an arbitrarily small constant; hidden constant factors in the approximation and update bounds may depend on $\eps$ and $\delta$. 
For unweighted 2D arbitrary squares and 2D halfplanes, the previous and new results are randomized.  In the entry marked ($\ast$),
the algorithm can only return the size of the solution, not the solution itself.}\label{table1}
\end{table}

\subsection{Techniques.}

We give six different methods to achieve these results.  
%With the possible
%exception of the unweighted arbitrary square result, new conceptual ideas that go beyond straightforward technical modifications of previous techniques:
Many of these methods require significant
new ideas that go beyond minor modifications of previous
techniques:

\begin{enumerate}
\item
For the unweighted 1D intervals, Agarwal et al.~\cite{agarwal2020dynamic} obtained their
result with $O(n^\delta)$ update time by a ``bootstrapping'' approach, but
extra factors accumulate in each round of bootstrapping.
To obtain polylogarithmic update time,
we refine their approach with a better recursion, whose analysis distinguishes
between ``one-sided'' and ``two-sided'' intervals.
\item
For the unweighted 2D unit squares, it suffices to solve the problem for
quadrants (i.e., 2-sided orthogonal ranges) due to a standard reduction.  We adopt
an interesting geometric divide-and-conquer approach  (different from
more common approaches like k-d trees or segment trees).  Roughly, we form
an $r\times r$ nonuniform grid, where each column/row has $O(n/r)$ points, and recursively build data structures 
for each grid cell and for each grid column and each grid row.  Agarwal et al.'s previous
data structure~\cite{agarwal2020dynamic} also used an $r\times r$ grid but did not use recursion per
column or row; the boundary of a quadrant intersects $O(r)$ out of the 
$r^2$ grid cells and so updating a quadrant causes $O(r)$ recursive calls, eventually
leading to $O(n^{1/2+\delta})$ update time.  With
our new ideas, updating a quadrant requires recursive calls
in only $O(1)$ grid columns/rows and grid cells, leading to $n^{o(1)}$ update time.
\item
For unweighted 2D arbitrary squares, our method resembles Chan and He's
previous method~\cite{chan2021dynamic}, dividing the problem into two cases: when the optimal value
$\opt$ is small or when $\opt$ is large.  Their small $\opt$ algorithm
was obtained by modifying a known static approximation 
algorithm based on multiplicative weight updates~\cite{agarwal2014near,bronnimann1995almost,ChanH20,clarkson1993algorithms}, and achieved $\tO(\opt^2)$
update time.\footnote{
The $\tO$ notation hides polylogarithmic factors.
}
Their large $\opt$ algorithm employed quadtrees and achieved
$\tO(n^{1/2+\delta} + n/\opt)$ update time.  Combining the two algorithms
yielded $\tO(n^{2/3})$ update time, as the critical case occurs when $\opt$ is near $n^{1/3}$.  We modify their large $\opt$ algorithm by incorporating some extra
technical ideas (treating so-called ``light'' vs.\ ``heavy'' canonical rectangles differently, and carefully tuning parameters);
this allows us to improve the update time to $O(n^{1/2+\delta})$ uniformly for all $\opt$, pushing the approach to its natural limit.
\item
For unweighted 2D halfplanes, we handle the small $\opt$ case by adapting
Chan and He's previous method~\cite{chan2021dynamic}, but we present a new method for the large $\opt$ case.  We propose a geometric divide-and-conquer approach based on
the well-known Partition Theorem of Matou\v sek~\cite{matouvsek1992efficient}.  The Partition Theorem
was originally formulated for the design of range searching data structures, but its applicability
to decompose geometric set cover instances is less apparent.  The key to the approximation factor analysis is a simple observation
that the boundary of the union of the halfplanes in the optimal solution is a convex 
chain with $O(\opt)$ edges, and so in a partition of the plane into $b$ disjoint cells, the number of intersecting pairs of edges and cells is $O(\opt + b)$.

%The unweighted 2D unit disk case reduces to unweighted 2D ``pseudo-halfplanes'' and is solved similarly.
\end{enumerate}

For weighted dynamic geometric set cover, none of the previous approaches generalizes.
Essentially all previous approaches for the unweighted setting make use of the dichotomy
of small vs.\ large $\opt$: in the small $\opt$ case, we can generate a solution   quickly from scratch; on the other hand, in the large $\opt$ case, we can tolerate a
large additive error (in particular, this enables divide-and-conquer with a large number of parts).
However, all this breaks down in the weighted setting because the cardinality of the optimal solution is no longer related to its value.  A different way to bound approximation factors is required.
\begin{enumerate}
\item[5.]
For weighted 1D intervals, our key new idea is to incorporate dynamic programming (DP)
into the divide-and-conquer.  In addition, we use a common trick of grouping
weights by powers of a constant, so that the number of distinct weight groups is
logarithmic. %(assuming that weights are polynomially bounded).
\item[6.]
For weighted 2D unit squares, we again use a geometric divide-and-conquer based on the $r\times r$ grid, but the recursion gets even more interesting as we incorporate DP\@. (We also group weights by powers of a constant.) 
%The geometric divide-and-conquer via the $r\times r$ grid now becomes more intricate,
To keep the approximation factor $O(1)$, the number of 
levels of recursion needs to be $O(1)$, but we can still achieve $O(n^\delta)$ update time.
\end{enumerate}

\subsection{Preliminaries.}

Throughout the paper, we use $\opt$  to denote the size of the optimal set cover (in the unweighted case), and $[r]$ to denote the set $\{1,\ldots,r\}$.
In a \emph{size query}, we want to output an approximation to the size $\opt$.  In a \emph{membership query}, we want to determine whether a given object is in the approximate solution maintained by the data structure.  In a \emph{reporting query}, we want to report all elements in the approximate solution (in time sensitive to the output size). 
As in the previous work \cite{agarwal2020dynamic,chan2021dynamic}, in all of our results, the set cover solution we maintain is a \textit{multi-set} of ranges (i.e., each range may have multiple duplicates).
We denote by $A \sqcup B$ the disjoint union of two multi-sets $A$ and $B$.
%where the multiplicity is the sum

%\medskip \noindent \textbf{Due to limited space, some proofs and details are omitted in this short version of the paper. The full version of the paper can be found in the appendix.}

%\TIMOTHY{anything else to add here?  can remove Sec 2...}
%\Jie{Yes, we plan to remove this section and move the definition of different queries to Sec1?}

%\TIMOTHY{Do we need to assume polynomially bounded weights, at least for weighted unit squares?}
%\Qizheng{We can deal with this issue, by performing binary search on the weights on the maximal and long quadrants used. Do you want to discuss on skype?}

%\section{Preliminaries} \label{sec-preliminary}
%\input{preliminary}

%\section{Short version for SODA submission} \label{sec-overview}
%\input{overview}

%\Qizheng{--------------short version ends here----------------}

%\section{Unweighted dynamic geometric set cover} \label{sec-unweighted}

%\TIMOTHY{should we make Sec 8.1-8.4 separate sections, to be consistent with the short version?  same for 9.1 and 9.2?}

%\section{Weighted dynamic geometric set cover} \label{sec-weighted}

\section{Unweighted Interval Set Cover} \label{sec-unweightedint}

%Our data structure for dynamic (unweighted) interval set cover essentially follows the bootstrapping framework of \cite{agarwal2020dynamic}.
Let $(S,\mathcal{I})$ be a dynamic (unweighted) interval set cover instance where $S$ is the set of points in $\mathbb{R}$ and $\mathcal{I}$ is the set of intervals, and let $\varepsilon >0$ be the approximation factor.
%We use $n$ to denote the size of the current $(S,\mathcal{I})$, i.e., $n = |S| + |\mathcal{I}|$, and use $n_0$ to denote the size of the initial $(S,\mathcal{I})$.
Our goal is to design a data structure $\mathcal{D}$ that maintains a $(1+\varepsilon)$-approximate set cover solution for the current instance $(S,\mathcal{I})$ and supports the desired queries (i.e., size, membership, report queries) to the solution.
Without loss of generality, we may assume that the \textit{point range} of $(S,\mathcal{I})$ is $[0,1]$, i.e., the points in $S$ are always in the range $[0,1]$.

%Our data structure itself is essentially the same as the one in \cite{agarwal2020dynamic}.
%The main difference occurs in the analyses.
%We analyze the data structure more carefully based on some new observations.
%For completeness, we present our data structure in a self-contained way.

Let $r$ and $\alpha<1$ be parameters to be determined.
Consider the initial instance $(S,\mathcal{I})$ and let $n = |S|+|\mathcal{I}|$.
We partition the range $[0,1]$ into $r$ connected portions (i.e., intervals) $J_1,\dots,J_r$ such that each portion $J_i$ contains $O(n/r)$ points in $S$ and $O(n/r)$ endpoints of intervals in $\mathcal{I}$.
%As in \cite{agarwal2020dynamic}, we build a (static) point-location data structure $\mathcal{L}$ which can report, for a given query point $q \in \mathbb{R}$, the portion $J_i$ that contains $q$.
Define $S_i = S \cap J_i$ and $\mathcal{I}_i = \{I \in \mathcal{I}: I \cap J_i \neq \emptyset \text{ and } J_i \nsubseteq I\}$.
When the instance $(S,\mathcal{I})$ changes, the portions $J_1,\dots,J_r$ remain unchanged while the $S_i$'s and $\mathcal{I}_i$'s will change along with $S$ and $\mathcal{I}$.
Thus, we can view each $(S_i,\mathcal{I}_i)$ as a dynamic interval set cover instance with point range $J_i$.
We then recursively build a dynamic interval set cover data structure $\mathcal{D}_i$ which maintains a $(1+\tilde{\varepsilon})$-approximate set cover solution for $(S_i,\mathcal{I}_i)$, where $\tilde{\varepsilon} = \alpha \varepsilon$.
We call $(S_1,\mathcal{I}_1),\dots,(S_r,\mathcal{I}_r)$ \textit{sub-instances} and call $\mathcal{D}_1,\dots,\mathcal{D}_r$ \textit{sub-structures}.
Besides the recursively built sub-structures, we also need three simple support data structures.
The first one is the data structure $\mathcal{A}$ in the following lemma that can help compute an optimal interval set cover in \textit{output-sensitive} time.
\begin{lemma}[\cite{agarwal2020dynamic}] \label{lem-intoslong}
One can store a dynamic (unweighted) interval set cover instance $(S,\mathcal{I})$ in a data structure $\mathcal{A}$ with $O(n\log n)$ construction time and $O(\log n)$ update time such that at any point, an optimal solution for $(S,\mathcal{I})$ can be computed in $O(\mathsf{opt} \cdot \log n)$ time with the access to $\mathcal{A}$.
\end{lemma}

\noindent
The second one is a dynamic data structure $\mathcal{B}$ built on $\mathcal{I}$ which can report, for a given query interval $J$, an interval $I \in \mathcal{I}$ that contains $J$ (if such an interval exists); as shown in \cite{agarwal2020dynamic}, there exists such a data structure with $O(\log n)$ update time, $O(\log n)$ query time, and $O(n \log n)$ construction time.
The third one is a (static) data structure $\mathcal{L}$ which can report, for a given query point $q \in \mathbb{R}$, the portion $J_i$ that contains $q$; for this one, we can simply use a binary search tree built on $J_1,\dots,J_r$ which has $O(\log r)$ query time.
Our data structure $\mathcal{D}$ simply consists of the sub-structures $\mathcal{D}_1,\dots,\mathcal{D}_r$ and the support data structures.
It is easy to construct $\mathcal{D}$ in $O(n \log^2 n)$ time.
To see this, we define $|(S,\mathcal{I})|$ as the total number of points in $S$ and endpoints of intervals in $\mathcal{I}$ that are contained in the point range $[0,1]$ of $(S,\mathcal{I})$.
%Note that $|(S,\mathcal{I})|$
We have $|(S,\mathcal{I})| \leq \sum_{i=1}^r |(S_i,\mathcal{I}_i)|$ and $|(S_i,\mathcal{I}_i)| \leq |(S,\mathcal{I})|/2$ for all $i \in [r]$ (as $r$ is sufficiently large).
Now let $C(m)$ denote the time for constructing the data structure on an instance $(S,\mathcal{I})$ with $|(S,\mathcal{I})|=m$.
We then have the recurrence $C(m) = \sum_{i=1}^r C(m_i) + O(m \log m)$, where $m \leq \sum_{i=1}^r m_i$ and $m_i \leq m/2$ for all $i \in [r]$.
This recurrence solves to $C(m) = O(m \log^2 m)$.
Since $|(S,\mathcal{I})| = O(n)$, $\mathcal{D}$ can be constructed in $O(n \log^2 n)$ time, i.e., in $\widetilde{O}(n)$ time.
%Our data structure $\mathcal{D}$

\subsubsection*{Updating the sub-structures and reconstruction.}
Whenever the instance $(S,\mathcal{I})$ changes due to an insertion/deletion on $S$ or $\mathcal{I}$, we first update the support data structures.
After that, we update the sub-structures $\mathcal{D}_i$ for $i \in [r]$ that $(S_i,\mathcal{I}_i)$ changes.
An insertion/deletion on $S$ only changes one $S_i$ and an insertion/deletion on $\mathcal{I}$ changes at most two $\mathcal{I}_i$'s (because an interval has two endpoints).
Also, we observe that if the inserted/deleted interval $I$ is ``one-sided'' in the sense that one endpoint of $I$ is outside the point range $[0,1]$, then that insertion/deletion only changes one $\mathcal{I}_i$.
This observation is critical in the analysis of our data structure.
%Using the point location data structure $\mathcal{L}$, we can find the indices $i \in [r]$ for which $(S_i,\mathcal{I}_i)$ changes.
Besides the update, our data structure $\mathcal{D}$ will be periodically reconstructed.
Specifically, the $(i+1)$-th reconstruction happens after processing $n_i/r$ updates from the $i$-th reconstruction, where $n_i$ denotes the size of $(S,\mathcal{I})$ at the point of the $i$-th reconstruction.
(The $0$-th reconstruction is just the initial construction of $\mathcal{D}$.)

\subsubsection*{Constructing a solution.}
We now describe how to construct an approximately optimal set cover $\mathcal{I}_\text{appx}$ for the current $(S,\mathcal{I})$ using our data structure $\mathcal{D}$.
%This part is similar to that in \cite{agarwal2020dynamic}.
Denote by $\mathsf{opt}$ the size of an optimal set cover for the current $(S,\mathcal{I})$; we define $\mathsf{opt} = \infty$ if $(S,\mathcal{I})$ does not have a set cover.
Set $\delta = \min\{n,c \cdot (r+\varepsilon r)/(\varepsilon - \alpha \varepsilon)\}$ for a sufficiently large constant $c$.
If $\mathsf{opt} \leq \delta$, then we are able to use the algorithm of Lemma~\ref{lem-intoslong} to compute an optimal set cover for $(S,\mathcal{I})$ in $O(\delta \cdot \log n)$ time (with the help of the support data structure $\mathcal{A}$).
Therefore, we simulate that algorithm within that amount of time.
If the algorithm successfully computes a solution, we use it as our $\mathcal{I}_\text{appx}$.
Otherwise, we construct $\mathcal{I}_\text{appx}$ as follows.
For each $i \in [r]$, if $J_i$ can be covered by an interval $I \in \mathcal{I}$, we define $\mathcal{I}_i^* = \{I\}$, otherwise let $\mathcal{I}_i^*$ be the $(1+\tilde{\varepsilon})$-approximate solution for $(S_i,\mathcal{I}_i)$ maintained in the sub-structure $\mathcal{D}_i$.
(If for some $i \in [r]$, $J_i$ cannot be covered by any interval in $\mathcal{I}$ and the sub-structure $\mathcal{D}_i$ tells us that the current $(S_i,\mathcal{I}_i)$ does not have a set cover, then we immediately decide that the current $(S,\mathcal{I})$ has no feasible set cover.)
Then we define $\mathcal{I}_\text{appx} = \bigsqcup_{i=1}^r \mathcal{I}_i^*$, which is clearly a set cover of $(S,\mathcal{I})$.
%We say an interval $J$ is \textit{coverable} if there exists $I \in \mathcal{I}$ such that $J \subseteq I$.
%Let $P = \{i \in [r]: J_i \text{ is coverable}\}$ and $P' = \{i \in [r]: J_i \text{ is not coverable}\}$.
%For each $i \in P$, we find an interval in $\mathcal{I}$ that covers $J_i$; denote by $\mathcal{I}^*$ the collection of these intervals.
%Then we consider the indices in $P'$.
%If for some $i \in P'$, the data structure $\mathcal{D}_i$ tells us that the current $(S_i,\mathcal{I}_i)$ does not have a set cover, then we immediately decide that the current $(S,\mathcal{I})$ has no feasible set cover.
%Otherwise, each $\mathcal{D}_i$ for $i \in P'$ maintains a $(1+ \varepsilon - (\varepsilon/\log n))$-approximation solution $\mathcal{I}_i^*$ for the set cover instance $(S_i,\mathcal{I}_i)$.
%We then define $\mathcal{I}_\text{appx} = \mathcal{I}^* \sqcup (\bigsqcup_{i \in P'} \mathcal{I}_i^*)$ as the solution for $(S,\mathcal{I})$.
Note that for each $i \in [r]$, we can find in $O(\log n)$ time an interval $I \in \mathcal{I}$ that covers $J_i$ using the support data structure $\mathcal{B}$ (if such an interval exists).

\subsubsection*{Answering queries to the solution.}
We show how to store the solution $\mathcal{I}_\text{appx}$ properly so that the desired queries for $\mathcal{I}_\text{appx}$ can be answered efficiently.
If $\mathcal{I}_\text{appx}$ is computed by the algorithm of Lemma~\ref{lem-intoslong}, then the size of $\mathcal{I}_\text{appx}$ is at most $\delta$ and we have all elements of $\mathcal{I}_\text{appx}$ in hand.
In this case, we simply build a binary search tree on $\mathcal{I}_\text{appx}$ which can answer the desired queries with the required time costs.
On the other hand, if $\mathcal{I}_\text{appx}$ is defined as $\mathcal{I}_\text{appx} = \bigsqcup_{i=1}^r \mathcal{I}_i^*$, the size of $\mathcal{I}_\text{appx}$ can be large and we are not able to retrieve all elements of $\mathcal{I}_\text{appx}$.
However, in this case, each $\mathcal{I}_i^*$ either consists of a single interval that covers $J_i$ or is the solution maintained in the sub-structure $\mathcal{D}_i$.
To support the size query, we only need to compute $|\mathcal{I}_i^*|$ (which can be done by recursively making size queries to the sub-structures) and calculate $|\mathcal{I}_\text{appx}| = \sum_{i=1}^r |\mathcal{I}^*|$; we then simply store this quantity so that a size query can be answered in $O(1)$ time.
To support membership queries, we compute an index set $P \subseteq [r]$ consisting of the indices $i \in [r]$ such that $\mathcal{I}_i^*$ consists of a single interval covering $J_i$.
Then we collect all intervals in the $\mathcal{I}_i^*$'s for $i \in P$, the number of which is at most $r$.
We store these intervals in a binary search tree $T$ which can answer membership queries in $O(\log r)$ time.
To answer a membership query $I \in \mathcal{I}$, we first check if $I$ is stored in $T$.
After that, we find the (up to) two instances $\mathcal{I}_i$'s that contains $I$, and make membership queries to the sub-structures $\mathcal{D}_i$ to check whether $I \in \mathcal{I}_i^*$ (if $i \in [r] \backslash P$).
Finally, to answer the reporting query, we first report the intervals stored in $T$ and then for every $i \in [r] \backslash P$, we make recursively a reporting query to $\mathcal{D}_i$, which reports the intervals in $\mathcal{I}_i^*$.

Now we analyze the query time.
If the solution $\mathcal{I}_\text{appx}$ is computed by the algorithm of Lemma~\ref{lem-intoslong}, then it is stored in a binary search tree and we can answer a size query, a membership query, and a reporting query in $O(1)$ time, $O(\log |\mathcal{I}_\text{appx}|)$ time, and $O(|\mathcal{I}_\text{appx}|)$ time, respectively.
So it suffices to consider the case where we construct the solution as $\mathcal{I}_\text{appx} = \bigsqcup_{i=1}^r \mathcal{I}_i^*$.
In this case, answering a size query still takes $O(1)$, because we explicitly compute $|\mathcal{I}_\text{appx}|$.
To analyze the time cost for a membership query, we need to distinguish \textit{one-sided} and \textit{two-sided} queries.
%We say a membership query $I \in \mathcal{I}$ is a \textit{one-sided} query if one endpoint of $I$ is outside the point range $[0,1]$ of $(S,\mathcal{I})$ and is a \textit{two-sided} query if both endpoints of $I$ are in the point range.
We use $Q_1(n)$ and $Q_2(n)$ to denote the time cost for a one-sided membership query (i.e., one endpoint of the query interval is outside the point range) and a two-sided membership query (i.e., both endpoints of the query interval are inside the point range), respectively, when the size of the current instance is $n$.
Then for $Q_1(n)$, we have the recurrence $Q_1(n) \leq Q_1(O(n/r)) + O(\log r)$, which solves to $Q_1(n) = O(\log n)$, as we only need to recursively query on one $\mathcal{D}_i$ (which is again a one-sided query).
For $Q_2(n)$, we have the recurrence $Q_2(n) \leq \max\{Q_2(O(n/r)),2Q_1(O(n/r))\} + O(\log r)$, which also solves to $Q_2(n) = O(\log n)$, as we may need to have a recursive two-sided query on one $\mathcal{D}_i$ or have recursive one-sided queries on two $\mathcal{D}_i$'s.
Therefore, a membership query can be answered in $O(\log n)$ time.
Finally, to answer a reporting query, we first report the intervals stored in $T$ and recursively query the data structures $\mathcal{D}_i$ for all $i \in [r] \backslash P$ such that $\mathcal{I}_i^* \neq \emptyset$.
Thus, in the recurrence tree, the number of leaves is bounded by $|\mathcal{I}_\text{appx}|$ since at each leaf node we need to report at least one element.
Since the height of the recurrence tree is $O(\log_r n)$ and at each node of the recurrence tree the work can be done in $O(\log r)$ time plus $O(1)$ per outputted element, the overall time cost for a reporting query is $O(|\mathcal{I}_\text{appx}| \cdot \log n)$.

\subsubsection*{Correctness.}
%The correctness of our data structure $\mathcal{D}$ follows almost directly from the analysis in \cite{agarwal2020dynamic}.
First, we observe that $\mathcal{D}$ makes a no-solution decision iff the current instance $(S,\mathcal{I})$ has no set cover.
Indeed, if we make a no-solution decision, then $J_i$ is not covered by any interval in $\mathcal{I}$ and the sub-instance $(S_i,\mathcal{I}_i)$ has no set cover for some $i \in [r]$; in this case, $(S,\mathcal{I})$ has no set cover because the points in $S_i$ can only be covered by the intervals in $\mathcal{I}_i$ or by an interval that covers $J_i$.
On the other hand, if we do not make a no-solution decision, then the set $\mathcal{I}_\text{appx}$ we construct is a feasible solution for $(S,\mathcal{I})$.
%if $(S,\mathcal{I})$ has no set cover, then we must have $(S_i,\mathcal{I}_i)$ has no set cover for some $i \in P'$, and hence $\mathcal{D}$ makes a no-solution decision.
Now it suffices to show that the solution $\mathcal{I}_\text{appx}$ is a $(1+\varepsilon)$-approximation of an optimal set cover for $(S,\mathcal{I})$.
Let $\mathcal{I}_\text{opt}$ be an optimal set cover for $(S,\mathcal{I})$.
We have to show $|\mathcal{I}_\text{appx}| \leq (1+\varepsilon) \cdot |\mathcal{I}_\text{opt}|$.
If $\mathcal{I}_\text{appx}$ is computed by the algorithm of Lemma~\ref{lem-intoslong}, then $|\mathcal{I}_\text{appx}| = |\mathcal{I}_\text{opt}|$.
Otherwise, we know that $|\mathcal{I}_\text{opt}| > \delta$, which implies $|\mathcal{I}_\text{opt}| > c \cdot (r+\varepsilon r)/(\varepsilon - \alpha \varepsilon)$ for a sufficiently large constant $c$, because we cannot have $|\mathcal{I}_\text{opt}| > n$.
In this case, we show the following.
\begin{fact}
$|\mathcal{I}_\textnormal{appx}| \leq (1+\tilde{\varepsilon}) \cdot |\mathcal{I}_\textnormal{opt}| + O(r)$.
\end{fact}
\begin{myproof}
For $i \in [r]$, let $\mathsf{opt}_i$ be the size of an optimal set cover of $(S_i,\mathcal{I}_i)$ if $\mathcal{I}_i^*$ is the solution of $(S_i,\mathcal{I}_i)$ maintained by $\mathcal{D}_i$, and let $\mathsf{opt}_i = 1$ otherwise.
Then for all $i \in [r]$, we have $|\mathcal{I}_i^*| \leq (1+\tilde{\varepsilon}) \cdot \mathsf{opt}_i$.
Since $|\mathcal{I}_\text{appx}| = \sum_{i=1}^r |\mathcal{I}_i^*|$, we have $|\mathcal{I}_\text{appx}| \leq (1+\tilde{\varepsilon}) \cdot \sum_{i=1}^r \mathsf{opt}_i$.
It suffices to show that $\sum_{i=1}^r \mathsf{opt}_i = |\mathcal{I}_\text{opt}| + O(r)$.
Let $n_i$ be the number of intervals in $\mathcal{I}_\text{opt}$ that are contained in $J_i$ for $i \in [r]$.
Clearly, $|\mathcal{I}_\text{opt}| \geq \sum_{i=1}^r n_i$.
We claim that $\mathsf{opt}_i \leq n_i + 2$, which implies $\sum_{i=1}^r \mathsf{opt}_i = |\mathcal{I}_\text{opt}| + O(r)$.
If $J_i$ can be covered by some interval in $\mathcal{I}$, then $\mathsf{opt}_i = 1 \leq n_i + 2$.
Otherwise, we take all $n_i$ intervals in $\mathcal{I}_\text{opt}$ that are contained in $J_i$ and the (at most) two intervals in $\mathcal{I}_\text{opt}$ with one endpoint in $J_i$ which have maximal intersections with $J_i$ (i.e., the interval containing the left end of $J_i$ with the rightmost right endpoint and the interval containing the right end of $J_i$ with the leftmost left endpoint).
These $n_i+2$ intervals form a set cover of $(S_i,\mathcal{I}_i)$ and thus $\mathsf{opt}_i\leq n_i + 2$.
\end{myproof}

%It was shown in \cite{agarwal2020dynamic} that $\sum_{i \in P'} |\mathcal{I}_\text{opt} \cap \mathcal{I}_i| \leq |\mathcal{I}_\text{opt}| + 2r$.
%Note that for each $i \in P'$, $\mathcal{I}_\text{opt} \cap \mathcal{I}_i$ is a set cover for $(S_i,\mathcal{I}_i)$.
%Therefore, $\sum_{i \in P'} |\mathcal{I}_i^*| \leq (1+ \varepsilon - (\varepsilon/\log n)) \cdot \sum_{i \in P'} |\mathcal{I}_\text{opt} \cap \mathcal{I}_i|$.
\noindent
Using the above observation and the fact $|\mathcal{I}_\text{opt}| > c \cdot (r+\varepsilon r) / (\varepsilon - \alpha \varepsilon) = c \cdot (r+\varepsilon r)/ (\varepsilon - \tilde{\varepsilon})$, we conclude that $|\mathcal{I}_\text{appx}| \leq (1+\varepsilon) \cdot |\mathcal{I}_\text{opt}|$.
%Because $|\mathcal{I}_\text{opt}| > (3+2\varepsilon) \cdot r\log n/\varepsilon$, we have $(\varepsilon/\log n) \cdot |\mathcal{I}_\text{opt}| > (3 + 2\varepsilon) \cdot r$ and hence the above inequality implies $|\mathcal{I}_\text{appx}| \leq (1+\varepsilon) \cdot |\mathcal{I}_\text{opt}|$.

\subsubsection*{Update time.}
To analyze the update time of our data structure $\mathcal{D}$, it suffices to consider the first period (including the first reconstruction).
The first period consists of $n_0/r$ operations, where $n_0$ is the size of the initial $(S,\mathcal{I})$.
The size of $(S,\mathcal{I})$ during the first period is always in between $(1-1/r) n_0$ and $(1+1/r) n_0$ and is hence $\Theta(n_0)$, since $r$ is a sufficiently large constant.
We first observe that, excluding the recursive updates for the sub-structures, each update of $\mathcal{D}$ takes $O(r \log n/(\varepsilon - \alpha \varepsilon) + r \log^2 n)$ (amortized) time, where $n$ is the size of the current instance $(S,\mathcal{I})$.
Updating the support data structures takes $O(\log n)$ time.
When constructing the solution $\mathcal{I}_\text{appx}$, we need to simulate the algorithm of Lemma~\ref{lem-intoslong} within $O(\delta \cdot \log n)$ time, i.e., $O(r \log n/(\varepsilon - \alpha \varepsilon))$ time.
%Also, we may need to compute the sets $P$, $P'$, and $\mathcal{I}^*$, which takes $O(\log n)$ time.
The time for storing the solution $\mathcal{I}_\text{appx}$ is also bounded by $O(r \log n/(\varepsilon - \alpha \varepsilon))$, because we only need to explicitly store $\mathcal{I}_\text{appx}$ when it is computed by the algorithm of Lemma~\ref{lem-intoslong}, in which case its size is at most $\delta = O(r/(\varepsilon - \alpha \varepsilon))$.
Finally, the reconstruction takes $O(r \log^2 n)$ amortized time, because the time cost of the (first) reconstruction is $O(n_1 \log^2 n_1) = O(n_0 \log^2 n_0)$ and the first period consists of $n_0/r$ operations.

Next, we consider the recursive updates for the sub-structures.
The depth of the recursion is $O(\log_r n)$.
If we set $\alpha = 1-1/\log_r n$, the approximation parameter is $\Theta(\varepsilon)$ in any level of the recurrence.
%Here, our analysis differs from that in \cite{agarwal2020dynamic}.
We distinguish three types of updates according to the current operation.
The first type is caused by an insertion/deletion of a point in $S$ (we call it \textit{point update}).
The second type is caused by an insertion/deletion of an interval in $\mathcal{I}$ whose one endpoint is outside the point range $[0,1]$ of $(S,\mathcal{I})$ (we call it \textit{one-sided interval update}).
The third type is caused by an insertion/deletion of an interval in $\mathcal{I}$ whose both endpoints are inside the point range (we call it \textit{two-sided interval update}).
In a point update, we only need to recursively update one sub-structure (which is again a point update), because an insertion/deletion on $S$ only changes one $S_i$.
Similarly, in a one-sided interval update, we only need to do a recursive one-sided interval update on one sub-structure, because the inserted/deleted interval belongs to one $\mathcal{I}_i$.
Finally, in a two-sided interval update, we may need to do a recursive two-sided interval update on one sub-structure (when the two endpoints of the inserted/deleted interval belong to the same range $J_i$) or two recursive one-sided interval updates on two sub-structures (when the two endpoints belong to different $J_i$'s).
Let $U(n)$, $U_1(n)$, $U_2(n)$ denote the time costs of a point update, a one-sided interval update, a two-sided interval update, respectively, when the size of the current instance is $n$.
%Because $r$ is sufficiently large, the size of each instance $(S_i,\mathcal{I}_i)$ is always smaller than $n/2$ during the entire period.
Then for $U(n)$, we have the recurrence
\begin{equation*}
    U(n) \leq U(O(n/r)) + O(r \log^2 n/\varepsilon),
\end{equation*}
which solves to $U(n) = O(r \log_r n \log^2 n/\varepsilon)$.
Similarly, for $U_1(n)$, we have the same recurrence, solving to $U_1(n) = O(r \log_r n \log^2 n/\varepsilon)$.
Finally, the recurrence for $U_2(n)$ is
\begin{equation*}
    \begin{aligned}
        U_2(n) &\leq \max\{U_2(O(n/r)),2U_1(O(n/r))\} + O(r \log^2 n/\varepsilon)\\
        &= \max\{U_2(O(n/r)),O(r \log_r n \log^2 n/\varepsilon)\} + O(r \log^2 n/\varepsilon).
    \end{aligned}
\end{equation*}
A simple induction argument shows that $U_2(n) = O(r \log_r n \log^2 n/\varepsilon)$.
Setting $r$ to be a sufficiently large constant, our data structure $\mathcal{D}$ can be updated in $O(\log^3 n/\varepsilon)$ amortized time.
%Also, the query time for a membership query is $O(\log n)$ and for the reporting query is $O(|\mathcal{I}_\text{appx}| \log n)$.

%\unweightedint*

\begin{theorem}
There exists a dynamic data structure for $(1+\varepsilon)$-approximate unweighted interval set cover with $O(\log^3 n/\varepsilon)$ amortized update time and $O(n\log^2 n)$ construction time, which can answer size, membership, and reporting queries in $O(1)$, $O(\log n)$, and $O(k \log n)$ time, respectively, where $n$ is the size of the instance and $k$ is the size of the maintained solution.
\end{theorem}

%the reconstructions takes $O(\log n)$ amortized time because the $i$-th reconstruction takes $O(n_i \log n_i)$ time and the $(i-1)$-th period consists of $\Omega(n_i)$ operations (since $r$ is a constant and is sufficiently large).

%The update time of our data structure $\mathcal{D}$ consists of the time for recursively updating the data structures $\mathcal{D}_i$ (and the support data structures)

\section{Unweighted Unit-Square Set Cover}\label{unweighted unit-square (long version)}
It was shown in \cite{agarwal2020dynamic} that dynamic unit-square set cover can be reduced to dynamic \textit{quadrant} set cover.
Specifically, dynamic unit-square set cover can be solved with the same update time as dynamic quadrant set cover, by losing only a constant factor on the approximation ratio.
Therefore, it suffices to consider dynamic quadrant set cover. Note that the problem is still challenging, as we need to simultaneously deal with all four types of quadrants.

Similar to interval set cover, quadrant set cover also admits an output-sensitive algorithm:
%which was given by \cite{agarwal2020dynamic}.
\begin{lemma}[\cite{agarwal2020dynamic}] \label{lem-quadoslong}
One can store a dynamic (unweighted) quadrant set cover instance $(S,\mathcal{Q})$ in a data structure $\mathcal{A}$ with $\widetilde{O}(n)$ construction time and $\widetilde{O}(1)$ update time such that at any point, a constant-approximate solution for $(S,\mathcal{Q})$ can be computed in $\widetilde{O}(\mathsf{opt})$ time with the access to $\mathcal{A}$.
\end{lemma}

Let $(S,\mathcal{Q})$ be a dynamic (unweighted) quadrant set cover instance where $S$ is the set of points in $\mathbb{R}^2$ and $\mathcal{Q}$ is the set of quadrants.
%We use $n$ to denote the size of the current $(S,\mathcal{I})$, i.e., $n = |S| + |\mathcal{I}|$, and use $n_0$ to denote the size of the initial $(S,\mathcal{I})$.
Suppose $\mu = O(1)$ is the approximation factor of the algorithm of Lemma~\ref{lem-quadoslong}.
Our goal is to design a data structure $\mathcal{D}$ that maintains a $(\mu+\varepsilon)$-approximate set cover solution for the current instance $(S,\mathcal{Q})$ and supports the desired queries to the solution, for a given parameter $\varepsilon > 0$.
Without loss of generality, we may assume that the \textit{point range} of $(S,\mathcal{Q})$ is $[0,1]^2$, i.e., the points in $S$ are always in the range $[0,1]^2$.
We say a quadrant in $\mathcal{Q}$ is \textit{trivial} (resp., \textit{nontrivial}) if its vertex is outside (resp., inside) the point range $[0,1]^2$.
Note that a trivial quadrant is ``equivalent'' to a horizontal/vertical halfplane in terms of the coverage in $[0,1]^2$.

Let $r$ and $\alpha<1$ be parameters to be determined.
Consider the initial instance $(S,\mathcal{Q})$ and let $n = |S| + |\mathcal{Q}|$.
We partition the point range $[0,1]^2$ into $r \times r$ rectangular cells using $r-1$ horizontal lines and $r-1$ vertical lines such that each row (resp., column) of $r$ cells contains $O(n/r)$ points in $S$ and $O(n/r)$ vertices of the quadrants in $\mathcal{Q}$.
Let $\Box_{i,j}$ be the cell in the $i$-th row and $j$-th column for $(i,j) \in [r]^2$.
We denote by $R_i$ the $i$-th row (i.e., $R_i = \bigcup_{j=1}^r \Box_{i,j}$) for $i \in [r]$ and by $C_j$ the $j$-th column (i.e., $C_j = \bigcup_{i=1}^r \Box_{i,j}$) for $j \in [r]$.
Define $S_{i,j} = S \cap \Box_{i,j}$, $S_{i,\bullet} = S \cap R_i$, and $S_{\bullet,j} = S \cap C_j$, for $i,j \in [r]$.
Next, we decompose $\mathcal{Q}$ into small subsets as follows.
We say a quadrant $Q$ \textit{left intersects} a rectangle $R$ if $R \nsubseteq Q$ and $Q$ contains the left boundary of $R$.
Among a set of quadrants that left intersect a rectangle $R$, the \textit{maximal} one refers to the quadrant whose vertex is the rightmost, or equivalently, whose intersection with $R$ is maximal.
Similarly, we can define the notions of ``right intersect'', ``top intersect'', and ``bottom intersect''.
For $i,j \in [r]$, we define $\mathcal{Q}_{i,j} \subseteq \mathcal{Q}$ be the subset consisting of all \textit{nontrivial} quadrants whose vertices lie in $\Box_{i,j}$ and the (up to) four \textit{nontrivial} maximal quadrants that left, right, top, bottom intersect $\Box_{i,j}$; we call the latter the four \textit{special} quadrants in $\mathcal{Q}_{i,j}$.
%We denote by $Q_\leftarrow^R$, $Q_\rightarrow^R$, $Q_\uparrow^R$, and $Q_\downarrow^R$ be the maximal quadrants in $\mathcal{Q}$ that left, right, top, bottom intersect the rectangle $R$.
%Then we define $\mathcal{Q}_{i,j} = \mathcal{Q}_{i,j}' \cup \{Q_\leftarrow^{\Box_{i,j}},Q_\rightarrow^{\Box_{i,j}},Q_\uparrow^{\Box_{i,j}},Q_\downarrow^{\Box_{i,j}}\}$, where $\mathcal{Q}_{i,j}' \subseteq \mathcal{Q}$ consists of all \textit{nontrivial} quadrants whose vertices are in $\Box{i,j}$.
%We call $Q_\leftarrow^{\Box_{i,j}},Q_\rightarrow^{\Box_{i,j}},Q_\uparrow^{\Box_{i,j}},Q_\downarrow^{\Box_{i,j}}$ the \textit{special} quadrants in $\mathcal{Q}_{i,j}$.
Similarly, for $i \in [r]$ (resp., $j \in [r]$), we define $\mathcal{Q}_{i,\bullet} \subseteq \mathcal{Q}$ (resp., $\mathcal{Q}_{\bullet,j} \subseteq \mathcal{Q}$) be the subset consisting of all nontrivial quadrants whose vertices lie in $R_i$ (resp., $C_j$) and the four nontrivial maximal quadrants that left, right, top, bottom intersect $R_i$ (resp., $C_j$); we call the latter the four \textit{special} quadrants in $\mathcal{Q}_{i,\bullet}$ (resp., $\mathcal{Q}_{\bullet,j}$).
%Also, we define $\mathcal{Q}_{i,\bullet} = \mathcal{Q}_{i,\bullet}' \cup \{Q_\leftarrow^{R_i},Q_\rightarrow^{R_i},Q_\uparrow^{R_i},Q_\downarrow^{R_i}\}$ and $\mathcal{Q}_{\bullet,j} = \mathcal{Q}_{\bullet,j}' \cup \{Q_\leftarrow^{C_j},Q_\rightarrow^{C_j},Q_\uparrow^{C_j},Q_\downarrow^{C_j}\}$, where $\mathcal{Q}_{i,\bullet}' \subseteq \mathcal{Q}$ (resp., $\mathcal{Q}_{\bullet,j}' \subseteq \mathcal{Q}$) consists of all \textit{nontrivial} quadrants whose vertices are in $R_i$ (resp., $C_j$).
When the instance $(S,\mathcal{Q})$ changes, the cells $\Box_{i,j}$ (as well as the rows $R_i$ and columns $C_j$) remain unchanged while the sets $S_{i,j}$, $S_{i,\bullet}$, $S_{\bullet,j}$ (resp., $\mathcal{Q}_{i,j}$, $\mathcal{Q}_{i,\bullet}$, $\mathcal{Q}_{\bullet,j}$) will change along with $S$ (resp., $\mathcal{Q}$).
We view each $(S_{i,j},\mathcal{Q}_{i,j})$ as a dynamic quadrant set cover instance with point range $\Box_{i,j}$, and recursively build a sub-structure $\mathcal{D}_{i,j}$ that maintains a $(\mu+\tilde{\varepsilon})$-approximate set cover solution for $(S_{i,j},\mathcal{Q}_{i,j})$, where $\tilde{\varepsilon} = \alpha \varepsilon$.
Similarly, we view each $(S_{i,\bullet},\mathcal{Q}_{i,\bullet})$ (resp., $(S_{\bullet,j},\mathcal{Q}_{\bullet,j})$) as a dynamic quadrant set cover instance with point range $R_i$ (resp., $C_j$), and recursively build a sub-structure $\mathcal{D}_{i,\bullet}$ (resp., $\mathcal{D}_{\bullet,j}$) that maintains a $(\mu+\tilde{\varepsilon})$-approximate set cover solution for $(S_{i,\bullet},\mathcal{Q}_{i,\bullet})$ (resp., $(S_{\bullet,j},\mathcal{Q}_{\bullet,j})$).
For convenience, we call $(S_{i,j},\mathcal{Q}_{i,j})$ the \textit{cell sub-instances}, $(S_{i,\bullet},\mathcal{Q}_{i,\bullet})$ the \textit{row sub-instances}, and $(S_{\bullet,j},\mathcal{Q}_{\bullet,j})$ the \textit{column sub-instances}.
Besides the data structures recursively built on the sub-instances, we also need some simple support data structures.
The first one is the data structure $\mathcal{A}$ required for the output-sensitive algorithm for quadrant set cover (Lemma~\ref{lem-quadoslong}).
The second one is a dynamic data structure $\mathcal{B}$ built on $\mathcal{Q}$, which can report, for a given query rectangle $R$, the maximal quadrant in $\mathcal{Q}$ that left/right/top/bottom intersects $R$.
The third one is a dynamic data structure $\mathcal{C}$ built on $\mathcal{Q}$, which can report, for a given query rectangle $R$, a quadrant in $\mathcal{Q}$ that contains $R$ (if such a quadrant exists).
The fourth one is a plane point-location data structure $\mathcal{L}$, which can report, for a given query point $q \in [0,1]^2$, the cell $\Box_{i,j}$ that contains $q$.
As shown in \cite{agarwal2020dynamic}, all these support data structures can be built in $\widetilde{O}(n)$ time and updated in $\widetilde{O}(1)$ time.
Our data structure $\mathcal{D}$ consists of the recursively built sub-structures $\mathcal{D}_{i,j}$, $\mathcal{D}_{i,\bullet}$, $\mathcal{D}_{\bullet,j}$ and the support data structures $\mathcal{A}$, $\mathcal{B}$, $\mathcal{C}$, $\mathcal{L}$.
It is easy to construct $\mathcal{D}$ in $\widetilde{O}(2^{O(\log_r n)} \cdot n)$ time.
To see this, we notice that the size of each sub-instance is of size $O(n/r)$.
Also, the total size of all (row, column, cell) sub-instances is bounded by $O(n)$.
Therefore, if we denote by $C(n)$ the construction time of the data structure when the size of the instance is $n$, we have the recurrence $C(n) = \sum_{i=1}^r \sum_{j=1}^r C(n_{i,j}) + \sum_{i=1}^r C(n_{i,\bullet}) + \sum_{j=1}^r C(n_{\bullet,j}) + \widetilde{O}(n)$ for some $n_{i,j},n_{i,\bullet},n_{\bullet,j}$ satisfying $\sum_{i=1}^r \sum_{j=1}^r n_{i,j} + \sum_{i=1}^r n_{i,\bullet} + \sum_{j=1}^r n_{\bullet,j} = O(n)$ and $n_{i,j} = O(n/r)$, $n_{i,\bullet} = O(n/r)$, $n_{\bullet,j} = O(n/r)$ for all $i,j \in [r]$.
The recurrence solves to $C(n) = \widetilde{O}(2^{O(\log_r n)} \cdot n)$.

\subsubsection*{Update of the sub-structures and reconstruction.}
Whenever the instance $(S,\mathcal{Q})$ changes due to an insertion/deletion on $S$ or $\mathcal{Q}$, we first update the support data structures.
After that, we update the sub-structures $\mathcal{D}_{i,j}$, $\mathcal{D}_{i,\bullet}$, $\mathcal{D}_{\bullet,j}$ for which the underlying sub-instances change.
Observe that an insertion/deletion on $S$ only changes one $S_{i,j}$, one $S_{i,\bullet}$, and one $S_{\bullet,j}$ (so at most three sub-instances).
An insertion/deletion of a trivial quadrant does not change any sub-instances, while an insertion/deletion of a nontrivial quadrant changes at most $O(r)$ sub-instances.
Besides the update, our data structure $\mathcal{D}$ will be periodically reconstructed.
Specifically, the $(i+1)$-th reconstruction happens after processing $n_i/r$ updates from the $i$-th reconstruction, where $n_i$ denotes the size of $(S,\mathcal{Q})$ at the point of the $i$-th reconstruction.
(The $0$-th reconstruction is just the initial construction of $\mathcal{D}$.)

\subsubsection*{Constructing a solution.}
We now describe how to construct an approximately optimal set cover $\mathcal{Q}_\text{appx}$ for the current $(S,\mathcal{Q})$ using our data structure $\mathcal{D}$.
%This part is similar to that in \cite{agarwal2020dynamic}.
Denote by $\mathsf{opt}$ the size of an optimal set cover for the current $(S,\mathcal{Q})$; we define $\mathsf{opt} = \infty$ if $(S,\mathcal{Q})$ does not have a set cover.
Set $\delta = \min\{n, c (r^2 + \varepsilon r^2) /(\varepsilon - \alpha \varepsilon)\}$, where $c$ is a sufficiently large constant.
If $\mathsf{opt} \leq \delta$, then we are able to use the algorithm of Lemma~\ref{lem-quadoslong} to compute a $\mu$-approximate set cover solution for $(S,\mathcal{Q})$ in $\widetilde{O}(\delta)$ time.
Therefore, we simulate that algorithm within that amount of time.
If the algorithm successfully computes a solution, we use it as our $\mathcal{Q}_\text{appx}$.
Otherwise, we know that $\mathsf{opt} > \delta$.
In this case, we construct $\mathcal{Q}_\text{appx}$ by combining the solutions maintained by the sub-structures as follows.

Consider the trivial quadrants in $\mathcal{Q}$.
There are (up to) four maximal trivial quadrants that left, right, top, bottom intersect the point range $[0,1]^2$, which we denote by $Q_\uparrow,Q_\downarrow,Q_\leftarrow,Q_\rightarrow$, respectively.
%Let $Q_\leftarrow$, $Q_\rightarrow$, $Q_\uparrow$, $Q_\downarrow$ be the maximal quadrants in $\mathcal{Q}$ that left, right, top, bottom intersect the point range $[0,1]^2$.
Let $i^- \in [r]$ (resp., $j^- \in [r]$) be the smallest index such that $R_{i^-} \nsubseteq Q_\uparrow$ (resp., $C_{j^-} \nsubseteq Q_\leftarrow$), and $i^+ \in [r]$ (resp., $j^+ \in [r]$) be the largest index such that $R_{i^+} \nsubseteq Q_\downarrow$ (resp., $C_{j^+} \nsubseteq Q_\rightarrow$).
Note that $i^- \leq i^+$, because otherwise $S \subseteq [0,1]^2 \subseteq Q_\uparrow \cup Q_\downarrow$ and thus $\mathsf{opt} \leq 2$ (which contradicts with the fact $\mathsf{opt} > \delta$). 
For the same reason, $j^- \leq j^+$.
We include $Q_\leftarrow,Q_\rightarrow,Q_\uparrow,Q_\downarrow$ in our solution $\mathcal{Q}_\text{appx}$.
By doing this, all points in $R_i$ (resp., $C_j$) for $i < i^-$ or $i > i^+$ (resp., $j < j^-$ or $j > j^+$) are covered.
The remaining task is to cover the points in the complement $X$ of the $Q_\leftarrow \cup Q_\rightarrow \cup Q_\uparrow \cup Q_\downarrow$ in $[0,1]^2$; these points lie in the cells $\Box_{i,j}$ for $i^- \leq i \leq i^+$ and $j^- \leq j \leq j^+$.

\begin{figure}[!htbp]
    \centering
    \includegraphics[height=2.8cm]{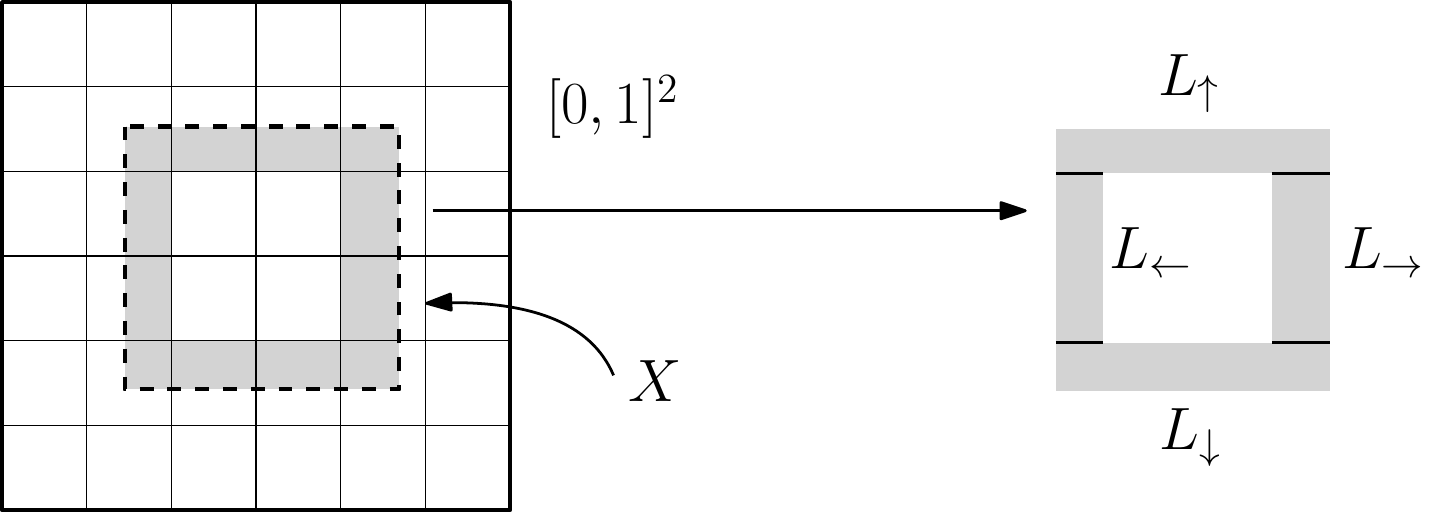}
    \caption{The rectangular annulus (the grey area) are partitioned into four rectangles.}
    \label{fig-annuluslong}
\end{figure}

We cover the points in $X$ using two collections of quadrants.
The first collection covers all points in the cells $\Box_{i,j}$ contained in $X$, i.e., the cells $\Box_{i,j}$ for $i^- < i < i^+$ and $j^- < j < j^+$.
Specifically, if the cell $\Box_{i,j}$ can be covered by a single quadrant $Q \in \mathcal{Q}$, we define $\mathcal{Q}_{i,j}^*=\{Q\}$, otherwise we define $\mathcal{Q}_{i,j}^* \subseteq \mathcal{Q}_{i,j}$ as the $(\mu+\tilde{\varepsilon})$-approximate set cover solution for the sub-instance $(S_{i,j},\mathcal{Q}_{i,j})$ maintained by $\mathcal{D}_{i,j}$.
(If there exists a cell $\Box_{i,j}$ for $i^- < i < i^+$ and $j^- < j < j^+$ that is not covered by any single quadrant $Q \in \mathcal{Q}$ and the sub-structure $\mathcal{D}_{i,j}$ tells us that the sub-instance $(S_{i,j},\mathcal{Q}_{i,j})$ has no solution, then we make a no-solution decision for $(S,\mathcal{Q})$.)
We include in our solution $\mathcal{Q}_\text{appx}$ all quadrants in $\bigcup_{i=i^-+1}^{i^+-1} \bigcup_{j=j^-+1}^{j^+-1} \mathcal{Q}_{i,j}^*$, which cover the points in $\Box_{i,j}$ for $i^- < i < i^+$ and $j^- < j < j^+$.
%By doing this, all points in $S$ are covered except those in a rectangular annulus which is the complement of the union of the cells $\Box_{i,j} \subseteq X$ in $X$ (see Figure~\ref{fig-annuluslong}).
Now the only points uncovered are those lie in the rectangular annulus, which is the complement of the union of the cells $\Box_{i,j} \subseteq X$ in $X$ (see Figure~\ref{fig-annuluslong}).
We partition this rectangular annulus into four rectangles $L_\uparrow,L_\downarrow,L_\leftarrow,L_\rightarrow$ (again see Figure~\ref{fig-annuluslong}), which are contained in $R_{i^-},R_{i^+},C_{j^-},C_{j^+}$, respectively.
We obtain a set cover for the points in each of $L_\uparrow,L_\downarrow,L_\leftarrow,L_\rightarrow$ using the corresponding row/column sub-structure as follows.
Consider $L_\uparrow$.
We temporarily insert the three \textit{virtual} quadrants $Q_\uparrow,Q_\leftarrow,Q_\rightarrow$ to the sub-instance $(S_{i^-,\bullet},\mathcal{Q}_{i^-,\bullet})$ (these quadrants will be deleted afterwards) and update the sub-structure $\mathcal{D}_{i^-,\bullet}$ so that $\mathcal{D}_{i^-,\bullet}$ now maintains a solution for $(S_{i^-,\bullet},\mathcal{Q}_{i^-,\bullet} \cup \{Q_\uparrow,Q_\leftarrow,Q_\rightarrow\})$.
This solution covers all points in $S_{i^-,\bullet}$.
We then remove the quadrants $Q_\uparrow,Q_\leftarrow,Q_\rightarrow$ from the solution (if any of them are used), and the set $\mathcal{Q}_\uparrow^*$ of the remaining quadrants should cover all points in $L_\uparrow$.
In a similar way, we can construct sets $\mathcal{Q}_\downarrow^*,\mathcal{Q}_\leftarrow^*,\mathcal{Q}_\rightarrow^*$ that cover the points in $L_\downarrow,L_\leftarrow,L_\rightarrow$, respectively, by using the sub-structures $\mathcal{D}_{i^+,\bullet},\mathcal{D}_{\bullet,j^-},\mathcal{D}_{\bullet,j^+}$.
(If any of those sub-structures tells us the corresponding sub-instance has no solution, then we make a no-solution decision for $(S,\mathcal{Q})$.)
We include in $\mathcal{Q}_\text{appx}$ all quadrants in $\mathcal{Q}^* = \mathcal{Q}_\uparrow^* \cup \mathcal{Q}_\downarrow^* \cup \mathcal{Q}_\leftarrow^* \cup \mathcal{Q}_\rightarrow^*$.
This completes the construction of $\mathcal{Q}_\text{appx}$.
To summarize, we define
\begin{equation} \label{eq-qappx}
    \mathcal{Q}_\text{appx} = \{Q_\uparrow,Q_\downarrow,Q_\leftarrow,Q_\rightarrow\} \sqcup \mathcal{Q}^* \sqcup \left( \bigsqcup_{(i,j) \in P} \mathcal{Q}_{i,j}^* \right),
\end{equation}
where $P = \{(i,j): i^-<i<i^+, j^-<j<j^+\}$.
From the construction, it is easy to verify that $\mathcal{Q}_\text{appx}$ is a set cover for $(S,\mathcal{Q})$.

\subsubsection*{Answering queries to the solution.}
We show how to store the solution $\mathcal{Q}_\text{appx}$ properly so that the desired queries for $\mathcal{Q}_\text{appx}$ can be answered efficiently.
%This part is similar to that for dynamic interval set cover in Section~\ref{sec-unweightedint}.
If $\mathcal{Q}_\text{appx}$ is computed using the output-sensitive algorithm of Lemma~\ref{lem-quadoslong}, then $|\mathcal{Q}_\text{appx}| \leq \delta$ and we have all elements of $\mathcal{Q}_\text{appx}$.
In this case, we simply build a binary search tree on $\mathcal{Q}_\text{appx}$, which can answer the desired queries with the required time costs.
On the other hand, if $\mathcal{Q}_\text{appx}$ is defined using Equation~\ref{eq-qappx}, we cannot compute $\mathcal{Q}_\text{appx}$ explicitly.
Instead, we simply compute the size of $\mathcal{Q}_\text{appx}$.
We have $|\mathcal{Q}_\text{appx}| = 4+|\mathcal{Q}^*|+ \sum_{(i,j) \in P} |\mathcal{Q}_{i,j}^*|$, where $|\mathcal{Q}^*|$ and $|\mathcal{Q}_{i,j}^*|$ can be obtained by querying the sub-structures $\mathcal{D}_{i^-,\bullet},\mathcal{D}_{i^+,\bullet},\mathcal{D}_{\bullet,j^-},\mathcal{D}_{\bullet,j^+}$ and $\mathcal{D}_{i,j}$'s.
By storing $|\mathcal{Q}_\text{appx}|$, we can answer the size query in $O(1)$ time.
In order to answer membership queries, we need some extra work.
The main difficulty is that one quadrant $Q \in \mathcal{Q}^*$ may belong to many $\mathcal{Q}_{i,j}^*$'s, but we cannot afford recursively querying all sub-structures $\mathcal{D}_{i,j}$.
To overcome this difficulty, the idea is to store the special quadrants in $\mathcal{Q}_{i,j}^*$'s separately.
Recall that $\mathcal{Q}_{i,j}$ consists of all nontrivial quadrants in $\mathcal{Q}$ whose vertices are in $\Box_{i,j}$ and four special quadrants $Q_\leftarrow^{\Box_{i,j}},Q_\rightarrow^{\Box_{i,j}},Q_\uparrow^{\Box_{i,j}},Q_\downarrow^{\Box_{i,j}}$.
We collect all special quadrants in $\mathcal{Q}_{i,j}^*$ for $(i,j) \in P$, the number of which is at most $4|P| = O(r^2)$.
We then store these special quadrants in a binary search tree $T$ which can support membership queries.
To answer a membership query $Q \in \mathcal{Q}$, we first compute its multiplicity in $\{Q_\uparrow,Q_\downarrow,Q_\leftarrow,Q_\rightarrow\} \sqcup \mathcal{Q}^*$, which can be done by $O(1)$ recursive membership queries on the sub-structures $\mathcal{D}_{i^-,\bullet},\mathcal{D}_{i^+,\bullet},\mathcal{D}_{\bullet,j^-},\mathcal{D}_{\bullet,j^+}$.
Then it suffices to compute the multiplicity of $Q$ in $\bigsqcup_{(i,j) \in P} \mathcal{Q}_{i,j}^*$.
Note that although there can be many $\mathcal{Q}_{i,j}$'s containing $Q$, all of them contain $Q$ as a special quadrant except the one containing the vertex of $Q$.
So we only need to query $T$ to obtain the multiplicity of $Q$ contained $\mathcal{Q}_{i,j}^*$'s as a special quadrant, and recursively query the cell sub-structure $\mathcal{D}_{i,j}$ for the cell $\Box_{i,j}$ that contains the vertex of $Q$.
Handling reporting queries is easy and is similar to that for dynamic interval set cover presented in Section~\ref{sec-unweightedint}.
We first report the four quadrants $Q_\uparrow,Q_\downarrow,Q_\leftarrow,Q_\rightarrow$, and then report the quadrants in $\mathcal{Q}^*$ and $\mathcal{Q}_{i,j}^*$'s by recursively querying the sub-structures which maintain nonempty solutions.

Now we analyze the query time.
If the solution $\mathcal{Q}_\text{appx}$ is computed by the algorithm of Lemma~\ref{lem-quadoslong}, then it is stored in a binary search tree and we can answer a size query, a membership query, and a reporting query in $O(1)$ time, $O(\log |\mathcal{Q}_\text{appx}|)$ time, and $O(|\mathcal{Q}_\text{appx}|)$ time, respectively.
So it suffices to consider the case where we construct $\mathcal{Q}_\text{appx}$ using Equation~\ref{eq-qappx}.
In this case, answering a size query still takes $O(1)$, because we explicitly compute $|\mathcal{Q}_\text{appx}|$.
To answer a membership query $Q \in \mathcal{Q}$, we need to do $O(1)$ recursive queries on the sub-structures (the cell $\Box_{i,j}$ containing the vertex of $Q$ can be found in $O(\log r)$ time using the point location data structure $\mathcal{L}$).
Besides, we need to query the binary search tree $T$ that stores special quadrants, which takes $O(\log r)$ time.
All the other work takes $O(1)$ time.
Note that the instances maintained in the sub-structures $\mathcal{D}_{i,\bullet},\mathcal{D}_{\bullet,j},\mathcal{D}_{i,j}$ have size $O(n/r)$.
So if we use $Q(n)$ to denote the time cost for a membership query when the size of the instance is $n$, we have the recurrence $Q(n) = O(1) \cdot Q(O(n/r)) + O(\log r)$, which solves to $Q(n) = 2^{O(\log_r n)} \log r$.
Finally, to answer a reporting query, we first report the elements of $\{Q_\uparrow,Q_\downarrow,Q_\leftarrow,Q_\rightarrow\}$ and recursively query the relevant sub-structures which maintain nonempty solutions.
Thus, in the recurrence tree, the number of leaves is bounded by $|\mathcal{Q}_\text{appx}|$ since at each leaf node we need to report at least one element.
Since the height of the recurrence tree is $O(\log_r n)$ and at each node of the recurrence tree the work can be done in $O(1)$ time, the overall time cost for a reporting query is $O(|\mathcal{Q}_\text{appx}| \cdot \log_r n)$.

\subsubsection*{Correctness.}
First, we show that $\mathcal{D}$ makes a no-solution decision iff the current instance $(S,\mathcal{Q})$ does not have a feasible set cover.
The ``if'' part is clear because the set $\mathcal{Q}_\text{appx}$ we construct is always a set cover for $(S,\mathcal{Q})$.
To see the ``only if'' part, we notice there are two points that $\mathcal{D}$ can make a no-solution decision.
The first point is when the sub-instance $(S_{i,j},\mathcal{Q}_{i,j})$ has no set cover for some cell $\Box_{i,j}$ for $i^- < i < i^+$ and $j^- < j < j^+$ that is not covered by any single quadrant $Q \in \mathcal{Q}$.
In this case, there is some point $a \in S_{i,j}$ that cannot be covered by any quadrant in $\mathcal{Q}_{i,j}$.
Note that $\mathcal{Q}_{i,j}$ contains all nontrivial quadrants in $\mathcal{Q}$ which partially intersect $\Box_{i,j}$ and the intersection is maximal (among all quadrants in $\mathcal{Q}$ that partially intersect $\Box_{i,j}$).
Therefore, $a$ cannot be covered by any nontrivial quadrant in $\mathcal{Q}$ which partially intersects $\Box_{i,j}$ (and all nontrivial quadrant in $\mathcal{Q}$ intersecting $\Box_{i,j}$ must intersect $\Box_{i,j}$ partially).
Also, $a$ cannot be covered by any trivial quadrant in $\mathcal{Q}$ because $a \in \Box_{i,j} \subseteq X$.
%Since $\Box_{i,j}$ is uncoverable, any quadrant in $\mathcal{Q}$ intersecting $\Box_{i,j}$ must intersect $\Box_{i,j}$ partially, which implies $a$ cannot be covered by any quadrant in $\mathcal{Q}$.
So the no-solution decision made here is correct.
The second point is when constructing $\mathcal{Q}^*$.
It is easy to see that if any of the sub-structures $\mathcal{D}_{i^-,\bullet},\mathcal{D}_{i^+,\bullet},\mathcal{D}_{\bullet,j^-},\mathcal{D}_{\bullet,j^+}$ reports ``no solution'', then $(S,\mathcal{Q})$ has no set cover.
For example, if $\mathcal{D}_{i^-,\bullet}$ reports ``no solution'', then the points in $S_{i^-,\bullet}$ cannot be covered by the quadrants in $\mathcal{Q}_{i^-,\bullet} \cup \{Q_\uparrow,Q_\leftarrow,Q_\rightarrow\}$ and thus the points in $L_\uparrow$ cannot be covered by the quadrants in $\mathcal{Q}$.
%If one of the instances $(S_{i^-,\bullet},\mathcal{Q}_{i^-,\bullet} \cup \{Q_\uparrow,Q_\leftarrow,Q_\rightarrow\})$, $(S_{i^+,\bullet},\mathcal{Q}_{i^+,\bullet} \cup \{Q_\downarrow,Q_\leftarrow,Q_\rightarrow\})$, $(S_{\bullet,j^-},\mathcal{Q}_{\bullet,j^-} \cup \{Q_\uparrow',Q_\downarrow',Q_\leftarrow\})$, $(S_{\bullet,j^+},\mathcal{Q}_{\bullet,j^+} \cup \{Q_\uparrow',Q_\downarrow',Q_\rightarrow\})$ does not have a feasible set cover, $\mathcal{D}$ makes a no-solution decision.
%Note that if one of these instances have no set cover, then $(S,\mathcal{Q})$ has no set cover.
%For example, if $(S_{i^-,\bullet},\mathcal{Q}_{i^-,\bullet} \cup \{Q_\uparrow,Q_\leftarrow,Q_\rightarrow\})$ has no set cover, then $S_{i^-,\bullet}$ cannot be covered by $\mathcal{Q}$, because $\mathcal{Q}_{i^-,\bullet} \cup \{Q_\uparrow,Q_\leftarrow,Q_\rightarrow\}$ contains all quadrants in $\mathcal{Q}$ whose intersection with $R_{i^-}$ is maximal.
Therefore, the no-solution decision made here is correct.

Now it suffices to show $\mathcal{Q}_\text{appx}$ is a $(\mu+\varepsilon)$-approximate solution for $(S,\mathcal{Q})$.
If $\mathcal{Q}_\text{appx}$ is constructed by the algorithm of Lemma~\ref{lem-quadoslong}, then it is a $\mu$-approximate solution.
So suppose $\mathcal{Q}_\text{appx}$ is constructed using Equation~\ref{eq-qappx}.
Let $\mathsf{opt}$ be the size of an optimal set cover for $(S,\mathcal{Q})$.
Our key observation is the following.
%We observe the following fact.

\begin{lemma}
$|\mathcal{Q}_\textnormal{appx}| \leq (\mu+\tilde{\varepsilon}) \cdot \mathsf{opt} + O(r^2)$.
\end{lemma}
\begin{myproof}
Let $\mathcal{Q}_\textnormal{opt}$ be an optimal set cover of $(S,\mathcal{Q})$.
Define $n_\uparrow,n_\downarrow,n_\leftarrow,n_\rightarrow$ as the number of quadrants in $\mathcal{Q}_\textnormal{opt}$ whose vertices are in $L_\uparrow,L_\downarrow,L_\leftarrow, L_\rightarrow$, respectively.
Also, define $n_{i,j}$ as the number of quadrants in $\mathcal{Q}_\textnormal{opt}$ whose vertices are in $\Box_{i,j}$.
We have $\mathsf{opt} = |\mathcal{Q}_\textnormal{opt}| \geq n_\uparrow + n_\downarrow + n_\leftarrow + n_\rightarrow + \sum_{i=i^-}^{i^+} \sum_{j=j^-}^{j^+} n_{i,j}$.
On the other hand, by Equation \ref{eq-qappx}, we have 
\begin{equation*}
    \begin{aligned}
        |\mathcal{Q}_\textnormal{appx}| & = 4 + |\mathcal{Q}^*| + \sum_{i=i^-}^{i^+}  \sum_{j=j^-}^{j^+} |\mathcal{Q}_{i,j}^*| \\
        & = 4 + |\mathcal{Q}_\uparrow^*| + |\mathcal{Q}_\downarrow^*| + |\mathcal{Q}_\leftarrow^*| + |\mathcal{Q}_\rightarrow^*| + \sum_{j=j^-}^{j^+} |\mathcal{Q}_{i,j}^*|.
    \end{aligned}
\end{equation*}
We show that $|\mathcal{Q}_\uparrow^*| \leq (\mu+\tilde{\varepsilon}) \cdot (n_\uparrow + 7)$, $|\mathcal{Q}_\downarrow^*| \leq (\mu+\tilde{\varepsilon}) \cdot (n_\downarrow + 7)$, $|\mathcal{Q}_\leftarrow^*| \leq (\mu+\tilde{\varepsilon}) \cdot (n_\leftarrow + 7)$, $|\mathcal{Q}_\rightarrow^*| \leq (\mu+\tilde{\varepsilon}) \cdot (n_\rightarrow + 7)$, and $|\mathcal{Q}_{i,j}^*| \leq (\mu+\tilde{\varepsilon}) \cdot (n_{i,j} + 4)$ for $i^- < i < i^+$ and $j^- < j < j^+$, which implies the inequality in the lemma.
All these inequalities are proved similarly, so we only show $|\mathcal{Q}_\uparrow^*| \leq (\mu+\tilde{\varepsilon}) \cdot (n_\uparrow + 7)$ here.
Recall that $\mathcal{Q}_\uparrow^*$ is the solution maintained by $\mathcal{D}_{i^-,\bullet}$ for $(S_{i^-,\bullet},\mathcal{Q}_{i^-,\bullet} \cup \{Q_\uparrow,Q_\leftarrow,Q_\rightarrow\})$, excluding the quadrants $Q_\uparrow,Q_\leftarrow,Q_\rightarrow$.
Now we create a set of at most $n_\uparrow + 7$ quadrants in $\mathcal{Q}_{i^-,\bullet} \cup \{Q_\uparrow,Q_\leftarrow,Q_\rightarrow\}$, which consists of all quadrants in $\mathcal{Q}_\textnormal{opt}$ whose vertices are in $L_\uparrow$, the (up to) four maximal nontrivial quadrants in $\mathcal{Q}$ that top, bottom, left, right intersect $R_{i^-}$, and the three quadrants $Q_\uparrow,Q_\leftarrow,Q_\rightarrow$.
These $n_\uparrow + 7$ quadrants cover all points in $S_{i^-,\bullet}$, because $\mathcal{Q}_\textnormal{opt}$ is a set cover of $(S,\mathcal{Q})$.
Since $\mathcal{D}_{i^-,\bullet}$ maintains a $(\mu+\varepsilon)$-approximate solution for $(S_{i^-,\bullet},\mathcal{Q}_{i^-,\bullet} \cup \{Q_\uparrow,Q_\leftarrow,Q_\rightarrow\})$, we have $|\mathcal{Q}_\uparrow^*| \leq (\mu+\tilde{\varepsilon}) \cdot (n_\uparrow + 7)$.
\end{myproof}

\noindent
Recall that we construct $\mathcal{Q}_\text{appx}$ using Equation~\ref{eq-qappx} only when $\mathsf{opt} \geq \delta \geq c (r^2 + \varepsilon r^2) /(\varepsilon - \alpha \varepsilon)$.
By the above lemma and the fact that $c$ is sufficiently large, we have
\begin{equation*}
    |\mathcal{Q}_\text{appx}| - (\mu+\tilde{\varepsilon}) \cdot \mathsf{opt} \leq c'\cdot r^2 \leq (\varepsilon - \alpha \varepsilon) \mathsf{opt} = (\varepsilon - \tilde{\varepsilon}) \mathsf{opt},
\end{equation*}
which implies $|\mathcal{Q}_\textnormal{appx}| \leq (\mu+\varepsilon) \cdot \mathsf{opt}$. 

\subsubsection*{Update time.}
To analyze the update time of our data structure $\mathcal{D}$, it suffices to consider the first period (including the first reconstruction).
The first period consists of $n_0/r$ operations, where $n_0$ is the size of the initial $(S,\mathcal{Q})$.
The size of $(S,\mathcal{Q})$ during the first period is always in between $(1-1/r) n_0$ and $(1+1/r) n_0$ and is hence $\Theta(n_0)$ (later we shall choose a super-constant $r$).
We first observe that, excluding the recursive updates for the sub-structures, each update of $\mathcal{D}$ takes $\widetilde{O}(r^2/(\varepsilon - \alpha \varepsilon) + r^2 \cdot 2^{O(\log_r n)})$ (amortized) time.
Updating the support data structures can be done in $\widetilde{O}(1)$ time.
When constructing the solution $\mathcal{Q}_\textnormal{appx}$, we need to simulate the algorithm of Lemma~\ref{lem-quadoslong} within $\widetilde{O}(\delta) = \widetilde{O}(r^2/(\varepsilon - \alpha \varepsilon))$ time.
If $\mathcal{Q}_\textnormal{appx}$ is defined using Equation~\ref{eq-qappx}, we need to do some extra work.
First, we need to obtain the quadrants $Q_\uparrow,Q_\downarrow,Q_\leftarrow,Q_\rightarrow$, which can be done in $\widetilde{O}(1)$ time using the support data structure $\mathcal{B}$.
Then we need to compute $|\mathcal{Q}^*|$, which involves $O(1)$ size queries to the sub-structures $\mathcal{D}_{i^-,\bullet},\mathcal{D}_{i^+,\bullet},\mathcal{D}_{\bullet,j^-},\mathcal{D}_{\bullet,j^+}$ and hence takes $O(1)$ time.
Finally, we need to compute $|\mathcal{Q}_{i,j}^*|$ and retrieve the special quadrants in $\mathcal{Q}_{i,j}^*$ for all $(i,j) \in P$ (and build the query structure for these special quadrants).
To check whether a cell $\Box_{i,j}$ is coverable can be done in $\widetilde{O}(1)$ time using the support data structure $\mathcal{C}$.
After knowing whether each cell is coverable, to compute $|\mathcal{Q}_{i,j}^*|$ and retrieve the special quadrants can be done via $O(r^2)$ size and membership queries to the sub-structures $\mathcal{D}_{i,j}$, which takes $\widetilde{O}(r^2 \cdot 2^{O(\log_r n)})$ time.
The reconstruction of $\mathcal{D}$ takes $\widetilde{O}(2^{O(\log_r n)} \cdot r)$ amortized time, because the time cost of the (first) reconstruction is $\widetilde{O}(2^{O(\log_r n_1)} \cdot n_1)$, i.e., $\widetilde{O}(2^{O(\log_r n_0)} \cdot n_0)$, while the first period consists of $O(n_0/r)$ operations.
%This includes calculating $|\mathcal{Q}_\textnormal{appx}$

Next, we consider the recursive updates of the sub-structures.
Similarly to the analysis of our dynamic interval set cover data structure, we distinguish three types of updates according to the current operation.
The first type is \textit{point update}, which is caused by the insertion/deletion of a point in $S$.
The second type is \textit{trivial quadrant update} (or \textit{trivial update} for short), which is caused by the insertion/deletion of a trivial quadrant in $\mathcal{Q}$ (recall that a quadrant is trivial if its vertex is outside the point range $[0,1]^2$ of $(S,\mathcal{Q})$).
The third type is \textit{nontrivial quadrant update} (or \textit{nontrivial update} for short), which is caused by the insertion/deletion of a nontrivial quadrant in $\mathcal{Q}$.
We first consider the recursive updates required for all three types of updates.
Recall that when constructing the solution $\mathcal{Q}_\text{appx}$ using Equation~\ref{eq-qappx}, we need to temporarily insert some virtual quadrants to $\mathcal{Q}_{i^-,\bullet},\mathcal{Q}_{i^+,\bullet},\mathcal{Q}_{\bullet,j^-},\mathcal{Q}_{\bullet,j^+}$ (and delete them afterwards).
This involves a constant number of recursive updates, which are all trivial updates because the virtual quadrants inserted are all trivial.
Besides these recursive updates, a point update requires three recursive (point) updates, because the insertion/deletion of a point in $S$ changes one cell sub-instance, one row sub-instance, and one column sub-instance.
A trivial update does not require any other recursive updates, because the insertion/deletion of trivial quadrant in $\mathcal{Q}$ does not change any sub-instance.
Finally, we consider a nontrivial update.
Let $Q$ be the nontrivial quadrant inserted/deleted and suppose the vertex of $Q$ is contained in the cell $\Box_{i,j}$.
Then we may need to update the cell sub-structures $\mathcal{D}_{i,1},\dots,\mathcal{D}_{i,r}$ and $\mathcal{D}_{1,j},\dots,\mathcal{D}_{r,j}$, in which the update of $\mathcal{D}_{i,j}$ is a nontrivial update and the others are all trivial updates since the vertex of $Q$ is outside the point ranges of all cell sub-instances except $(S_{i,j},\mathcal{Q}_{i,j})$.
Also, we need to update the row (resp., column) sub-structures, in which the update of $\mathcal{D}_{i,\bullet}$ (resp., $\mathcal{D}_{\bullet,j}$) are nontrivial updates and the others are all trivial updates.
To summarize, a point update requires $O(1)$ recursive point updates and $O(1)$ recursive trivial updates, a trivial update requires $O(1)$ recursive trivial updates, and a nontrivial update requires $O(1)$ recursive nontrivial updates and $O(r)$ recursive trivial updates.
The depth of the recursion is $O(\log_r n)$.
If we set $\alpha = 1-1/\log_r n$, the approximation factor parameter is $\Theta(\varepsilon)$ in any level of the recurrence.
Let $U(n)$, $U_1(n)$, $U_2(n)$ denote the time costs of a point update, a trivial update, a nontrivial update, respectively, when the size of the current instance is $n$.
The we have the recurrences
\begin{equation*}
    \begin{aligned}
        U(n) &= O(1) \cdot U(O(n/r)) +  O(1) \cdot U_1(O(n/r)) + \widetilde{O}(r^2/\varepsilon + r^2 \cdot 2^{O(\log_r n)}), \\
        U_1(n) &= O(1) \cdot U_1(O(n/r)) + \widetilde{O}(r^2/\varepsilon + r^2 \cdot 2^{O(\log_r n)}), \\
        U_2(n) &= O(1) \cdot U_2(O(n/r)) +  O(r) \cdot U_1(O(n/r)) + \widetilde{O}(r^2/\varepsilon + r^2 \cdot 2^{O(\log_r n)}).
    \end{aligned}
\end{equation*}
%where $\tilde{\varepsilon} = \varepsilon - \varepsilon/\log n$.
%The depth of each recurrence is $O(\log_r n)$.
%The value of ``$\varepsilon$'' in the $i$-th level of the recursion is greater than $\varepsilon - \varepsilon i /\log n$.
%If we let $r$ be sufficiently large, the value of ``$\varepsilon$'' in every level of the recursion will be at least half of the initial $\varepsilon$.
The recurrence for $U_1(n)$ solves to $U_1(n) = \widetilde{O}((r^2/\varepsilon) \cdot 2^{O(\log_r n)})$.
Based on this, we further solve the recurrences for $U(n)$ and $U_2(n)$, and obtain $U(n) = \widetilde{O}((r^2/\varepsilon) \cdot 2^{O(\log_r n)})$ and $U_2(n) = \widetilde{O}((r^3/\varepsilon) \cdot 2^{O(\log_r n)})$.
%Now suppose $r$ is at least a sufficiently large constant, so the depth of the recurrences is $\log n$.
Setting $r = 2^{O(\sqrt{\log n})}$, the amortized update time of $\mathcal{D}$ is then $\widetilde{O}((r^3/\varepsilon) \cdot 2^{O(\log_r n)}) = 2^{O(\sqrt{\log n})}$.
The time cost of a membership query is $2^{O(\log_r n)} = 2^{O(\sqrt{\log n})}$, while the time cost of a reporting query is $O(|\mathcal{Q}_\text{appx}| \cdot \log_r n) = O(|\mathcal{Q}_\text{appx}| \cdot \sqrt{\log n})$.
Also, the construction time is $\widetilde{O}(2^{O(\log_r n)} \cdot n) = 2^{O(\sqrt{\log n})} \cdot n$.
We conclude the following.

%When inserting/deleting a point, we only change one $S_{i,j}$, one $S_{i,\bullet}$, and one $S_{\bullet,j}$.
%So we need to update at most three sub-structures.

%A \textit{maximal} 
%In addition, we define $\mathcal{Q}_{i,j}$ as the set of the nontrivial quadrants $Q \in \mathcal{Q}$ that \textit{partially intersect} $\Box_{i,j}$ (i.e., $Q \cap \Box_{i,j} \neq \emptyset$ and $\Box_{i,j} \nsubseteq Q$), and define $\mathcal{Q}_{i,\bullet}$ (resp., $\mathcal{Q}_{\bullet,j}$) as the set of the quadrants $Q \in \mathcal{Q}$.
%\unweightedusq*

\begin{theorem}
There exists a dynamic data structure for $O(1)$-approximate unweighted unit-square set cover with $2^{O(\sqrt{\log n})}$ amortized update time and $2^{O(\sqrt{\log n})} \cdot n$ construction time, which can answer size, membership, and reporting queries in $O(1)$, $2^{O(\sqrt{\log n})}$, and $O(k \sqrt{\log n})$ time, respectively, where $n$ is the size of the instance and $k$ is the size of the maintained solution.
\end{theorem}
\section{Unweighted Square Set Cover}
In this section, we present a data structure for $O(1)$-approximate dynamic (axis-aligned) square set cover, improving the previous near $O(n^{2/3})$ time result by Chan and He \cite{chan2021dynamic}.

Let $(S,\mathcal{I})$ be a dynamic (unweighted) square set cover instance where $S$ is the set of points in $\mathbb{R}^2$ and $\mathcal{I}$ is the set of squares. Let $n=|S|+|\mathcal{I}|$ denote the total number of points and squares.

\subsection{Algorithm for small opt.}\label{square:small opt} Based on the randomized multiplicative weight update (MWU) method~\cite{agarwal2014near,bronnimann1995almost,clarkson1993algorithms}, Chan and He \cite{chan2021dynamic} provided an $O(1)$-approximation algorithm with $\tilde{O}(\mathsf{opt}^2)$ query time and $\tilde{O}(1)$ update time, assuming the points and objects have been preprocessed in standard range searching data structures. When $\mathsf{opt}$ is small, this algorithm runs in sublinear time. We will use this algorithm as a subroutine later.

\begin{lemma}\label{lemma:small squares}
There exists a data structure for the dynamic set cover problem for $O(m)$ points and $O(n)$ squares in 2D that supports updates in $\tilde{O}(1)$ time and can find an $O(1)$-approximate solution with high probability (w.h.p.)\ in $\tilde{O}(\mathsf{opt}^2)$ time.
\end{lemma}

\subsection{Algorithm for large opt.}
When $\mathsf{opt}$ is large, we can afford a larger additive error. The previous paper \cite{chan2021dynamic} utilized this observation and used a quadtree to partition the problem into subproblems, paying $O(1)$ additive error per subproblem when combining the solutions. We refine their approach and further improve the update time to near $O(n^{1/2})$.

\subsubsection*{Previous data structures.} Our data structure is based on the previous data structure by \cite{chan2021dynamic}, so we will first briefly redescribe their approach, and then introduce our new ideas. For simplicity, assume all coordinates are integers bounded by $U=\mathrm{poly}(n)$. This assumption can be removed using the technique in \cite{chan2021dynamic}, namely by using the BBD tree. We also assume an $O(1)$-approximation $t$ of $\mathsf{opt}$ is known, by running our algorithm for all possible guesses $t=2^i$ in parallel (our algorithm is able to detect whether the guess is wrong).

Their key idea is to construct a standard quadtree, starting with a bounding square cell and recursively divide into four square cells. We stop subdividing when a leaf cell $\Gamma$ has size at most $b$, for a parameter $b$ to be set later, where the size of $\Gamma$ is defined as the total number of points in $S$ and vertices of squares in $\mathcal{I}$ that are inside $\Gamma$. This yields $O(\frac{n}{b})$ cells per level, and thus $O(\frac{n}{b}\log U)$ cells in total.

\begin{figure}[!htbp]\centering
\includegraphics[scale=0.7]{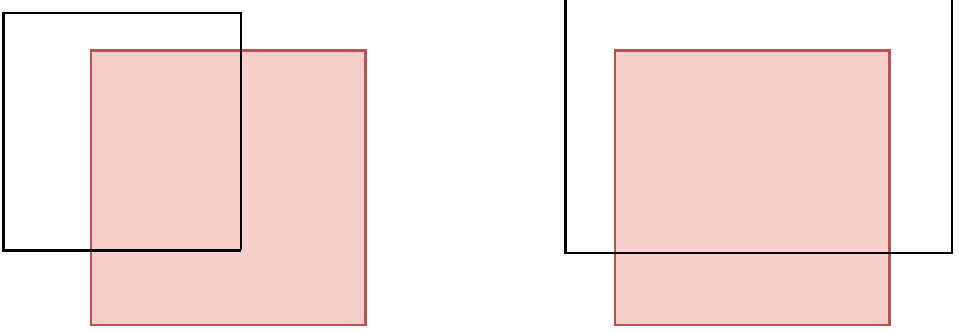}\\
 \caption{Short square (left) and long square (right). The quadtree cell is shaded.}\label{fig:long}
\end{figure}

Since the quadtree cells are also squares, there are only two types of intersections between a quadtree cell $\Gamma$ and an input square in $\mathcal{I}$. Call a square $s$ \emph{short} in the cell $\Gamma$, if at least one of its vertices is in $\Gamma$, otherwise \emph{long}, as shown in Fig.~\ref{fig:long}. A key observation is that it suffices to keep (at most) $4$ ``maximal'' long squares in each cell, since their union covers the union of all long squares of the cell. We use $\mathcal{M}_\Gamma$ to denote the set of maximal long squares in the cell $\Gamma$.

\begin{figure}[htbp]\centering
\includegraphics[scale=0.7]{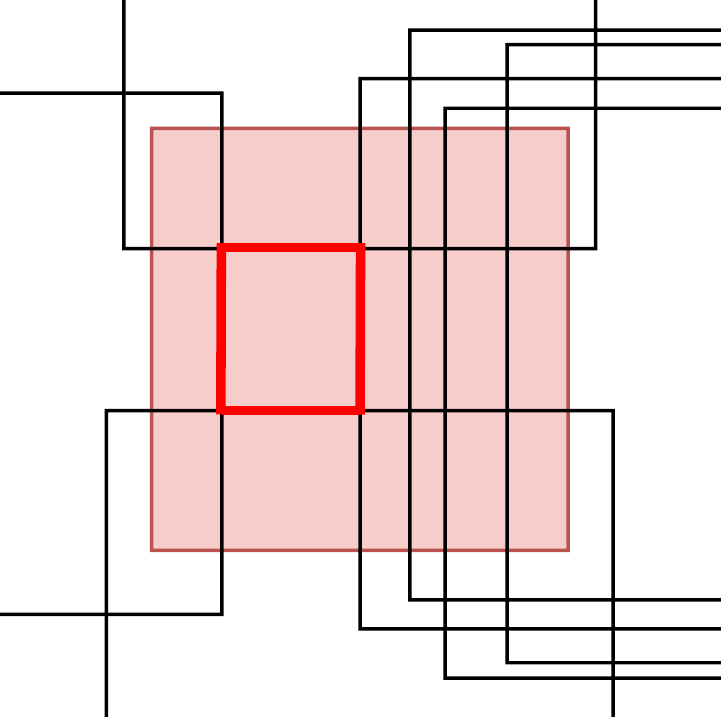}\\
\caption{The complement region of the (at most) $4$ maximal long squares of a cell.}\label{fig:long complement}
\end{figure}

Now we only need to maintain a data structure $\mathcal{D}_\Gamma$ for each leaf cell $\Gamma$, that supports the following type of query:
\begin{quote}
Given any query rectangle $r$ in $\Gamma$, compute an $O(1)$-approximate set cover solution for the points in $S\cap r$, using only the short squares in $\Gamma$.
\end{quote}
To compute an approximate set cover solution for $\Gamma$, it suffices to first include the $4$ maximal long squares $\mathcal{M}_\Gamma$ of $\Gamma$ in the approximate solution (thus paying $O(1)$ additive error), and then query for an $O(1)$-approximate solution $\mathcal{I}_{\mathrm{appx}_\Gamma}$ in the complement region $\Box_\Gamma$ of $\mathcal{M}_\Gamma$, which is a rectangle as shown in Figure~\ref{fig:long complement}, using only the short squares. $\mathcal{D}_\Gamma$ is implemented using 2D range trees \cite{AgaEriSURV,BerBOOK} with branching factor $a=b^{\delta}$ built on the points within $\Gamma$, where $\delta$ is a sufficiently small constant. In this way, we form a set of canonical rectangles with total size $O(a^{O(1)}b(\log_a b)^2)=O(b^{1+O(\delta)})$, such that any query rectangle in the cell $\Gamma$ can be decomposed into $O((\log_a b)^2)=(\frac{1}{\delta})^{O(1)}=O(1)$ canonical rectangles. Then $\mathcal{I}_{\mathrm{appx}_\Gamma}$ is obtained by taking the union of the $O(1)$-approximate solutions in the $O(1)$ canonical rectangles that $\Box_\Gamma$ decomposes to, and doing this only loses a constant factor.

The global approximate solution $\mathcal{I}_{\mathrm{appx}}$ is formed by taking the union of the approximate solutions in each leaf cell $\Gamma$, which is $\mathcal{I}_{\mathrm{appx}_\Gamma}$ plus the at most $4$ maximal long squares of $\Gamma$, i.e., we let $\mathcal{I}_{\mathrm{appx}}=\bigsqcup_{\Gamma} (\mathcal{I}_{\mathrm{appx}_\Gamma}\sqcup \mathcal{M}_\Gamma)$.

To analyze the approximation factor, let $\mathcal{I}_{\mathsf{opt}_{\Gamma}}$ contain the squares in the optimal solution that are short in the cell $\Gamma$. We have $\sum_\Gamma |\mathcal{I}_{\mathsf{opt}_{\Gamma}}|\leq 4\cdot \mathsf{opt}$, since an input square is short in at most $4$ leaf cells containing its $4$ vertices. The size of our approximate solution $\mathcal{I}_{\mathrm{appx}}$ can be upper-bounded as follows:

$$|\mathcal{I}_{\mathrm{appx}}|\leq \sum_{\Gamma}
\left(|\mathcal{I}_{\mathrm{appx}_\Gamma}|+|\mathcal{M}_\Gamma|\right)\leq \sum_{\Gamma}\left(O\left(|\mathcal{I}_{\mathsf{opt}_\Gamma}|\right)+4\right)\leq O(\mathsf{opt})+\tilde{O}\left(\frac{n}{b}\right).$$
%This gives additive error $\tilde{O}(\frac{n}{b})$, i.e.\ we have $|\mathcal{I}_{\mathrm{appx}}|\leq 4\cdot \mathsf{opt}+\tilde{O}(\frac{n}{b})$.

As long as we set $b=\tilde{\Omega}(\frac{n}{\mathsf{opt}})$, this is an $O(1)$-approximation.

\subsubsection*{New approach.} Now we describe the parts that we will change. In the previous algorithm of \cite{chan2021dynamic}, an $O(1)$-approximate solution within each canonical rectangle in each cell is maintained, using the static $O(1)$-approximate set cover algorithm with near-linear running time \cite{agarwal2014near,ChanH20}. An observation from \cite{chan2021dynamic} is that for a canonical rectangle $r$ with size $b_i$, although there may exist a lot of squares that cut across $r$, it suffices to keep only $O(b_i)$ ``maximal'' long squares with respect to $r$ among them, which can be found in $\tilde{O}(b_i)$ time using range searching. So the running time for the static algorithm is $\tilde{O}(b_i)$.

To further improve the update time, our new idea is to classify the canonical rectangles into two categories, based on their sizes. Call a canonical rectangle \emph{heavy}, if its size exceeds $g$, otherwise \emph{light}, where $g$ is a parameter to be set later.

For each heavy canonical rectangle $r$ with size $g'\geq g$, we maintain another level of subquadtree, using the previous algorithm as stated before, which subdivides it into $\tilde{O}(\lambda)$ subcells each with size $O(\frac{g'}{\lambda})$, where $\lambda$ is a parameter to be set later. For each subcell $\Lambda$ in the subquadtree, maintain the set $\mathcal{M}_\Lambda$ of the at most $4$ maximal long squares, and an $O(1)$-approximate set cover solution $\mathcal{I}_{\mathrm{appx}_\Lambda}$ for the complement region.
%Record the size of the approximate solution in $r$.

For the light canonical rectangles, we don't maintain the approximate solution, but rather choose to compute it from scratch during the query.

For each cell $\Gamma$, the data structure $\mathcal{D}_\Gamma$ can be constructed in $O(b^{1+O(\delta)})$ time, since the total size of the heavy canonical rectangles in the cell is at most $O(b^{1+O(\delta)})$, and the static approximate set cover algorithm runs in near-linear time.

%since there are at most $O(\frac{b^{1+O(\delta)}}{g})$ heavy canonical rectangles in the cell.

\subsubsection*{Update.} When we insert or delete a square $s$, for each leaf cell $\Gamma$ that contains a vertex of $s$, we update the subquadtree for each heavy canonical rectangle in $\Gamma$. For a heavy canonical rectangle with size $g'\geq g$, this takes $O((\lambda+\frac{g'}{\lambda})\cdot g'^{O(\delta)})$ time (as analyzed in the previous paper). Summing over all heavy canonical rectangles, the total time is at most $O((\lambda\cdot \frac{b}{g}+\frac{b}{\lambda})\cdot b^{O(\delta)})$, as the total size of the (heavy) canonical rectangles is bounded by $O(b^{1+O(\delta)})$. We also update the set of maximal long squares $\mathcal{M}_\Gamma$ for each leaf cell $\Gamma$, in $\tilde{O}(\frac{n}{b})$ time.
%in $\tilde{O}(b^{1+O(\delta)})$ time

When we insert or delete a point $p$, we update the data structure $\mathcal{D}_\Gamma$ for the leaf cell $\Gamma$ containing $p$. For each canonical rectangle $r$ in $\Gamma$ that contains $p$, we may need to update $O(1)$ maximal long squares that cut across $r$, which takes $O((\lambda+\frac{b}{\lambda})\cdot b^{O(\delta)})$ time for the heavy ones, and $O(b^{O(\delta)})$ time for the light ones.

\subsubsection*{Query.} When we perform a query, we need to compute the approximate solution $\mathcal{I}_{\mathrm{appx}_\Gamma}$ in the complement region $\Box_\Gamma$ of the maximal long squares for each cell $\Gamma$. The region $\Box_\Gamma$ can be decomposed into $O(1)$ canonical rectangles. For each light canonical rectangle $r$, let $\mathsf{opt}_i$ denote the size of the optimal solution, we compute an $O(1)$-approximate solution from scratch, using either the small $\mathsf{opt}$ algorithm as described earlier in Sec.~\ref{square:small opt} in $\tilde{O}(\mathsf{opt}_i^2)$ time, or the static algorithm \cite{agarwal2014near,ChanH20} in $\tilde{O}(g)$ time (since the size of $r$ is at most $g$), whichever is faster (by running them in parallel). For each heavy canonical rectangle, the size of the precomputed approximate solution (which is implicitly represented by the subquadtree) can be retrieved in $O(1)$ time, but it has additive error $O(\lambda)$. So if the result is $O(\lambda)$, we need to recompute an $O(1)$-approximate solution from scratch, using the small $\mathsf{opt}$ algorithm in $\tilde{O}(\mathsf{opt}_i^2)$ time, which is bounded by $\tilde{O}(\mathsf{opt}_i\cdot \lambda)$ since $\mathsf{opt}_i=O(\lambda)$.

The total query time is
$$\tilde{O}\left(\sum_i \min\{\mathsf{opt}_i^2,g\}+\sum_i \mathsf{opt}_i\cdot \lambda+\frac{n}{b}\right),$$
which is at most $\tilde{O}(\mathsf{opt}\cdot \sqrt{g}+\mathsf{opt}\cdot \lambda+\frac{n}{b})$, since $\sum_i \mathsf{opt}_i=\mathsf{opt}$.

%\paragraph{Combining the algorithms.} 
%Let $t$ be a guess on $\mathsf{opt}$. We will run our algorithm for each possible $t=2^i$ in parallel.

To balance the query and update times, when $\mathsf{opt}\leq \sqrt{n}$, set $b=\frac{n}{\mathsf{opt}}$, $g=\frac{n}{\mathsf{opt}^2}$ and $\lambda=\frac{\sqrt{n}}{\mathsf{opt}}$ (note that the requirements $b\geq g\geq \lambda\geq 1$ and $b=\tilde{\Omega}(\frac{n}{\mathsf{opt}})$ are satisfied); both the query and update time are $O(n^{1/2+\delta})$---interesting, we get the same bound uniformly for all the terms. When $\mathsf{opt}>\sqrt{n}$, the previous algorithm by \cite{chan2021dynamic} already obtains $O(n^{1/2+\delta})$ query and update time, by setting $b=\sqrt{n}$.
%set $b=\sqrt{n}$, $g=$ and $\lambda=$.
%combining the guesses.

The actual solution $\mathcal{I}_{\mathrm{appx}}$ can be reported by taking the union of all solutions in the leaf cells of the quadtree, which takes $\tilde{O}(\mathsf{opt})$ time.

%\paragraph{Reporting the solutions.}

\begin{theorem}
There exists a dynamic data structure for $O(1)$-approximate unweighted square set cover with $O(n^{1/2+\delta})$ query and update time and $O(n^{1+\delta})$ construction time w.h.p., for any constant $\delta>0$.
\end{theorem}

\section{Unweighted 2D Halfplane Set Cover}
In this section, we present a data structure for $O(1)$-approximate dynamic 2D halfplane set cover. Previously, Chan and He \cite{chan2021dynamic} provided a dynamic set cover structure for 3D halfspace with $O(n^{12/13+\delta})$ update time, which clearly also holds for 2D halfplanes, but their scheme is unable to actually find a set cover---it only reports its size. This is because their idea for the ``large $\mathsf{opt}$'' case is based on estimating the size of the solution by summing over a small random sample of the terms. We not only improve the update time, but can also find the approximate set cover solution.

%In this section, we present a data structure for $O(1)$-approximate dynamic 2D halfplane set cover. Previously, Chan and He \cite{chan2021dynamic} provided a data structure for $O(1)$-approximate dynamic 3D halfspace set cover with update time near $O(n^{12/13})$, which clearly can be adapted to work for the 2D case. However, their approach is only able to report the size of the approximate solution, since their idea for the ``large $\mathsf{opt}$'' case is based on estimating the size of the solution by summing over a small random sample of the terms. We not only improve the update time to $O(n^{2/3+\delta})$ (for any constant $\delta>0$), but can also report the actual solution.
%\Qizheng{do we need to work out this exponent? @Timothy}
%Their update time for 2D halfplanes is not explicitly stated in the previous paper, but upon close inspection, the update time seems to be no better than $O(n^{?})$.

Let $(S,\mathcal{H})$ be a dynamic (unweighted) 2D halfplanes set cover instance where $S$ is a set of points in $\mathbb{R}^2$ and $\mathcal{H}$ is a set of halfplanes, with $m=|S|$ and $n=|\mathcal{H}|$. We use $N=n+m$ to denote the global upper bound on the instance size.
%and $M$

\subsection{Algorithm for small opt.}\label{sec:small opt 2D halfplanes}
We first note that there is an algorithm that is efficient when $\mathsf{opt}$ is small, which will be used later as a subroutine. The idea is to modify the small $\mathsf{opt}$ algorithm for axis-aligned squares by Chan and He \cite{chan2021dynamic}, which is based on an efficient implementation of the randomized multiplicative weight update (MWU) method~\cite{agarwal2014near,bronnimann1995almost,ChanH20,clarkson1993algorithms}, using $(\leq b)$-levels and various geometric data structures. Here we briefly redescribe their algorithm, and note the changes that we make in order to work for 2D halfplanes.
%assuming the points and objects have already been processed in standard range searching data structures

%\TIMOTHY{add something about epsilon-nets and halfplanes...}

The following pseudocode shows how the randomized MWU algorithm works.
Here, the \emph{depth} of a point refers to the number of ranges containing it.
An \emph{$\frac 1t$-net} of $S$ is a subset of $S$ that covers all points of depth at least $\frac{|S|}t$.
(The reason that this algorithm yields an $O(1)$-approximation for the 2D halfplane and 3D halfspace cases is that there exists $\frac 1t$-nets of size $O(t)$~\cite{matouvsek1990net}.)

%by Chan and He \cite{ChanH20}
\begin{algorithm}[H]
\caption{MWU for set cover}\label{algo:1}
\begin{algorithmic}[1]
\State Guess a value $t\in [\mathsf{opt},2\,\mathsf{opt}]$.
\State Define a multiset $\hat{\mathcal{H}}$ where each object $i$ in $\mathcal{H}$ initially has multiplicity $m_i=1$.
\Loop \Comment{call this the start of a new \emph{round}}
    \State Fix $\rho:=\frac{c_0 t\log (n+m)}{|\hat{\mathcal{H}}|}$ and take a random sample $R$ of $\hat{\mathcal{H}}$ with sampling probability $\rho$.
        \While {there exists a point $p\in S$ with depth in $R$ at most $\frac{c_0}{2}\log (n+m)$}
                \For {each object $i$ containing $p$} \Comment{call lines 6--8 a \emph{multiplicity-doubling step}}
                    \State Double its multiplicity $m_i$, i.e., insert $m_i$ new copies of object $i$ into $\hat{\mathcal{H}}$. 
                    \State For each copy, independently decide to insert it into $R$ with probability $\rho$.
                \EndFor
                \If {the number of multiplicity-doubling steps in this round exceeds $t$}
                    \State Go to line~3 and start a new round.
                \EndIf
        \EndWhile
    \State Terminate and return a $\frac1{8t}$-net of $R$.
\EndLoop
\end{algorithmic}
\end{algorithm}
To efficiently implement this MWU algorithm, we need to solve two subproblems: 1) finding a low-depth point $p$, and 2) weighted range sampling.

\subsubsection*{Finding a low-depth point.} Let $b:=\frac{c_0}{2}\log (n+m)$ where $c_0$ is a sufficiently large constant. To find a low-depth point in line~5, we compute (from scratch) the $(\leq b)$-level $\mathcal{L}_{\leq b}(R)$ of $R$, which is the collection of all cells in the arrangement of the halfplanes in $R$ of depth at most $b$. It is known that $\mathcal{L}_{\leq b}(R)$ has $O(|R|b)$ cells (after triangulation) and can be constructed in $\tilde{O}(|R|b)$ time \cite{clarkson1989applications}, which is $\tilde{O}(t)$.

To find a point $p\in S$ with depth in $R$ at most $b$, we perform a triangle range query for each (triangulated) cell of $\mathcal{L}_{\leq b}(R)$, to test if the cell contains a point $p\in S$. It is known that we can construct a 2D triangle range searching structure $\tilde{\mathcal{D}}$ \cite{matouvsek1992efficient} on the point set $S$, with $O(\frac{m^{1/2+\delta}}{z^{1/2}})$ query time for emptiness/counting/sampling and $\tilde{O}(z)$ insertion/deletion time, for a given trade-off parameter $z\in [1,m]$. As there are $\tilde{O}(t)$ multiplicity-doubling steps, the total cost is $\tilde{O}(t^2\cdot \frac{m^{1/2+\delta}}{z^{1/2}})$.
%note. the running time can be improved to $\tilde{O}(opt^2)$, because we can decompose the <=b levels into trapezoids unbounded from below/above. each query takes polylog time.

\subsubsection*{Weighted range sampling.} The algorithm is similar to the previous part, but this time we work in the dual. We use $h^*$ to denote the dual point of a halfplane $h$, and $p^*$ denote the dual halfplane of a point $p$.

Let $Q$ be the set of all points $p$ for which we have performed multiplicity-doubling steps thus far. Note that $|Q|=\tilde{O}(t)$. Each time we perform a multiplicity-doubling step, we compute (from scratch) the $(\leq b)$-level $\mathcal{L}_{\leq b}(Q^*)$. The multiplicity of a halfplane $h\in \mathcal{H}$ is equal to $2^{\mbox{\scriptsize depth of $h^*$ in $Q^*$}}$, and all dual points $h^*$ in a cell of ${\cal L}_{\le b}(Q^*)$ share the same multiplicity. It is known that the multiplicities are bounded by $(n+m)^{O(1)}$, so each $h^*$ is covered by $\mathcal{L}_{\leq b}(Q^*)$.
%The multiplicity of $h$ is determined by which cell of ${\cal L}_{\le b}(Q^*)$ the point $h^*$ is in.  

To generate a multiplicity-weighted sample of the halfplanes containing $p$ for line 8, after $p$ has been inserted to $Q$, we examine all cells of ${\cal L}_{\le b}(Q^*)$ contained in $p^*$. For each such cell $\gamma$, we use triangle range counting to compute its size, using a 2D triangle range searching structure $\tilde{\mathcal{D}}^*$ built on the dual point set $\mathcal{H}^*$ as described before. Knowing the sizes and multiplicities for all such $\tilde{O}(t)$ cells, we can then generate the weighted sample in time $O(\frac{n^{1/2+\delta}}{z^{1/2}})$ times the size of the sample, again using the data structure $\tilde{\mathcal{D}}^*$. The random sample $R$ in line 4 with size $\tilde{O}(t)$ is generated similarly.

As we perform $\tilde{O}(t^2)$ triangle range counting queries, and the total size of the samples is $\tilde{O}(t)$, the total cost is $\tilde{O}(t^2\cdot \frac{n^{1/2+\delta}}{z^{1/2}}+t\cdot \frac{n^{1/2+\delta}}{z^{1/2}})=\tilde{O}(t^2\cdot \frac{n^{1/2+\delta}}{z^{1/2}})$.

%Set the trade-off parameter $z=n^{1/3}$. 
%which supports query and update in $O(n^{1/3+\delta})$ time.

\begin{lemma}\label{lemma:small 2D halfplanes v1}
There exists a data structure for the dynamic set cover problem for $O(m)$ points and $O(n)$ 2D halfplanes that supports updates in $\tilde{O}(z)$ time and can find an $O(1)$-approximate solution w.h.p.\ in $\tilde{O}(\mathsf{opt}^2\cdot \frac{(n+m)^{1/2+\delta}}{z^{1/2}})$ time, for any constant $\delta>0$ and trade-off parameter $z\in [1,n+m]$. The data structure can be constructed in $\tilde{O}((n+m)\cdot z)$ time.
\end{lemma}

\subsubsection*{Alternative algorithm.} As an alternative to implement the small $\mathsf{opt}$ algorithm, we can also slightly modify the small $\mathsf{opt}$ algorithm by Chan and He \cite{chan2021dynamic} for halfspaces in 3D (which uses partition trees), and obtain an algorithm for 2D halfplanes. The only difference is that we use partition trees in 2D instead of 3D.

\begin{lemma}\label{lemma:small 2D halfplanes v2}
There exists a data structure for the dynamic set cover problem for $O(m)$ points and $O(n)$ 2D halfplanes that supports updates in $\tilde{O}(1)$ time and can find an $O(1)$-approximate solution w.h.p.\ in $\tilde{O}(\mathsf{opt}\cdot (n+m)^{1/2+\delta})$ time, for any constant $\delta>0$. The data structure can be constructed in $\tilde{O}(n+m)$ time.
\end{lemma}

%\paragraph{Remark.} There are other alternatives to implement the small $\mathsf{opt}$ algorithm. For example, we can also slightly modify the small $\mathsf{opt}$ algorithm by Chan and He \cite{chan2021dynamic} for halfspaces in 3D (which uses partition trees), and obtain an algorithm for 2D halfplanes. The only difference is that we use partition trees in 2D instead of 3D, and obtain update and query time $\tilde{O}(\mathsf{opt}\cdot (n+m)^{1/2+\delta})$ for any constant $\delta>0$. In this way, the running time dependency on $\mathsf{opt}$ is improved. However, in our application we will apply the small $\mathsf{opt}$ algorithm for very small $\mathsf{opt}$ ($\approx N^\delta$), so we care more about the dependency on $n$ and $m$.

\subsection{Main algorithm.}
Here we first present a solution that only supports halfplane insertions and deletions as well as point deletions; the ways to support point insertions are more technical, and will be explained in the last part.

\subsubsection*{Data structures.} To construct the data structure, our idea is to recursively apply Matou\v sek's Partition Theorem \cite{matouvsek1992efficient} as stated below, and decompose the problem into subproblems.

\begin{theorem}[Matou\v{s}ek's Partition Theorem]\label{thm:partition}
Given a set $S$ of $m$ points in $\R^2$, for any positive integer $b\le m$, we can partition $S$ into $b$ subsets $S_i$ each with size $O(\frac{m}{b})$ and find $b$ disjoint triangular cells $\Delta_i\supset S_i$, where each cell is a triangle, such that any halfplane crosses (i.e., the boundary intersects) $O(\sqrt{b})$ cells. The partition can be constructed in $O(m^{1+\delta})$ time.
%($\tilde{O}(m)$?)
\end{theorem}
%It is known that we can construct the partition in $O(m^{1+\delta})$ time.
%for t=n^\delta
%polyhedron with constant complexity

The original version of Matou\v sek's theorem does not guarantee disjointness of cells, but a later version by
Chan~\cite{Chan12DCG} does.

Intuitively, with subproblems defined with this partition, any inserted/deleted halfplane $h$ only affects a small fraction of the subproblems. More precisely, we use Theorem~\ref{thm:partition} 
%\TIMOTHY{consistency: to abbreviate or not?}
to partition the set of points $S$ into $b$ disjoint cells $\Delta_1,\dots,\Delta_b$, each containing a subset $S_i$ of $O(\frac{m}{b})$ points, where $b$ is a parameter to be set later. Any halfplane $h$ will cross $O(\sqrt{b})$ cells, so each cell is crossed by $O(\frac{n\cdot \sqrt{b}}{b})=O(\frac{n}{\sqrt{b}})$ halfplanes in $\mathcal{H}$ on average. Call a cell \emph{good} if it crosses $\leq g\cdot \frac{n}{\sqrt{b}}$ halfplanes in $\mathcal{H}$ (for a sufficiently large constant $g$), otherwise \emph{bad}. The halfplanes in $\mathcal{H}$ cross $O(n\cdot \sqrt{b})$ cells in total, so the total number of bad cells is $O(\frac{n\sqrt{b}}{gn/\sqrt{b}})=O(\frac{b}{g})$.
 
We construct the standard range searching data structures $\tilde{\mathcal{D}}$ and $\tilde{\mathcal{D}}^*$ required by the small $\mathsf{opt}$ algorithm in Section~\ref{sec:small opt 2D halfplanes} on the problem instance $(S,\mathcal{H})$ in $\tilde{O}((n+m)\cdot z)$ time, and maintain an (implicit) $O(1)$-approximate solution $\mathcal{H}_{\mathrm{appx}}$ for $(S,\mathcal{H})$. For each good cell $\Delta_i$, let $\mathcal{H}_i$ be the set of halfplanes crossing $\Delta_i$, we recursively construct a data structure $\mathcal{D}_i$ for the subproblem $(S_i,\mathcal{H}_i)$. If there exists a halfplane that completely contains $\Delta_i$, then record any one of them. We also construct a data structure $\mathcal{D}_{\mathrm{bad}}$ for the subproblem $(S_{\mathrm{bad}},\mathcal{H})$ where $S_{\mathrm{bad}}=\bigcup_{i:~\text{cell}~i~\text{bad}} S_i$, which contains all points in the union of the bad cells. In the worst case, the union of bad cells may cross all halfplanes in $\mathcal{H}$, so the number of halfplanes in the subproblem will not necessarily decrease. However the total number of points decreases by a constant fraction, since $|S_{\mathrm{bad}}|=\sum_{i:~\text{cell}~i~\text{bad}}|S_i|\leq O(\frac{b}{g}\cdot \frac{m}{b})=O(\frac{m}{g})$.
%($b=O(n^\epsilon)$)
%The base case is when $m\leq b$.

The construction time satisfies the recurrence
$$T\left(m,n\right)=b\cdot T\left(O\left(\frac{m}{b}\right),\frac{gn}{\sqrt{b}}\right)+T\left(O\left(\frac{m}{g}\right),n\right)+\tilde{O}(m+nb)+\tilde{O}((n+m)\cdot z).$$
Set $b=N^{\delta_0}$ where $\delta_0>0$ is a sufficiently small constant, and set $z=(n+m)^{1/3}$ to later balance the terms in the update time. For simplicity of analysis we assume that initially we have $m\leq n$, and then this condition holds for all subproblems as we recurse. The running time is dominated by the costs at the lowest level, so the recurrence solves to $T(m,n)=\tilde{O}(m\cdot (\frac{n}{\sqrt{m}})^{4/3})=O(m^{1/3}n^{4/3}\cdot N^{O(\delta)})$.
%$T(m,n)=O(m^{1/3}n^{4/3}\cdot N^{O(\delta)})$
%$T(m,n)=O((\frac{m}{t})^{(1-\beta)/2} n^{1+\beta}\cdot N^{O(\delta)})$
%$z=(n+m)^{\beta}$ for a constant parameter $\beta$ to be set later
%$T(m,n)=O(m^{1/2+\alpha/(4-2\alpha)}n^{1-\alpha/(2-\alpha)}\cdot N^{O(\delta)})$

\subsubsection*{Update.} When we insert or delete a halfplane $h$, we recurse in the $O(\sqrt{b})$ good cells $\Delta_1,\dots,\Delta_{\tau}$ crossed by $h$, in order to recompute the approximate solutions $\mathcal{H}_{\mathrm{appx}}(S_i)$ for $1\leq i\leq \tau$. We also recurse in the subproblem $(S_{\mathrm{bad}},\mathcal{H})$ containing the union of bad cells, and recompute the approximate solution $\mathcal{H}_{\mathrm{appx}}(S_{\mathrm{bad}})$. We update the range searching data structure $\tilde{\mathcal{D}}^*$ required by the small $\mathsf{opt}$ algorithm, using $\tilde{O}(z)$ time.

To reconstruct the approximate solution $\mathcal{H}_{\mathrm{appx}}$ we proceed as follows. If $\mathsf{opt}\triangleq |\mathcal{H}_{\mathsf{opt}}|\leq c_2\cdot b\log N$ for a sufficiently large constant $c_2$, we use the $\tilde{O}(\mathsf{opt}^2\cdot \frac{(n+m)^{1/2+\delta}}{z^{1/2}})$ time algorithm for small $\mathsf{opt}$ (Lemma~\ref{lemma:small 2D halfplanes v1}) to recompute $\mathcal{H}_{\mathrm{appx}}$ from scratch. The condition $\mathsf{opt}\overset{?}{\leq } c_2\cdot b\log N$ can be tested by the small $\mathsf{opt}$ algorithm itself. Otherwise $\mathsf{opt}>c_2\cdot b\log N$, and we can afford a larger additive error, so we return the union of the $O(1)$-approximate solutions for the good cells $S_i$ and the union of the bad cells $S_\mathrm{bad}$, i.e.\ let $\mathcal{H}_{\mathrm{appx}}=\bigsqcup_{i:~\text{cell}~i~\text{good}}\mathcal{H}_{\mathrm{appx}}(S_i)\sqcup \mathcal{H}_{\mathrm{appx}}(S_{\mathrm{bad}})$, which is stored implicitly. (One special case is when there exists a halfplane $h$ that contains the cell $\Delta_i$, we include $h$ instead of $\mathcal{H}_{\mathrm{appx}}(S_i)$ in the solution $\mathcal{H}_{\mathrm{appx}}$.)

We rebuild the entire data structure after every $g\cdot \frac{n}{\sqrt{b}}$ halfplane updates, so that good cells will not become bad. The amortized cost per update is $\frac{T(m,n)}{gn/\sqrt{b}}+\tilde{O}(b^2\cdot \frac{(n+m)^{1/2+\delta}}{z^{1/2}})+\tilde{O}(z)=O(m^{1/3}n^{1/3}\cdot N^{O(\delta)})$.

%For the base case $m\leq n^{\alpha}$, we compute $\mathcal{H}_{\mathrm{appx}}$ from scratch, using the second small $\mathsf{opt}$ algorithm (Lemma~\ref{lemma:small 2D halfplanes v2}) in $\tilde{O}(\mathsf{opt}\cdot (n+m)^{1/2+\delta})=\tilde{O}(n^{1/2+\alpha+\delta})$ time, since $\mathsf{opt}\leq m$.

%(since we always have $m\leq n$, we will maintain this invariant)

%(The size of the optimal solution can change by at most $1$ for point updates, so we mainly focus on the halfplane updates.)

%To balance the query time $\tilde{O}((b\log N)^2\cdot \frac{(n+m)^{1/2+\delta}}{z^{1/2}})$ and update time $\tilde{O}(z)$ for the small $\mathsf{opt}$ algorithm, we set $z=(n+m)^{1/3}$ (i.e.\ set $\beta=1/3$).

When we delete a point $p$, if $p$ is contained in a good cell $\Delta_i$, we recurse in the subproblem within $\Delta_i$ in order to recompute the approximate solution $\mathcal{H}_{\mathrm{appx}}(S_i)$. Otherwise $p$ is contained in a bad cell, so we recurse in the subproblem $(S_{\mathrm{bad}},\mathcal{H})$ and recompute the approximate solution $\mathcal{H}_{\mathrm{appx}}(S_{\mathrm{bad}})$. Then we recompute $\mathcal{H}_{\mathrm{appx}}$, using the procedure described above. We update the range searching data structure $\tilde{\mathcal{D}}$ required by the small $\mathsf{opt}$ algorithm, using $\tilde{O}(z)$ time.
%The amortized cost per update is $\frac{T(m,n)}{m/b}+O((n+m)^{1/2+\delta})=O(m^{-1/2+\alpha/(4-2\alpha)}n^{1-\alpha/(2-\alpha)}\cdot N^{O(\delta)})$.
%To balance the two terms $\tilde{O}(n^{1/2+\alpha+\delta})$ and $O(m^{-1/2+\alpha/(4-2\alpha)}n^{1-\alpha/(2-\alpha)}\cdot N^{O(\delta)})$ at the critical case $m=n^\alpha$, set $\alpha=1/4$ be the root of the equation $1/2+\alpha=\alpha(-1/2+\alpha/(4-2\alpha))+1-\alpha/(2-\alpha)$.
%We reconstruct the entire data structure after every $\frac{m}{b}$ point updates, so that the cell sizes are always bounded by $O(\frac{m}{b})$.

Let $U(m,n)$ denote the update time for an instance with $m$ points and $n$ halfplanes. Since point deletions are easier, we mainly focus on halfplane updates. It satisfies the recurrence
$$U\left(m,n\right)=O\left(\sqrt{b}\right)\cdot U\left(O\left(\frac{m}{b}\right),\frac{gn}{\sqrt{b}}\right)+U\left(O\left(\frac{m}{g}\right),n\right)+O\left(m^{1/3}n^{1/3}\cdot N^{O(\delta)}\right),$$
which solves to $U(m,n)=O(m^{1/3}n^{1/3}\cdot N^{O(\delta)})$.

\begin{figure}[htbp]
\centering
\begin{minipage}[t]{0.48\textwidth}
\centering
\includegraphics[width=.9\textwidth]{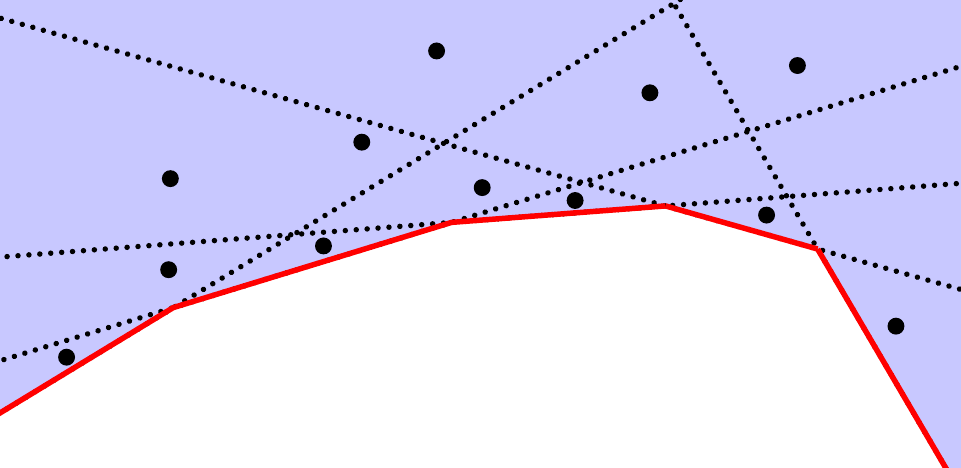}
\end{minipage}
\begin{minipage}[t]{0.48\textwidth}
\centering
\includegraphics[width=.9\textwidth]{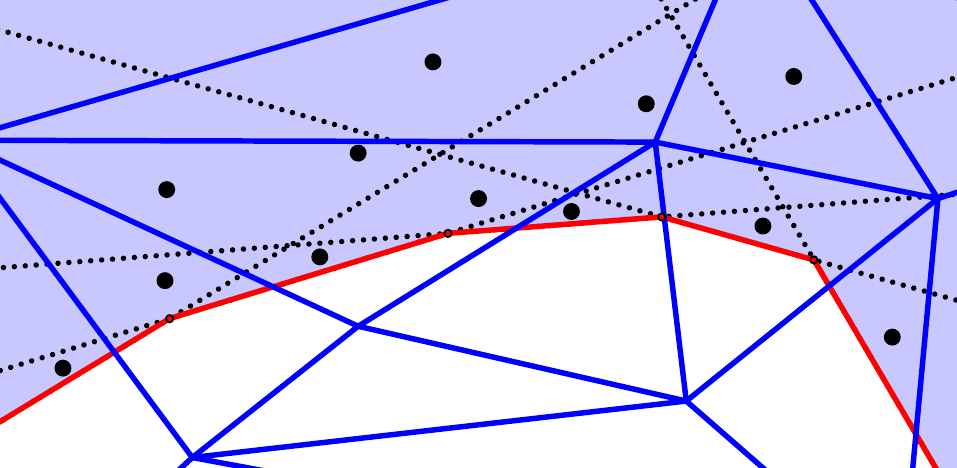}
%\caption{}
\end{minipage}
\caption{The boundary of the optimal solution must form a convex chain (see left). For simplicity we only include the upper halfplanes; the lower halfplanes are similar. Each triangular cell of the partition only intersects the convex chain $O(1)$ times (see right).
%The convex chain and the triangular cells of the partition.
}
\label{fig:convex chain}
\end{figure}

\subsubsection*{Approximation factor analysis.} The key observation here is that the boundary of the optimal solution $\mathcal{H}_{\mathsf{opt}}$ for $(S,\mathcal{H})$ must form a convex chain, as shown in Figure~\ref{fig:convex chain}. As each cell is a triangle, it can only intersect the optimal convex chain $O(1)$ times, thus the $b$ disjoint cells $\Delta_1,\dots,\Delta_b$ will partition $\mathcal{H}_{\mathsf{opt}}$ into $O(b)$ disjoint pieces. (Unfortunately, such nice property does not hold in 3D.) If we take the union of the optimal solutions in all the cells, the additive error is at most $O(b)$, i.e., we have $\sum_{1\leq i\leq b}|\mathcal{H}_{\mathsf{opt}}(S_i)|\leq \sum_{1\leq i\leq b}(|\mathcal{H}_{\mathsf{opt}}\cap \Delta_i|+O(1))\leq |\mathcal{H}_{\mathsf{opt}}|+O(b)$. Similarly, we have 
$$\sum_{i:~\text{cell}~i~\text{good}}|\mathcal{H}_{\mathsf{opt}}(S_i)|+|\mathcal{H}_{\mathsf{opt}}(S_{\mathrm{bad}})|\ \leq\ |\mathcal{H}_{\mathsf{opt}}|+c_1 b$$
for some constant $c_1$.

Let $c_0$ denote the constant multiplicative error factor guaranteed by the small $\mathsf{opt}$ algorithm. Suppose that for some constant multiplicative error factor $c\geq \max\{c_0,c_1\}$, we have $|\mathcal{H}_{\mathrm{appx}}(S_i)|\leq c\cdot |\mathcal{H}_{\mathsf{opt}}(S_i)|$ for all good cells $i$, and also $|\mathcal{H}_{\mathrm{appx}}(S_{\mathrm{bad}})|\leq c\cdot |\mathcal{H}_{\mathsf{opt}}(S_{\mathrm{bad}})|$. Then if $\mathsf{opt}>c_2\cdot b\log N$, we set $\mathcal{H}_{\mathrm{appx}}$ to be the union of the approximate solutions in the subproblems, and obtain
\begin{align*}
|\mathcal{H}_{\mathrm{appx}}|\ &\leq \sum_{i:~\text{cell}~i~\text{good}}|\mathcal{H}_{\mathrm{appx}}(S_i)|+|\mathcal{H}_{\mathrm{appx}}(S_{\mathrm{bad}})|\\
&\leq\ c\cdot \left(\sum_{i:~\text{cell}~i~\text{good}}|\mathcal{H}_{\mathsf{opt}}(S_i)|+|\mathcal{H}_{\mathsf{opt}}(S_{\mathrm{bad}})|\right)\\
&\leq\ c\cdot (|\mathcal{H}_{\mathsf{opt}}|+c_1 b)\\
&\leq\ \left(1+\tfrac{1}{\log N}\right)\cdot c\cdot |\mathcal{H}_{\mathsf{opt}}|.
%\nonumber
\end{align*}
Otherwise we will use the small $\mathsf{opt}$ algorithm to compute $\mathcal{H}_{\mathrm{appx}}$ from scratch, which guarantees that $|\mathcal{H}_{\mathrm{appx}}|\leq c_0\cdot |\mathcal{H}_{\mathsf{opt}}|$.
%This bound also holds for the base case, where we use the second small $\mathsf{opt}$ algorithm to recompute $\mathcal{H}_{\mathrm{appx}}$.

The recursion depth of the data structure is $O(\log_g N)=O(\log N)$. The multiplicative error factor multiplies by a factor of at most $1+\frac{1}{\log N}$ at each level, so the global multiplicative error is $c\cdot (1+\frac{1}{\log N})^{O(\log N)}=O(1)$.

%\paragraph{Running time analysis.} Let $T(m,n)$ denote the construction time for the data structure. We have the following recurrence:
%$$T\left(m,n\right)=b\cdot T\left(O\left(\frac{m}{b}\right),\frac{gn}{\sqrt{b}}\right)+T\left(O\left(\frac{m}{g}\right),n\right)+\tilde{O}\left(b\cdot (n+m)^{1+\delta}\right),$$
%which gives
%$$T\left(m,n\right)=\sum_i T\left(m_i,\frac{gn}{\sqrt{b}}\right)+T\left(O\left(\frac{m}{g}\right),n\right)+\tilde{O}\left(b\cdot (n+m)^{1+\delta}\right),$$
%where $\sum_i m_i=m$ and $m_i\leq O(\frac{m}{b})$.
%$T(m,n)=O((n+m)^{1+\delta})$.

%old: Use the small $\mathsf{opt}$ algorithm for squares, and for each cell use $\tilde{O}(n^{1/3})$ query and update time tradeoff. The total running time is $O(m^{1/3}n^{1/3}N^\delta)$.

\subsubsection*{Reporting the actual solution.} If the $O(1)$-approximate solution $\mathcal{H}_{\mathrm{appx}}$ is explicitly stored (computed by the small $\mathsf{opt}$ algorithm), then we can directly report the solution in $O(\mathsf{opt})$ time. Otherwise $\mathcal{H}_{\mathrm{appx}}$ is implicitly stored as the union of the approximate solutions in the subproblems, and we recursively report the approximate solutions $\mathcal{H}_{\mathrm{appx}}(S_i)$ for each good cell $i$, as well as $\mathcal{H}_{\mathrm{appx}}(S_{\mathrm{bad}})$ for the bad cells. The running time is proportional to the output size, i.e., $O(\mathsf{opt})$.

\begin{lemma}
There exists a data structure for $O(1)$-approximate dynamic 2D halfplane set cover that supports halfplane insertion/deletions and point deletions in amortized $O(n^{2/3+\delta})$ time w.h.p.\ for any constant $\delta>0$, and can answer size and reporting queries in $O(1)$ and $O(\mathsf{opt})$ time, respectively.
\end{lemma}

\subsubsection*{Point insertions.} The issue with point insertions is we need to rebuild the whole structure every $O(\frac{m}{b})$ point insertions to ensure the cell sizes are bounded by $O(\frac{m}{b})$, which is costly when $m$ is too small. We resolve this issue by reducing the fully dynamic problem to partially dynamic, using the \emph{logarithmic method} \cite{bentley1980decomposable} but with a larger base $N^\delta$ and a non-trivial base case, and finally obtain sublinear update time for all four types of updates.

Specifically, we use the logarithmic method to partition the set of points $S$ into $O(\frac{1}{\delta})$ subsets $S_i$, where the $i$-th subset contains $m_i\in [\frac N{(N^\delta)^{i+1}},\frac N{(N^\delta)^i})$ points ($i=0,\dots,f$)
for some $f=O(\frac 1\delta)$, for a sufficiently small constant $\delta$. We build a partially dynamic data structure $\mathcal{D}_i$ that supports halfplane insertion/deletions and point deletions, on the instance $(S_i,\mathcal{H})$; an exception is the last data structure $\mathcal{D}_f$, which will be fully dynamic.
We obtain the global approximate set cover solution by taking the union of the approximate solutions for each of the subproblems, losing only a constant approximation factor since there are only $O(\frac{1}{\delta})=O(1)$ subproblems. The logarithmic method guarantees that the subset $S_i$ is rebuilt only after every $\Theta(\frac{m_i}{N^\delta})$ point insertions.
%the overall running time only increases by a factor of $O(N^\delta)$.

%We use different methods to build the partially dynamic data structure $\mathcal{D}_i$ for the subproblem $(S_i,\mathcal{H})$, depending on the size $m_i$. Let $m_f$ be a parameter to be set later. For $m_i\leq m_f$, 

To implement the fully dynamic data structure $\mathcal{D}_f$,
we use the small $\mathsf{opt}$ algorithm (Lemma~\ref{lemma:small 2D halfplanes v1}), with $\tO(m_f^2\cdot \sqrt{\frac{n}{z}}+z)=\tO(m_f^{4/3}n^{1/3})$ update time, by setting $z=m_f^{4/3}n^{1/3}$ (here and in the following we ignore the $N^{O(\delta)}$ factors by a slight abuse of the $\widetilde{O}$ notation).

To implement the partially dynamic data structure $\mathcal{D}_i\ (i<f)$ for the subproblem $(S_i,\mathcal{H})$, we suggest two different methods depending on the size $m_i$.

For $m_f<m_i\leq n^{2/3}$, we modify our partially dynamic solution which we introduced earlier. In particular, at each node of the recursion tree we build the data structure required by the second small $\mathsf{opt}$ algorithm (Lemma~\ref{lemma:small 2D halfplanes v2}), instead of the first small $\mathsf{opt}$ algorithm. Another change is that we stop recursing when the current number of halfplanes $n'$ becomes at most $\frac{n}{\sqrt{B_i}}$ (which also ensures the the current number of points $m'\leq \frac{m}{B_i}$), for a parameter $B_i$ to be set later. Following a similar analysis, we obtain update time $\widetilde{O}(\sqrt{B_i}\cdot \frac{m}{B_i}\cdot \sqrt{\frac{n}{\sqrt{B_i}}})=\widetilde{O}(\frac{m\sqrt{n}}{B_i^{3/4}})$ (the cost is dominated by the lowest level) and construction time $\widetilde{O}(B_i\cdot \frac{n}{\sqrt{B_i}})=\widetilde{O}(n\sqrt{B_i})$. 
Recall that the whole data structure needs to be rebuilt after every $\Theta(\frac{m_i}{N^\delta})$ point insertions. By setting $B_i=\frac{m_i^{8/5}}{n^{2/5}}+1$ (verify that $1\leq B_i\leq m_i$), the amortized update time is $\widetilde{O}(\frac{m\sqrt{n}}{B_i^{3/4}}+\frac{n\sqrt{B_i}}{m_i})=\widetilde{O}(\frac{n^{4/5}}{m_f^{1/5}} + \frac{n}{m_f})$.

For $m_i>n^{2/3}$, we use our partially dynamic solution, but this time using a common parameter $z_i$ in the small $\mathsf{opt}$ algorithms at all nodes of the recursion tree. The construction time becomes $T(m',n')=\widetilde{O}(m'\cdot \frac{n}{\sqrt{m'}}\cdot z_i)$, and the update time becomes $U(m',n')=\widetilde{O}(\sqrt{m'}\cdot \sqrt{\frac{n'/\sqrt{m'}}{z_i}}+z_i+\sqrt{m'}\cdot z_i)$. Adding the amortized cost for rebuilding (after every $\Theta(\frac{m_i}{N^\delta})$ point insertions), the amortized update time is $\widetilde{O}(\frac{T(m_i,n)}{m_i}+U(m_i,n))=\widetilde{O}(\frac{m_i^{1/4}n^{1/2}}{\sqrt{z_i}}+(\sqrt{m_i}+\frac{n}{\sqrt{m_i}})\cdot z_i)$, which is $\widetilde{O}(n^{2/3})$ by setting $z_i=\frac{m_i^{1/2}}{n^{1/3}}\geq 1$.

% and stop recursing when $n'\leq \frac{n}{\sqrt{B_i}}$

Finally setting $m_f$ near $n^{7/23}$ (up to $N^\delta$ factors)
to balance the $\tO(m_f^{4/3}n^{1/3})$ and $\widetilde{O}(\frac{n^{4/5}}{m_f^{1/5}} + \frac{n}{m_f})$  terms, we obtain amortized update time $O(n^{17/23+O(\delta)})$.

%size query: maintain the size during updates. $O(1)$.

%membership query time: to query whether a halfplane $h$ is in $\mathcal{H}_{\mathrm{appx}}$, if $\mathcal{H}_{\mathrm{appx}}$ is explicitly stored, directly perform the query in $O(1)$ time. Otherwise recursively query in the $O(\sqrt{b})$ cells crossing $h$. The running time is $T(n)=O(\sqrt{b})\cdot T(\frac{n}{b})+T(\frac{n}{g})$, which gives $T(n)=O(n^{1/2+\delta})$. (improve?)
%$O(n^\epsilon)$.
%cheating query?

\begin{theorem}
There exists a data structure for $O(1)$-approximate dynamic 2D halfplane set cover with amortized $O(n^{17/23+\delta})$ update time w.h.p.\ for any constant $\delta>0$, and can answer size and reporting queries in $O(1)$ and $O(\mathsf{opt})$ time, respectively.
\end{theorem}
%(1+\epsilon)-approx? small OPT?
%the update time seems improvable to O(n^{5/9}).

The exponent $17/23<0.74$ is likely further improvable with more work.  The main message is that
sublinear update time is achievable for the fully dynamic problem while supporting efficient reporting queries.

\section{Weighted Interval Set Cover}
In this section, we present the first dynamic data structure with sub-linear update time and constant factor approximation of weighted interval set cover.
Let $(S,\mathcal{I})$ be a dynamic weighted interval set cover instance where $S$ is the set of points in $\mathbb{R}$ and $\mathcal{I}$ is the set of weighted intervals.
For each interval $I \in \mathcal{I}$, we use $w(I) \geq 0$ to denote the weight of $I$.
We assume that the \textit{point range} of $(S,\mathcal{I})$ is $[0,1]$, i.e., the points in $S$ are always in the range $[0,1]$.
We say an interval $I \in \mathcal{I}$ is \textit{two-sided} if both of the endpoints of $I$ lie in the interior of the point range $[0,1]$, and \textit{one-sided} if at least one endpoint of $I$ is outside $[0,1]$.

%Let $r$ be a parameter to be determined.
We first observe that the approach we used for unweighted dynamic interval set cover (Section~\ref{sec-unweightedint}) can be easily extended to the weighted case to obtain an $n^{O(1)}$-approximation.
Recall that in Section~\ref{sec-unweightedint}, we partitioned $[0,1]$ into $r$ connected portions $J_1,\dots,J_r$ each of which contains $O(n/r)$ points in $S$ and $O(n/r)$ endpoints of intervals in $\mathcal{I}$.
Then we defined $S_i = S \cap J_i$ and $\mathcal{I}_i = \{I \in \mathcal{I}: I \cap J_i \neq \emptyset \text{ and } J_i \nsubseteq I\}$.
Each $(S_i,\mathcal{I}_i)$ was viewed as a dynamic interval set cover instance (called a sub-instance) with point range $J_i$, and we recursively built a sub-structure $\mathcal{D}_i$ for $(S_i,\mathcal{I}_i)$.
In Section~\ref{sec-unweightedint}, we construct a set cover $\mathcal{I}_\text{appx}$ for $(S,\mathcal{I})$ by distinguishing two cases: when the optimum is small, we compute $\mathcal{I}_\text{appx}$ using the output-sensitive algorithm of Lemma~\ref{lem-intoslong}; when the optimum is large, $\mathcal{I}_\text{appx}$ is constructed by (essentially) taking the union of the solution maintained in the $\mathcal{D}_i$'s.
The output-sensitive algorithm of Lemma~\ref{lem-intoslong}, unfortunately, does not work for the weighted case.
Therefore, here we always construct $\mathcal{I}_\text{appx}$ in a way similar to that used for the large-optimum case.
Specifically, for each $J_i$, we find a minimum-weight interval $I \in \mathcal{I}$ such that $J_i \subseteq I$ (if it exist) and let $w_i$ be the cost of the set cover of $(S_i,\mathcal{I}_i)$ maintained by $\mathcal{D}_i$.
If $w_i \leq w(I)$, we define $\mathcal{I}_i^*$ as the set cover of $(S_i,\mathcal{I}_i)$ maintained by $\mathcal{D}_i$, otherwise we define $\mathcal{I}_i^* = \{I\}$.
We then define $\mathcal{I}_\text{appx} = \bigsqcup_{i=1}^r \mathcal{I}_i^*$.
We observe the following fact.
\begin{fact}
If each sub-structure $\mathcal{D}_i$ maintains a $t$-approximate set cover of the sub-instance $(S_i,\mathcal{I}_i)$, then $\mathcal{I}_\textnormal{appx}$ is an $rt$-approximate set cover of the instance $(S,\mathcal{I})$.
\end{fact}
\begin{myproof}
For $i \in [r]$, let $\mathsf{opt}_i$ denote the cost of an optimal set cover of $(S_i,\mathcal{I})$.
Clearly, $\mathsf{opt}_i \leq \mathsf{opt}$ for all $i \in [r]$ and thus $\sum_{i=1}^r \mathsf{opt}_i \leq r \cdot \mathsf{opt}$.
We then show that $\mathsf{cost}(\mathcal{I}_i^*) \leq t\cdot \mathsf{opt}_i$, which implies that $\mathsf{cost}(\mathcal{I}_\text{appx}) = \sum_{i=1}^r \mathsf{cost}(\mathcal{I}_i^*) \leq rt \cdot \mathsf{opt}$.
If an optimal set cover of $(S_i,\mathcal{I})$ consists of a single interval in $\mathcal{I}$ that covers $J_i$, then we have $\mathsf{cost}(\mathcal{I}_i^*) = \mathsf{opt}_i$.
Otherwise, an optimal set cover of $(S_i,\mathcal{I})$ is a set cover of $(S_i,\mathcal{I}_i)$, and hence $\mathsf{cost}(\mathcal{I}_i^*) \leq t\cdot\mathsf{opt}_i$.
\end{myproof}

The above fact shows that the approximation ratio of our data structure satisfies the recurrence $A(n) = r \cdot A(O(n/r))$, which solves to $A(n) = n^{O(1)}$.
Furthermore, as analyzed in Section~\ref{sec-unweightedint}, the data structure can be updated in $\widetilde{O}(r)$ amortized time.
Setting $r$ to be a constant, we get an $n^{O(1)}$-approximation data structure for dynamic weighted interval set cover with $\widetilde{O}(1)$ amortized update time.
In particular, we can maintain an estimation $\mathsf{opt}^\sim$ of the optimum $\mathsf{opt}$ of $(S,\mathcal{I})$ in $\widetilde{O}(1)$ amortized update time, which satisfies $\mathsf{opt} \leq \mathsf{opt}^\sim \leq n^{O(1)} \mathsf{opt}$.
With this observation, we now discuss our $(3+\varepsilon)$-approximation data structure.
Since our data structure here is somehow involved (compared to the unweighted one in Section~\ref{sec-unweightedint}), we shall first (informally) describe the underlying basic ideas, followed by the formal definitions and analysis.

\subsection{Main ideas.}
The main reason why the data structure in Section~\ref{sec-unweightedint} only achieves an $n^{O(1)}$-approximation is that it decomposes the entire problem into $r$ sub-problems and combines the solutions of the sub-problems in a trivial way.
However, these sub-problems are not independent: one interval in $\mathcal{I}$ can be used in all of the $r$ sub-problems in the worst case.
As such, each level of the recursion can possibly increase the approximation ratio by a factor of $r$.
In order to handle this issue, our first key idea is to combine the solutions of the sub-problems using \textit{dynamic programming}.
To see why DP is helpful, let us assume at this point that each sub-structure $\mathcal{D}_i$ maintains an \textit{optimal} solution for the sub-instance $(S_i,\mathcal{I}_i)$.
Under this assumption, we show how DP can be applied to obtain a 3-approximate solution for the instance $(S,\mathcal{I})$.

Let $x_1,\dots,x_{r+1}$ be the endpoints of $J_1,\dots,J_r$ sorted from left to right (so the endpoints of $J_i$ are $x_i$ and $x_{i+1}$).
Consider an interval $I \in \mathcal{I}$.
If $I$ contains at least one point in $\{x_1,\dots,x_{r+1}\}$, we ``chop'' $I$ into at most three pieces as follows.
Let $i^-$ (resp., $i^+$) be the smallest (resp., largest) index such that $x_{i^-} \in I$ (resp., $x_{i^+} \in I$).
Then $x_{i^-}$ and $x_{i^+}$ partition $I$ into three pieces: the \textit{left} piece (the part to the left of $x_{i^-}$), the \textit{middle} piece (the part in between $x_{i^-}$ and $x_{i^+}$), and the \textit{right} piece (the part to the right of $x_{i^+}$).
We give each piece a weight equal to $w(I)$.
%In the case where $x_{i^-}$ and $x_{i^+}$ do not exist, i.e., $I$ contains no point in $\{x_1,\dots,x_{r+1}\}$, we do not chop $I$.
Let $\mathcal{I}'$ be the resulting set of intervals after chopping the intervals in $\mathcal{I}$, i.e., $\mathcal{I}'$ consists of all pieces of the chopped intervals in $\mathcal{I}$ and all unchopped intervals in $\mathcal{I}$.
It is clear that the optimum of the instance $(S,\mathcal{I}')$ is within $[\mathsf{opt},3\mathsf{opt}]$, where $\mathsf{opt}$ is the optimum of $(S,\mathcal{I})$.

Now we observe a good property of the interval set $\mathcal{I}'$: each interval in $\mathcal{I}'$ is either contained in $J_i$ for some $i \in [r]$ (e.g., the unchopped intervals and the left/right pieces) or is equal to $[x_{i^-},x_{i^+}]$ for some $i^-,i^+ \in [r+1]$ (e.g., the middle pieces); we call the intervals of the first type \textit{short intervals} and those of the second type \textit{long intervals}.
Let $\mathcal{I}_\text{long}' \subseteq \mathcal{I}'$ be the set of long intervals and $\mathcal{I}_i' \subseteq \mathcal{I}'$ be the set of short intervals contained in $J_i$.
Then in any set cover of $(S,\mathcal{I}')$, for each $i \in [r]$, either $J_i$ is covered by a long interval or the points in $S_i$ are covered by short intervals in $\mathcal{I}_i'$.
Furthermore, in an optimal set cover of $(S,\mathcal{I}')$, if the points in $S_i$ are covered by short intervals in $\mathcal{I}_i'$, then those short intervals must be an optimal set cover of $(S_i,\mathcal{I}_i')$.
Note that the instance $(S_i,\mathcal{I}_i')$ is in fact equivalent to the sub-instance $(S_i,\mathcal{I}_i)$, because $\mathcal{I}_i' = \{I \cap J_i: I \in \mathcal{I}_i\}$ (where the weight of $I \cap J_i$ is equal to the weight of $I$) and the points in $S_i$ are all contained in $J_i$.
Thus, by assumption, an optimal set cover of $(S_i,\mathcal{I}_i')$ is already maintained in the sub-structure $\mathcal{D}_i$.
Based on this observation, we can use DP to compute an optimal set cover of $(S,\mathcal{I}')$ as follows.
%We compute two tables $\text{OPT}[1,\dots,r]$ and $\text{SC}[1,\dots,r]$, where $\text{SC}[i]$ stores an optimal set cover of $(\bigcup_{j=1}^i S_i, \mathcal{I}')$ and $\text{OPT}[i]$ stores the cost of the set cover $\text{SC}[i]$.
For a long interval $I = [x_{i^-},x_{i^+}] \in \mathcal{I}_\text{long}'$, we write $\pi(I) = i^- - 1$.
%Let $\text{OPT}[0] = 0$ and $\text{SC}[0] = \emptyset$.
For each $i$ from $1$ to $r$, we compute an optimal set cover for $(\bigcup_{j=1}^i S_j, \mathcal{I}')$.
To this end, we consider how the points in $S_i$ are covered.
Clearly, we can cover the points in $S_i$ using a long interval $I \in \mathcal{I}_\text{long}'$ satisfying $J_i \subseteq I$.
In this case, the best solution is the union of $\{I\}$ and an optimal set cover of $(\bigcup_{j=1}^{\pi(I)} S_j, \mathcal{I}')$ which has already been computed as $\pi(I) < i$.
%We enumerate all $I \in \mathcal{I}_\text{long}'$ satisfying $J_i \subseteq I$ and take the best solution found.
Alternatively, we can cover the points in $S_i$ using the short intervals in $\mathcal{I}_i'$.
In this case, the best solution is the union of an optimal set cover for $(\bigcup_{j=1}^{i-1} S_j, \mathcal{I}')$ and an optimal set cover for $(S_i,\mathcal{I}_i')$, where the former has already been computed and the latter is maintained in the sub-structure $\mathcal{D}_i$.
We try all these possibilities and take the best solution found, which is an optimal set cover for $(\bigcup_{j=1}^i S_j, \mathcal{I}')$.
When the DP procedure completes, we get an optimal set cover of $(S,\mathcal{I}')$, which in turn gives us a 3-approximation of an optimal set cover of $(S,\mathcal{I})$.

Although the above approach seems promising, there are two issues we need to resolve.
First, the above DP procedure takes $O(r \cdot |\mathcal{I}_\text{long}'|)$ time, but $|\mathcal{I}_\text{long}'| = \Omega(n)$ in the worst case.
This issue can be easily handled by observing that there are only $O(r^2)$ different intervals in $\mathcal{I}_\text{long}'$.
Indeed, every interval in $\mathcal{I}_\text{long}'$ is equal to $[x_{i^-},x_{i^+}]$ for some $i^-,i^+ \in [r+1]$.
Among a set of identical intervals in $\mathcal{I}_\text{long}'$, only the one with the minimum weight is useful.
Therefore, we only need to keep $O(r^2)$ minimum-weight intervals in $\mathcal{I}_\text{long}'$.
Furthermore, these minimum-weight intervals can be computed in $\widetilde{O}(r^2)$ time using a range-min data structure \textit{without} computing $\mathcal{I}_\text{long}'$.
Specifically, we identify each interval $I = [a,b] \in \mathcal{I}$ with a weighted point $(a,b) \in \mathbb{R}^2$ with weight $w(I)$.
The minimum-weight $[x_{i^-},x_{i^+}]$ in $\mathcal{I}_\text{long}'$ is just the middle piece of the minimum-weight interval whose left endpoint lies in $J_{i^--1}$ and right point lies in $J_{i^++1}$, which corresponds to the minimum-weight point in the rectangular range $[x_{i^--1},x_{i^-}] \times [x_{i^+},x_{i^++1}]$.
Thus, if we maintain the corresponding weighted points of the intervals in $\mathcal{I}$ in a dynamic 2D range-min data structure, the minimum-weight intervals in $\mathcal{I}_\text{long}'$ can be computed in $\widetilde{O}(r^2)$ time and the DP procedure can be done in $\widetilde{O}(r^3)$ time.

The second issue is more serious.
We assumed that each sub-structure $\mathcal{D}_i$ maintains an \textit{optimal} solution for the sub-instance $(S_i,\mathcal{I}_i)$.
Clearly, this is not the case, as the sub-structures are recursively built and hence can only maintain approximate solutions for the sub-instances.
%It is easy to see that if we maintain a $t$-approximation solution for each sub-instance $(S_i,\mathcal{I}_i)$, then the DP above can give us a $3t$-approximation solution of $(S,\mathcal{I})$.
In this case, the approximation ratio may increase by a constant factor at each level of the recursion: an interval in $\mathcal{I}$ is chopped into three pieces and its left/right pieces can be further chopped by the sub-structures in lower levels.
To handle this issue, we need to prevent the sub-structures from chopping the intervals that are already chopped in higher levels of the recursion.
A key observation is the following: if an interval is chopped in the current level, then its left/right pieces are both \textit{one-sided} intervals in the sub-instances.
Therefore, if we only chop the two-sided intervals, we should be able to avoid the issue that an interval is chopped more than once.
%Specifically, we now define $\mathcal{I}_i$ as the set consisting of all \textit{two-sided} intervals $I \in \mathcal{I}$ that $J_i \cap I \neq \emptyset$ and $J_i \nsubseteq I$.
However, this strategy brings us a new difficulty, i.e., handling the one-sided intervals when constructing the set cover.

We overcome this difficulty as follows.
We call a one-sided interval in $\mathcal{I}$ \textit{left} (resp., \textit{right}) one-sided interval if it covers the left (resp., right) end of the point range $[0,1]$.
First, observe that we need at most one left one-sided interval and one right one-sided interval in our solution, simply because the coverage of the left (resp., right) one-sided intervals is nested and thus only the rightmost (resp., leftmost) one in the solution is useful.
So a na{\"i}ve idea is to enumerate the left/right one-sided interval used in our solution.
(Clearly, we cannot afford to do this because there might be $\Omega(n)$ one-sided intervals.
But at this point let us ignore the issue about running time -- we will take care of it later.)
If $L \in \mathcal{I}$ and $R  \in \mathcal{I}$ are the left and right one-sided intervals in our solution, then the remaining task is to cover the points in $S \backslash (L \cup R)$.
It turns out that we can still apply the DP approach above to compute a set cover for the points in $S \backslash (L \cup R)$ using the intervals in $\mathcal{I}'$.
%assuming that $\mathcal{D}_i$ maintains an optimal solution for the sub-instance $(S_i,\mathcal{I}_i)$.
To see this, suppose the right endpoint of $L$ lies in $J_{i^-}$ and the left endpoint of $R$ lies in $J_{i^+}$.
Then the points to be covered are those lying in the portions $J_{i^-} \backslash L, J_{i^-+1},\dots,J_{i^+-1},J_{i^+} \backslash R$.
Same as before, in a set cover of $(S \backslash (L \cup R),\mathcal{I}')$, for each portion $J_i$ where $i^- < i < i^+$, either $J_i$ itself is covered by a long interval in $\mathcal{I}_\text{long}'$ or the points in $S_i$ are covered by short intervals in $\mathcal{I}_i'$; in the latter case we can use the solution of $(S_i,\mathcal{I}_i)$ maintained in the sub-structure $\mathcal{D}_i$.
The only difference occurs in the portions $J_{i^-} \backslash L$ and $J_{i^+} \backslash R$.
We can either cover $J_{i^-}$ (resp., $J_{i^+}$) using a long interval in $\mathcal{I}_\text{long}'$ or cover the points in $S_{i^-} \backslash L$ (resp., $S_{i^+} \backslash R$) using short intervals in $\mathcal{I}_{i^-}'$ (resp., $\mathcal{I}_{i^+}'$).
However, we do not have a good set cover for $(S_{i^-} \backslash L,\mathcal{I}_{i^-}')$ (resp., $(S_{i^+} \backslash R,\mathcal{I}_{i^+}')$) in hand: the solution maintained in the sub-structure $\mathcal{D}_{i^-}$ (resp., $\mathcal{D}_{i^+}$) is for covering all points in $S_{i^-}$ (resp., $S_{i^+}$) and hence might be much more expensive than an optimal solution of $(S_{i^-} \backslash L,\mathcal{I}_{i^-}')$ (resp., $(S_{i^+} \backslash R,\mathcal{I}_{i^+}')$).
We resolve this by temporarily inserting the interval $L$ (resp., $R$) with weight 0 to $\mathcal{I}_{i^-}'$ (resp., $\mathcal{I}_{i^+}'$) and update the sub-structure $\mathcal{D}_{i^-}$ (resp., $\mathcal{D}_{i^+}$).
Note that with the weight-0 interval $L$ (resp., $R$), the points in $S_{i^-} \cap L$ (resp., $S_{i^+} \cap R$) can be covered ``for free'' and thus the solution maintained in $\mathcal{D}_{i^-}$ (resp., $\mathcal{D}_{i^+}$) should be a good set cover of $(S_{i^-} \backslash L,\mathcal{I}_{i^-}')$ (resp., $(S_{i^+} \backslash R,\mathcal{I}_{i^+}')$).
%One easily verifies that if $\mathcal{I}^*$ is a good set cover of $(S_{i^-},\mathcal{I}_{i^-}' \cup \{L\})$ where $w(L) = 0$, then $\mathcal{I}^* \backslash \{L\}$ is a good set cover of $(S_{i^-} \backslash L,\mathcal{I}_{i^-}')$; the same holds for $(S_{i^+},\mathcal{I}_{i^+}' \cup \{R\})$ and $(S_{i^+} \backslash R,\mathcal{I}_{i^+}')$.
Once we have the set covers for the points in $J_{i^-} \backslash L, J_{i^-+1},\dots,J_{i^+-1},J_{i^+} \backslash R$ using short intervals, we can use the same DP as above to compute a set cover of $(S \backslash (L \cup R),\mathcal{I}')$, which together with $L$ and $R$ gives us a set cover solution of $(S,\mathcal{I})$.
One can verify that if the sub-structures $\mathcal{D}_1,\dots,\mathcal{D}_r$ are recursively built, then the set cover we obtain is a 3-approximate solution of $(S,\mathcal{I})$, essentially because when an interval is chopped (into up to three pieces) in the current level, its left/right pieces become one-sided intervals in the next level of recursion and can no longer cause any error.

Next, we discuss how to avoid enumerating all the left/right one-sided intervals.
The key idea is that if we have a set of left (resp., right) one-sided intervals whose weights are similar, say in a range $[w,(1+\varepsilon)w]$, then we can simply keep the one that has the maximum coverage, i.e., the rightmost (resp., leftmost) one, and discard the others.
Indeed, instead of using a left/right one-sided interval we discard, we can always use the one we keep, which increases the total weight by at most $\varepsilon w$.
Using the estimation $\mathsf{opt}^\sim$ of the optimum, we can actually classify the one-sided intervals in $\mathcal{I}$ into $\widetilde{O}(1/\varepsilon)$ groups where the intervals in each group have similar weights.
In each group, we only keep the interval with the maximum coverage.
In this way, we obtain a set of $\widetilde{O}(1/\varepsilon)$ \textit{candidate} one-sided intervals, and we only need to enumerate these candidate intervals, which can be done much more efficiently.

\subsection{The data structure.}
Now we are ready to formally present our data structure and analysis.
Let $\varepsilon > 0$ be the approximation factor.
Our goal is to design a data structure $\mathcal{D}$ that maintains a $(3+\varepsilon)$-approximate set cover solution for the dynamic weighted interval set cover instance $(S,\mathcal{I})$ and supports the size, membership, and report queries to the solution.
%For convenience of analysis, we introduce a new approximation criterion called $(c_1,c_2)$-\textit{approximation}.
%As before, we partitioned the point range $[0,1]$ into $r$ connected portions $J_1,\dots,J_r$ each of which contains $O(n/r)$ points in $S$ and $O(n/r)$ endpoints of intervals in $\mathcal{I}$.
Let $J_1,\dots,J_r$, $S_1,\dots,S_r$, and $\mathcal{I}_1,\dots,\mathcal{I}_r$ be defined in Section~\ref{sec-unweightedint}.
%Then we defined $S_i = S \cap J_i$ and $\mathcal{I}_i = \{I \in \mathcal{I}: I \cap J_i \neq \emptyset \text{ and } J_i \nsubseteq I\}$.
For each $i \in [r]$, we recursively build a sub-structure $\mathcal{D}_i$ on the sub-instance $(S_i,\mathcal{I}_i)$ with approximation factor $\tilde{\varepsilon} = \alpha \varepsilon$ for some parameter $\alpha<1$.
Next, we compute two sets $\mathcal{L}$ and $\mathcal{R}$ of one-sided intervals in $\mathcal{I}$ as follows.
Recall that we have the estimation $\mathsf{opt}^\sim$ satisfying $\mathsf{opt} \leq \mathsf{opt}^\sim \leq n^{O(1)} \mathsf{opt}$.
Set $\mathsf{opt}^- = (\varepsilon/4) \cdot \mathsf{opt}^\sim/n^c$ for a sufficiently large constant $c$ so that we have $\mathsf{opt}^- \leq ((\varepsilon - \alpha \varepsilon)/4) \cdot \mathsf{opt}$ assuming $\alpha = \Omega(1)$ (which is the case when we choose $\alpha$).
Define $\delta_0 = 0$ and $\delta_i = \mathsf{opt}^- \cdot (1+\tilde{\varepsilon}/2)^{i-1}$ for $i \geq 1$.
Let $m$ be the smallest number such that $\delta_m \geq (3+\varepsilon) \mathsf{opt}^\sim$.
Note that $m = \widetilde{O}(\frac{1}{\tilde{\varepsilon}} \log \frac{1}{\tilde{\varepsilon}})$.
For $i \in [m]$, let $L_i \in \mathcal{I}$ be the left one-sided interval with the rightmost right endpoint satisfying $w(L_i) \in [\delta_{i-1},\delta_i]$.
Then we define $\mathcal{L} = \{L_1,\dots,L_m\}$.
Similarly, let $R_i \in \mathcal{I}$ be the right one-sided interval with the leftmost left endpoint satisfying $w(R_i) \in [\delta_{i-1},\delta_i]$, and define $\mathcal{R} = \{R_1,\dots,R_m\}$.
The sets $\mathcal{L}$ and $\mathcal{R}$ can be computed in $\widetilde{O}(m)$ time using a (dynamic) 2D range-max/range-min data structure.
Indeed, if we map each interval $I = [a,b] \in \mathcal{I}$ into the point $(a,w(I)) \in \mathbb{R}^2$ with weight $b$, then the interval $L_i$ just corresponds to the maximum-weight point in the range $(-\infty,0] \times [\delta_{i-1},\delta_i]$.
Besides $\mathcal{L}$ and $\mathcal{R}$, we need another set $\mathcal{I}_\text{long} \subseteq \mathcal{I}$ defined as follows.
Recall that $x_1,\dots,x_{r+1}$ are the endpoints of $J_1,\dots,J_r$ sorted from left to right.
For an interval $I \in \mathcal{I}$ that contains at least one point in $\{x_1,\dots,x_{r+1}\}$, its \textit{middle piece} refers to the interval $[x_{i^-},x_{i^+}]$ where $x_{i^-}$ (resp., $x_{i^+}$) is the leftmost (resp., rightmost) point in $\{x_1,\dots,x_{r+1}\}$ that is contained in $I$.
For every $i^-,i^+ \in [r+1]$ where $i^- < i^+$, we include in $\mathcal{I}_\text{long}$ the minimum-weight interval in $\mathcal{I}$ whose middle piece is $[x_{i^-},x_{i^+}]$.
Note that $|\mathcal{I}_\text{long}| = O(r^2)$.
Also, we can compute $\mathcal{I}_\text{long}$ in $\widetilde{O}(r^2)$ time using a (dynamic) 2D range-min data structure.
Indeed, if we map each interval $I = [a,b] \in \mathcal{I}$ into the point $(a,b) \in \mathbb{R}^2$ with weight $w(I)$, then the minimum-weight interval in $\mathcal{I}$ whose middle piece is $[x_{i^-},x_{i^+}]$ just corresponds to the minimum-weight point in the range $[x_{i^--1},x_{i^-}] \times [x_{i^+},x_{i^++1}]$.
%What if the sub-structures $\mathcal{D}_1,\dots,\mathcal{D}_r$ do not maintain optimal solutions for the sub-instances and are instead built recursively?
%It turns out that the above approach can still give us a 3-approximation solution of $(S,\mathcal{I})$.
%To see this, we introduce a notion called $(p_1,p_2)$-\textit{approximation}.
%We define the $(p_1,p_2)$-\textit{cost} of a set cover $\mathcal{I}^*$ of $(S,\mathcal{I})$ as the total weight of the one-sided intervals in $\mathcal{I}^*$ times $p_1$ plus the total weight of the two-sided intervals in $\mathcal{I}^*$ times $p_2$.
%We say a set cover of $(S,\mathcal{I})$ is a $(p_1,p_2)$-\textit{approximation} solution if its (normal) cost is smaller than or equal to the $(p_1,p_2)$-cost of any set cover of $(S,\mathcal{I})$.

\subsubsection*{Update of the sub-structures and reconstruction.}
Whenever the instance $(S,\mathcal{I})$ changes, we need to update the sub-structures for which the underlying sub-instances change.
An insertion/deletion on $S$ or $\mathcal{I}$ can change at most two sub-instances.
We also need to re-compute the sets $\mathcal{L}$, $\mathcal{R}$, and $\mathcal{I}_\text{long}$.
As before, our data structure will be periodically reconstructed. 
Specifically, the $(i+1)$-th reconstruction happens after processing $n_i/r$ updates from the $i$-th reconstruction, where $n_i$ denotes the size of $(S,\mathcal{I})$ at the point of the $i$-th reconstruction.
(The 0-th reconstruction is just the initial construction of $\mathcal{D}$.)

\subsubsection*{Constructing a solution.}
For each pair $(L,R)$ where $L \in \mathcal{L}$ and $R \in \mathcal{R}$, we construct a set cover $\mathcal{I}^*(L,R)$ of $(S,\mathcal{I})$ that includes $L$ and $R$ as follows.
Suppose the right (resp., left) endpoint of $L$ (resp., $R$) lies in $J_{i^-}$ (resp., $J_{i^+}$).
If $i^- > i^+$, we simply let $\mathcal{I}^*(L,R) = \{L,R\}$.
If $i^- = i^+$, we temporarily insert the intervals $L$ and $R$ with weight 0 to the sub-instance $(S_i,\mathcal{I}_i)$ where $i = i^- = i^+$ and let $\mathcal{I}_i^*$ be the set cover of $(S_i,\mathcal{I}_i \cup \{L,R\})$ maintained by $\mathcal{D}_i$ excluding the weight-0 intervals $L$ and $R$.
We then define $\mathcal{I}^*(L,R) = \{L,R\} \cup \mathcal{I}_i^*$.
Now assume $i^- < i^+$.
We temporarily insert the interval $L$ (resp., $R$) with weight 0 to the sub-instance $(S_{i^-},\mathcal{I}_{i^-})$ (resp., $(S_{i^+},\mathcal{I}_{i^+})$) and let $\mathcal{I}_{i^-}^*$ (resp., $\mathcal{I}_{i^+}^*$) be the set cover of $(S_{i^-},\mathcal{I}_{i^-} \cup \{L\})$ (resp., $(S_{i^+},\mathcal{I}_{i^+} \cup \{R\})$) maintained by $\mathcal{D}_{i^-}$ (resp., $\mathcal{D}_{i^+}$) excluding the weight-0 interval $L$ (resp., $R$).
For $i^- < i < i^+$, let $\mathcal{I}_i^*$ be the set cover of $(S_i,\mathcal{I}_i)$ maintained by $\mathcal{D}_i$.
We construct $\mathcal{I}^*(L,R)$ using the DP procedure described before.
Let $\text{OPT}[0,\dots,i^+],\mathcal{I}_\text{long}^*[0,\dots,i^+],P[0,\dots,i^+]$ be three tables to be computed.
Set $\text{OPT}[i] = 0$, $\mathcal{I}_\text{long}^*[i] = \emptyset$, and $P[i] = \emptyset$ for all $i<i^-$.
For each $i$ from $i^-$ to $i^+$, we fill out the entries $\text{OPT}[i]$, $\mathcal{I}_\text{long}^*[i]$, $P[i]$ as follows.
We find the interval $I \in \mathcal{I}_\text{long}$ satisfying $J_i \subseteq I$ that minimizes $\text{OPT}[\pi(I)] + w(I)$ where $\pi(I) \in [r]$ is the index such that the left endpoint of the middle piece of $I$ is $x_{\pi(I)+1}$ (or equivalently, the left endpoint of $I$ contains in $J_{\pi(I)}$).
If $\text{OPT}[\pi(I)] + w(I) \leq \text{OPT}[i-1] + \mathsf{cost}(\mathcal{I}_i^*)$, then let $\text{OPT}[i] = \text{OPT}[\pi(I)] + w(I)$, $\mathcal{I}_\text{long}^*[i] = \mathcal{I}_\text{long}^*[\pi(I)] \cup \{I\}$, and $P[i] = P[\pi(I)]$.
Otherwise, let $\text{OPT}[i] = \text{OPT}[i-1] + \mathsf{cost}(\mathcal{I}_i^*)$, $\mathcal{I}_\text{long}^*[i] = \mathcal{I}_\text{long}^*[i-1]$, and $P[i] = P[i-1] \cup \{i\}$.
Then we define $\mathcal{I}^*(L,R) = \{L,R\} \sqcup \mathcal{I}_\text{long}^* \sqcup (\bigsqcup_{i \in P} \mathcal{I}_i^*)$ where $\mathcal{I}_\text{long}^* = \mathcal{I}_\text{long}^*[i^+]$ and $P = P[i^+]$.
It is clear that the cost of $\mathcal{I}^*(L,R)$ is equal to $w(L) + w(R) + \text{OPT}[i^+]$.
Also, as one can easily verify, the DP procedure guarantees the following property of $\mathcal{I}^*(L,R)$.
\begin{fact} \label{fact-DP}
Let $P \subseteq \{i^-,\dots,i^+\}$ and $\mathcal{I}' \subseteq \mathcal{I}$ such that for any $i \in \{i^-,\dots,i^+\} \backslash P$, $J_i \subseteq I$ for some $I \in \mathcal{I}'$.
Then $\mathsf{cost}(\mathcal{I}^*(L,R)) \leq w(L) + w(R) + \mathsf{cost}(\mathcal{I}') + \sum_{i \in P} \mathsf{cost}(\mathcal{I}_i^*)$.
\end{fact}

\noindent
We construct $\mathcal{I}^*(L,R)$ for all $L \in \mathcal{L}$ and $R \in \mathcal{R}$.
(Clearly, we cannot afford to construct $\mathcal{I}^*(L,R)$ explicitly as the size of $\mathcal{I}^*(L,R)$ can be large. So what we do is to only compute the DP tables, which implicitly represents $\mathcal{I}^*(L,R)$.)
Finally, among all $\mathcal{I}^*(L,R)$, we take the one of the smallest cost as the set cover solution $\mathcal{I}_\text{appx}$ for $(S,\mathcal{I})$.
%For $i \in [r]$, let $\mathcal{I}_i^*$ be the solution of $(S_i,\mathcal{I}_i)$ maintained by the sub-structure $\mathcal{D}_i$.
%We say a set cover solution $\mathcal{I}^*$ of $(S,\mathcal{I})$ is \textit{regular} if we can write $\mathcal{I}^* = \{L,R\} \sqcup \mathcal{I}_\text{long}^* \sqcup (\bigsqcup_{i \in P} \mathcal{I}_i^*)$ for some $L \in \mathcal{L}$, $R \in \mathcal{R}$, $P \subseteq [r]$, and $\mathcal{I}_\text{long}^* \subseteq \mathcal{I}_\text{long}$ such that for each $i \in [r] \backslash P$, there exists $I \in \mathcal{I}_\text{long}^*$ satisfying $J_i \subseteq I$.

\subsubsection*{Answering queries to the solution.}
How to store the solution $\mathcal{I}_\text{appx}$ for answering queries is essentially the same as the unweighted case in Section~\ref{sec-unweightedint}.
We explicitly calculate and store the cost of $\mathcal{I}_\text{appx}$, and the membership and reporting queries are handled by recursively querying the sub-structures.
By the same analysis as in Section~\ref{sec-unweightedint}, we can answer the size, membership, reporting queries in $O(1)$, $O(\log n)$, $O(|\mathcal{I}_\text{appx}| \log n)$ time, respectively.

\subsubsection*{Correctness.}
It is easy to see that $\mathcal{I}_\text{appx}$ is a set cover of $(S,\mathcal{I})$.
In order to show $w(\mathcal{I}_\text{appx}) \leq (3+\varepsilon) \cdot \mathsf{opt}$, we introduce a new approximation criterion called $(c_1,c_2)$-\textit{approximation}.
We define the $(c_1,c_2)$-\textit{cost} of a set cover $\mathcal{I}^*$ of $(S,\mathcal{I})$ as the total weight of the one-sided intervals in $\mathcal{I}^*$ times $c_1$ plus the total weight of the two-sided intervals in $\mathcal{I}^*$ times $c_2$.
We say a set cover of $(S,\mathcal{I})$ is a $(c_1,c_2)$-\textit{approximate} solution if its (normal) cost is smaller than or equal to the $(c_1,c_2)$-cost of any set cover of $(S,\mathcal{I})$.
%Clearly, a $(c_1,c_2)$-approximation solution is a $\max\{c_1,c_2\}$-approximation solution (in the common sense).
We shall show that $\mathcal{I}_\text{appx}$ is a $(1+\frac{\varepsilon}{2},3+\varepsilon)$-approximate solution for $(S,\mathcal{I})$, which implies $w(\mathcal{I}_\text{appx}) \leq (3+\varepsilon) \cdot \mathsf{opt}$.
By induction, we can assume that each sub-structure maintains a $(1+\frac{\tilde{\varepsilon}}{2},3+\tilde{\varepsilon})$-approximate solution for $(S_i,\mathcal{I}_i)$.
Consider a set cover $\mathcal{I}_\text{opt}$ of $(S,\mathcal{I})$ with minimum $(1+\frac{\varepsilon}{2},3+\varepsilon)$-cost.
Note that the $(1+\frac{\varepsilon}{2},3+\varepsilon)$-cost of $\mathcal{I}_\text{opt}$ is at most $(3+\varepsilon) \cdot \mathsf{opt}$ and hence the (normal) cost is at most $(3+\varepsilon) \cdot \mathsf{opt}$.
Let $L$ and $R$ be the left and right one-sided intervals used in $\mathcal{I}^*$.
We have $w(L) \leq (3+\varepsilon) \cdot \mathsf{opt} \leq \delta_m$, and thus $w(L) \in [\delta_{u-1},\delta_u]$ for some $u \in [m]$.
Similarly, $w(R) \in [\delta_{v-1},\delta_v]$ for some $v \in [m]$.
By construction, we have $L \cap [0,1] \subseteq L_u \cap [0,1]$ and $R \cap [0,1] \subseteq R_v \cap [0,1]$.
%Define $L = L_u$ and $R = R_v$.
Thus, $\mathcal{I}_\text{opt}' = (\mathcal{I}_\text{opt} \backslash \{L,R\}) \cup \{L_u,R_v\}$ is also a set cover of $(S,\mathcal{I})$.
Furthermore, we notice the following.
\begin{fact}
The $(1,3+\tilde{\varepsilon})$-cost of $\mathcal{I}_\textnormal{opt}'$ is at most the $(1+\frac{\varepsilon}{2},3+\varepsilon)$-cost of $\mathcal{I}_\textnormal{opt}$.
\end{fact}
\begin{myproof}
The $(1,3+\tilde{\varepsilon})$-cost of $\mathcal{I}_\textnormal{opt}'$ is equal to $w(L_u) + w(R_v) + (3+\tilde{\varepsilon}) \cdot \mathsf{cost}(\mathcal{I}_\text{opt} \backslash \{L,R\})$.
Clearly, $(3+\tilde{\varepsilon}) \cdot \mathsf{cost}(\mathcal{I}_\text{opt} \backslash \{L,R\})$ is the $(0,3+\tilde{\varepsilon})$-cost of $\mathcal{I}_\text{opt}$.
We shall show that $w(L_u) + w(R_v)$ is at most the $(1+\frac{\varepsilon}{2},(\varepsilon - \alpha \varepsilon)/2)$-cost of $\mathcal{I}_\text{opt}$, which implies the claim in the fact.
We have 
\begin{equation*}
    w(L_u) \leq (1+\tilde{\varepsilon}/2) \cdot w(L) + \delta_1 \leq (1+\tilde{\varepsilon}/2) \cdot w(L) + ((\varepsilon-\alpha\varepsilon)/4) \cdot \mathsf{opt},
\end{equation*}
and similarly $w(R_v) \leq (1+\tilde{\varepsilon}/2) \cdot w(R) + ((\varepsilon-\alpha\varepsilon)/4) \cdot \mathsf{opt}$.
It follows that $w(L_u) + w(R_v) \leq (1+\tilde{\varepsilon}/2) \cdot (w(L) + w(R)) + ((\varepsilon-\alpha\varepsilon)/2) \cdot \mathsf{opt}$.
Note that $((\varepsilon-\alpha\varepsilon)/2) \cdot \mathsf{opt} \leq ((\varepsilon-\alpha\varepsilon)/2) \cdot \mathsf{cost}(\mathcal{I}_\text{opt})$ and $((\varepsilon-\alpha\varepsilon)/2) \cdot \mathsf{cost}(\mathcal{I}_\text{opt})$ is the $((\varepsilon-\alpha\varepsilon)/2,(\varepsilon-\alpha\varepsilon)/2)$-cost of $\mathcal{I}_\text{opt}$.
Also, $(1+\tilde{\varepsilon}/2) \cdot (w(L) + w(R))$ is at most the $(1+\frac{\tilde{\varepsilon}}{2},0)$-cost of $\mathcal{I}_\text{opt}$.
Thus, $w(L_u) + w(R_v)$ is at most the $(1+\frac{\varepsilon}{2},(\varepsilon-\alpha\varepsilon)/2)$-cost of $\mathcal{I}_\text{opt}$.
\end{myproof}

Now it suffices to show that $\mathsf{cost}(\mathcal{I}^*(L_u,R_v))$ is at most the $(1,3+\tilde{\varepsilon})$-cost of $\mathcal{I}_\textnormal{opt}'$.
Suppose the right (resp., left) endpoint of $L_u$ (resp., $R_v$) lies in $J_{i^-}$ (resp., $J_{i^+}$).
If $i^- > i^+$, then $\mathcal{I}^*(L_u,R_v) = \{L_u,R_v\}$ and hence $\mathsf{cost}(\mathcal{I}^*(L_u,R_v))$ is at most the $(1,0)$-cost of $\mathcal{I}_\textnormal{opt}'$.
The remaining cases are $i^- < i^+$ and $i^- = i^+$.
Here we only analyze the case $i^- < i^+$, because the other case $i^- = i^+$ is similar and simpler.
Recall that when computing $\mathcal{I}^*(L_u,R_v)$, we temporarily inserted the interval $L_u$ (resp., $R_v$) with weight 0 to the sub-instance $(S_{i^-},\mathcal{I}_{i^-})$ (resp., $(S_{i^+},\mathcal{I}_{i^+})$) and let $\mathcal{I}_{i^-}^*$ (resp., $\mathcal{I}_{i^+}^*$) be the set cover of $(S_{i^-},\mathcal{I}_{i^-} \cup \{L\})$ (resp., $(S_{i^+},\mathcal{I}_{i^+} \cup \{R\})$) maintained by $\mathcal{D}_{i^-}$ (resp., $\mathcal{D}_{i^+}$) excluding the weight-0 interval $L$ (resp., $R$).
Also, for $i^- < i < i^+$, we let $\mathcal{I}_i^*$ be the set cover of $(S_i,\mathcal{I}_i)$ maintained by $\mathcal{D}_i$.
%Define $\mathcal{I}_\text{opt}' = (\mathcal{I}_\text{opt} \backslash \{L,R\}) \cup \{L_u,R_v\}$, which is also a set cover of $(S,\mathcal{I})$.
Let $P \subseteq \{i^-,\dots,i^+\}$ consist of all indices $i$ such that $J_i \nsubseteq I$ for all $I \in \mathcal{I}_\text{opt}'$ and $\mathcal{I}' \subseteq \mathcal{I}_\text{opt}'$ consist of all intervals that contain at least one point in $\{x_1,\dots,x_{r+1}\}$.
%and let $\mathcal{I}_i' = \mathcal{I}_\text{opt}' \cap \mathcal{I}_i$ for all $i \in P$.
Now let us define another set cover $\mathcal{I}_\text{opt}'' = \{L_u,R_v\} \sqcup \mathcal{I}' \sqcup (\bigsqcup_{i \in P} \mathcal{I}_i^*)$.
Note that the sets $P$ and $\mathcal{I}'$ satisfy the condition in Fact~\ref{fact-DP}.
Thus, by applying Fact~\ref{fact-DP}, we have
\begin{equation*}
    \mathsf{cost}(\mathcal{I}^*(L_u,R_v)) \leq w(L_u) + w(R_v) + \mathsf{cost}(\mathcal{I}') + \sum_{i \in P} \mathsf{cost}(\mathcal{I}_i^*) = \mathsf{cost}(\mathcal{I}_\text{opt}'').
\end{equation*}
With the above inequality, it suffices to show that $\mathsf{cost}(\mathcal{I}_\text{opt}'')$ is at most the $(1,3+\tilde{\varepsilon})$-cost of $\mathcal{I}_\text{opt}'$.
Equivalently, we show that $\mathsf{cost}(\mathcal{I}') + \sum_{i \in P} \mathsf{cost}(\mathcal{I}_i^*)$ is at most the $(0,3+\tilde{\varepsilon})$-cost of $\mathcal{I}_\text{opt}' \backslash \{L_u,R_v\}$.

Let $\mathcal{I}_i' = \mathcal{I}_\text{opt}' \cap \mathcal{I}_i$ for all $i \in P$.
Note that $\mathcal{I}_i'$ is a set cover of $(S_i,\mathcal{I}_i)$ for $i \in P \backslash \{i^-,i^+\}$.
By assumption, $\mathsf{cost}(\mathcal{I}_i^*)$ is at most the $(1+\tilde{\varepsilon}/2,3+\tilde{\varepsilon})$-cost of $\mathcal{I}_i'$ for $i \in P \backslash \{i^-,i^+\}$.
Also, it is easy to see that if $i^- \in P$ (resp., $i^+ \in P$), then $\mathsf{cost}(\mathcal{I}_{i^-}^*)$ (resp., $\mathsf{cost}(\mathcal{I}_{i^+}^*)$) is at most the $(1+\tilde{\varepsilon}/2,3+\tilde{\varepsilon})$-cost of $\mathcal{I}_{i^-}'$ (resp., $\mathcal{I}_{i^+}'$), because $\mathcal{I}_{i^-}' \cup L$ (resp., $\mathcal{I}_{i^+}' \cup R$) is a set cover of $(S_{i^-},\mathcal{I}_{i^-} \cup \{L\})$ (resp., $(S_{i^+},\mathcal{I}_{i^+} \cup \{R\})$) of the same cost as $\mathcal{I}_{i^-}'$ (resp., $\mathcal{I}_{i^+}'$) when $w(L) = 0$ (resp., $w(R) = 0$).
Let $\mathsf{cost}_i$ be the $(1+\tilde{\varepsilon}/2,3+\tilde{\varepsilon})$-cost of $\mathcal{I}_i'$ for $i \in P$.
By the above observation, we have $\sum_{i \in P} \mathsf{cost}(\mathcal{I}_i^*) \leq \sum_{i \in P} \mathsf{cost}_i$.
Each interval in $\mathcal{I}'$ belongs to (at most) two sub-instances as \textit{one-sided} intervals, so its weight is counted in $\sum_{i \in P} \mathsf{cost}_i$ with a multiplier at most $2 \cdot (1+\tilde{\varepsilon}/2) = 2+\tilde{\varepsilon}$.
Each interval in $\mathcal{I}_\text{opt}' \backslash (\mathcal{I}' \cup \{L_u,R_v\})$ belongs to one sub-instance, so its weight is counted in $\sum_{i \in P} \mathsf{cost}_i$ with a multiplier at most $3+\tilde{\varepsilon}$.
As a result, the weight of each interval in $\mathcal{I}_\text{opt}' \backslash \{L_u,R_v\}$ is counted in $\mathsf{cost}(\mathcal{I}') + \sum_{i \in P} \mathsf{cost}_i$ with a multiplier at most $3+\tilde{\varepsilon}$.
Because $\mathsf{cost}(\mathcal{I}') + \sum_{i \in P} \mathsf{cost}(\mathcal{I}_i^*) \leq \mathsf{cost}(\mathcal{I}') + \sum_{i \in P} \mathsf{cost}_i$, we know that $\mathsf{cost}(\mathcal{I}') + \sum_{i \in P} \mathsf{cost}(\mathcal{I}_i^*)$ is at most the $(0,3+\tilde{\varepsilon})$-cost of $\mathcal{I}_\text{opt}' \backslash \{L_u,R_v\}$.
It follows that $\mathsf{cost}(\mathcal{I}_\text{opt}'')$ is at most the $(1,3+\tilde{\varepsilon})$-cost of $\mathcal{I}_\text{opt}'$, which in turn implies $\mathcal{I}_\text{appx}$ is a $(1+\frac{\varepsilon}{2},3+\varepsilon)$-approximate solution of $(S,\mathcal{I})$.

\subsubsection*{Update time.}
We first observe that, except recursively updating the sub-structures, the (amortized) time cost of all the other work is $\widetilde{O}(r^3 m^2)$.
Specifically, computing the sets $\mathcal{L}$ and $\mathcal{R}$ can be done in $\widetilde{O}(m)$ time and computing $\mathcal{I}_\text{long}$ takes $\widetilde{O}(r^2)$ time.
Using DP to compute each $\mathcal{I}^*(L,R)$ can be done in $O(r^3)$ time, and hence constructing $\mathcal{I}_\text{appx}$ takes $O(r^3 m^2)$ time.
Storing $\mathcal{I}_\text{appx}$ for answering the queries can be done in $\widetilde{O}(r)$ time.
The reconstruction of the data structure takes $\widetilde{O}(r)$ amortized time.
Next, we consider the recursive updates of the sub-structures.
The depth of the recursion is $O(\log_r n)$.
If we set $\alpha = 1-1/\log_r n$, the approximation factor is $\Theta(\varepsilon)$ in any level of the recursion.
When inserting/deleting a point or a interval, we need to update at most two sub-structures whose underlying sub-instances change.
Besides, when computing $\mathcal{I}^*(L,R)$, we need to temporarily insert $L$ and $R$ with weight 0 to two sub-instances (and delete them afterwards), which involves a constant number of recursive updates.
So the total number of recursive updates is $O(m^2) = \widetilde{O}((\frac{1}{\tilde{\varepsilon}} \log \frac{1}{\tilde{\varepsilon}})^2) = \widetilde{O}((\frac{1}{\varepsilon} \log \frac{1}{\varepsilon})^2)$.
Therefore, if we use $U(n)$ to denote the update time when the instance size is $n$, we have the recurrence
\begin{equation*}
    U(n) = \widetilde{O}\left(\left(\frac{1}{\varepsilon} \log \frac{1}{\varepsilon}\right)^2\right) \cdot U(O(n/r)) + \widetilde{O}\left(r^3 \cdot \left(\frac{1}{\varepsilon} \log \frac{1}{\varepsilon}\right)^2\right),
\end{equation*}
which solves to $U(n) = (\log n \cdot \frac{1}{\varepsilon} \log \frac{1}{\varepsilon})^{O(\log_r n)} \cdot r^3$.
By setting $r = 2^{\sqrt{\log n \log\log n} + \sqrt{\log n \log(1/\varepsilon)}}$, we have $U(n) = 2^{O(\sqrt{\log n \log\log n} + \sqrt{\log n \log(1/\varepsilon)})}$.
%Also, the time for answering a reporting query becomes $O(|\mathcal{I}_\text{appx}| \sqrt{\log n/\log\log n})$.

%define $\mathcal{I}_i^* = \{I\}$ if $w(I)$ is smaller than or equal to the cost of the set cover of $(S_i,\mathcal{I}_i)$ maintained by $\mathcal{D}_i$.

%\weightedint*
\begin{theorem}
There exists a dynamic data structure for $(3+\varepsilon)$-approximate weighted interval set cover with $2^{O(\sqrt{\log n \log\log n} + \sqrt{\log n \log(1/\varepsilon)})}$ amortized update time and $\widetilde{O}(n)$ construction time, which can answer size, membership, and reporting queries in $O(1)$, $O(\log n)$, and $O(k \log n)$ time, respectively, where $n$ is the size of the instance and $k$ is the size of the maintained solution.
\end{theorem}

\section{Weighted Unit-Square Set Cover}
In this section, we present the first sublinear result for dynamic weighted unit-square set cover, which gets $O(1)$-approximation. It suffices to consider dynamic weighted quadrant set cover, since the reduction from dynamic unit-square set cover to dynamic quadrant set cover \cite{agarwal2020dynamic} still works in the weighted case.
%\begin{lemma}[\cite{agarwal2020dynamic}]
%Suppose there exists a $c$-approximate dynamic weighted quadrant set cover data structure with $f(n)$ amortized update time and $\tilde{O}(n_0)$ construction time. Then there exists an $O(c)$-approximate dynamic weighted unit-square set cover data structure with $\tilde{O}(f(n))$ amortized update time and $\tilde{O}(n_0)$ construction time.
%\end{lemma}
%dominance ranges with $4$ directions.

Let $(S,\mathcal{Q})$ be a dynamic weighted quadrant set cover instance where $S$ is the set of points in $\mathbb{R}^2$ and $\mathcal{Q}$ is the set of weighted quadrants, and let $n=|S|+|\mathcal{Q}|$ denote the instance size. We use $w(q)$ to denote the weight of a quadrant $q$, and $w(\mathcal{Q})$ for the total weight of a set $\mathcal{Q}$ of quadrants. W.l.o.g., assume the points in $S$ lie in the point range $[0,1]^2$. For simplicity, we assume the weights are positive integers bounded by $U=\mathrm{poly}(n)$.
%For simplicity, We will later comment on how to remove this assumption.

Our idea is based on our unweighted solution for quadrant set cover as explained in Sec.~\ref{unweighted unit-square (long version)}, which recursively solve for smaller sub-instances and properly combine them to obtain the global solution. In particular, we will again partition the space into $r\times r$ rectangular grid cells. However, our previous solution relies on an output-sensitive algorithm (Lemma~\ref{lem-quadoslong}), but such algorithm is not known in the weighted case. Therefore, we will introduce some new ideas.
%dynamic programming
%use a different approach

\subsubsection*{Data structures.} We construct a data structure $\mathcal{D}$ that supports a more powerful type of query:

\begin{quote}
Given a query rectangle $t$, compute an $O(1)$-approximate weighted set cover for the points in $S\cap t$, using the quadrants in $\mathcal{Q}$.
\end{quote}

For a quadrant $q\in \mathcal{Q}$ intersecting a rectangular range $\Gamma$, we say it is \emph{trivial} (resp., \emph{nontrivial}) with respect to $\Gamma$ if the vertex of $q$ is outside (resp., inside) $\Gamma$. Similar to the observation in the unweighted case, our idea is that it suffices to only keep a small subset of the trivial quadrants. In particular, among all trivial quadrants with weights $\in [1,2^{i})$, there are (at most) four maximal quadrants in $\Gamma$, which we denote as $\mathcal{M}_{\Gamma,i}$. (In the special case that there exist quadrants with weight in $[1,2^i)$ that completely contain $\Gamma$, $\mathcal{M}_{\Gamma,i}$ will contain any one among them.) We store $\mathcal{M}_{[0,1]^2,i}$ for $i=1,\dots,\log U$, and only keep the nontrivial quadrants with respect to $[0,1]^2$ in $\mathcal{Q}$. The intuition is suppose among all trivial quadrants in the optimal solution $\mathcal{Q}_{\mathrm{opt}}$ for $\Gamma$, the maximum weight is $w$, then for an $O(1)$-approximate solution we can just include $\mathcal{M}_{\Gamma,\lceil\log w\rceil}$, and then compute an $O(1)$-approximate solution in the complement region, using only the nontrivial quadrants. The union of the quadrants in $\mathcal{M}_{\Gamma,\lceil\log w\rceil}$ will contain the union of all trivial quadrants in $\mathcal{Q}_{\mathrm{opt}}$.

To build the data structure $\mathcal{D}$, we partition the space into $r\times r$ (nonuniform) grid cells using $r-1$ horizontal/vertical lines, such that each row (resp., column) has size $O(\frac{n}{r})$, where the size of a range $\Gamma$ is defined as the total number of points in $S$ and vertices of quadrants in $\mathcal{Q}$ inside $\Gamma$. Let $\Box_{i,j}$ be the cell in the $i$-th row and $j$-th column for $(i,j) \in [r]^2$. We define sub-instances for the rows/columns of the partition. In particular, let $R_i = \bigcup_{j=1}^r \Box_{i,j}$ denote the $i$-th row and $C_j = \bigcup_{i=1}^r \Box_{i,j}$ denote the $j$-th column of the partition. Create a sub-instance $(S_{i,\bullet},\mathcal{Q}_{i,\bullet})$ for each row $i$, where $S_{i,\bullet} = S \cap R_i$ contains all points in the row, and $\mathcal{Q}_{i,\bullet}$ contains all quadrants that are nontrivial with respect to the row. Similarly, create a sub-instance $(S_{\bullet,j},\mathcal{Q}_{\bullet,j})$ for each column $j$. Recursively construct the data structures $\mathcal{D}_{i,\bullet}$ for each of the $r$ rows, and similarly $\mathcal{D}_{\bullet,j}$ for each of the $r$ columns. Also store the sets of maximal quadrants $\mathcal{M}_{R_i,k}$ and $\mathcal{M}_{C_j,k}$ in the rows and columns, for $k=1,\dots,\log U$.

Let $\Box_{i,j,k,l}=\bigcup_{i'=i}^k\bigcup_{j'=j}^l\Box_{i',j'}$ denote the grid-aligned rectangular region from row $i$ to $k$ and from column $j$ to $l$. For each of these $O(r^4)$ grid-aligned rectangles $\Box_{i,j,k,l}$, we also maintain an (implicit) $O(1)$-approximate set cover solution $\mathcal{Q}_{\mathrm{appx}}(\Box_{i,j,k,l})$ within it, so that its weight $w(\mathcal{Q}_{\mathrm{appx}}(\Box_{i,j,k,l}))$ can be retrieved in $O(1)$ time.

\begin{figure}[htbp]\centering
\includegraphics[scale=0.8]{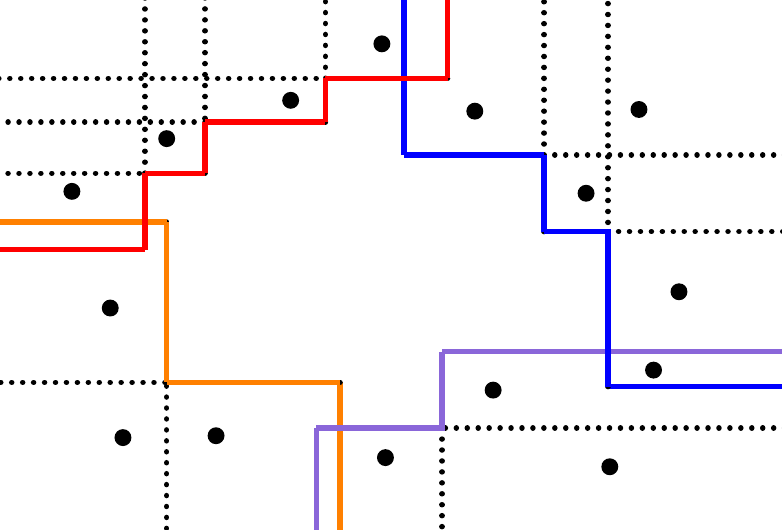}\\
 \caption{Decomposing a set cover solution into four subsets, based on the four directions of the quadrants. The boundary of the union of each of these subsets forms an orthogonal staircase curve.}\label{fig:staircase}
\end{figure}

\subsubsection*{Dynamic programming.} In the following, we show that given the substructures $\mathcal{D}_{\bullet,j}$ for the columns that support rectangular ranged queries, we can efficiently compute an $O(1)$-approximate solution $\mathcal{Q}_{\mathrm{appx}}(\Box_{i,j,k,l})$ for each of the $O(r^4)$ grid-aligned rectangles $\Box_{i,j,k,l}$, using dynamic programming.

Consider any grid-aligned rectangle $\Box_{i,j,k,l}$. Any set cover solution $\mathcal{Q}_{\mathrm{sol}}(\Box_{i,j,k,l})$ can be decomposed into four subsets $\mathcal{Q}_{\mathrm{sol}}^{\mathrm{NW}},\mathcal{Q}_{\mathrm{sol}}^{\mathrm{SW}},\mathcal{Q}_{\mathrm{sol}}^{\mathrm{NE}},\mathcal{Q}_{\mathrm{sol}}^{\mathrm{SE}}$, based on the four directions of the quadrants (northwest/southwest/
northeast/southeast). The boundary of the union of each of these subsets forms an orthogonal staircase curve (as shown in Figure~\ref{fig:staircase}), which we denote as $U_{\mathrm{sol}}^{\mathrm{NW}},U_{\mathrm{sol}}^{\mathrm{SW}},U_{\mathrm{sol}}^{\mathrm{NE}},U_{\mathrm{sol}}^{\mathrm{SE}}$, respectively.

Let $\ell_0,\dots,\ell_r$ denote the $r+1$ vertical lines that define the grid (including the boundary), and obtain vertical line $\ell_i^-$ by slightly shifting $\ell_i$ to the left. We use $f[i_0][q_{_{\nwarrow}}][q_{_{\swarrow}}][q_{_{\nearrow}}][q_{_{\searrow}}]$ to denote the weight of an $O(1)$-approximate set cover $\mathcal{Q}_{\mathrm{appx}}$ that covers all points in $\Box_{i,j,k,l}$ and to the left of $\ell_{i_0}$, such that when we decompose $\mathcal{Q}_{\mathrm{appx}}$ into four orthogonal staircase curves $U_{\mathrm{appx}}^{\mathrm{NW}},U_{\mathrm{appx}}^{\mathrm{SW}},U_{\mathrm{appx}}^{\mathrm{NE}},U_{\mathrm{appx}}^{\mathrm{SE}}$, $\ell_{i_0}^-$ intersects them at quadrants $q_{_{\nwarrow}},q_{_{\swarrow}},q_{_{\nearrow}},q_{_{\searrow}}$, respectively. (For the special case that $\ell_{i_0}'$ does not intersect $U_{\mathrm{appx}}^{\mathrm{NW}}$, we let $q_{_{\nwarrow}}$ to be a special ``null'' element. Similarly for $q_{_{\swarrow}},q_{_{\nearrow}},q_{_{\searrow}}$.)

We decompose each quadrant $q\in \mathcal{Q}$ into two parts: if $q$ has direction west (resp., east), the \emph{long} part is aligned with the rightmost (resp., leftmost) vertical grid line that intersects $q$, and the \emph{short} part covers the remaining space, which is fully contained in a column (i.e., nontrivial with respect to the column). In this way, each quadrant is duplicated twice, so the approximation factor will multiply by at most $2$.

For each vertical grid line $\ell_i$, among all long quadrants aligned with $\ell_i$, it suffices to keep the lowest (resp., highest) quadrant with direction north (resp., south) with weight $\in [2^k,2^{k+1})$, for each $k=1,\dots,\log U$. We keep $O(r\log U)$ long quadrants in total. The approximation factor will only multiply by a constant.

To compute $f[i_0][q_{_{\nwarrow}}'][q_{_{\swarrow}}'][q_{_{\nearrow}}'][q_{_{\searrow}}']$ (corresponding to the set cover solution $\mathcal{Q}_{\mathrm{appx}}'$) where the quadrants $q_{_{\nwarrow}}',q_{_{\swarrow}}',q_{_{\nearrow}}',q_{_{\searrow}}'$ are long (it suffices to only consider the long quadrants after the decomposition, since $\ell_{i_0}$ is a grid line), we guess that the four orthogonal staircase curves $U_{\mathrm{appx}}'^{\mathrm{NW}},U_{\mathrm{appx}}'^{\mathrm{SW}},U_{\mathrm{appx}}'^{\mathrm{NE}},U_{\mathrm{appx}}'^{\mathrm{SE}}$ corresponding to $\mathcal{Q}_{\mathrm{appx}}'$ intersect $\ell_{i_0-1}^-$ at the four long quadrants $q_{_{\nwarrow}},q_{_{\swarrow}},q_{_{\nearrow}},q_{_{\searrow}}$. The set cover solution $\mathcal{Q}_{\mathrm{appx}}$ corresponding to $f[i_0-1][q_{_{\nwarrow}}][q_{_{\swarrow}}][q_{_{\nearrow}}][q_{_{\searrow}}]$ already covers all points to the left of the vertical line $\ell_{i_0-1}$, so to obtain $\mathcal{Q}_{\mathrm{appx}}'$, we only need to cover the points between the two grid lines $\ell_{i_0-1}$ and $\ell_{i_0}$. In particular, the region within $\Box_{i,j,k,l}$ and between the vertical lines $\ell_{i_0-1}$ and $\ell_{i_0}$ that are not covered by the long quadrants $q_{_{\nwarrow}}',q_{_{\swarrow}}',q_{_{\nearrow}}',q_{_{\searrow}}'$ is a rectangle $t'$, and we need to cover $t'$ using the short quadrants in column $i_0$.

In other words, $f[i_0][q_{_{\nwarrow}}'][q_{_{\swarrow}}'][q_{_{\nearrow}}'][q_{_{\searrow}}']$ can be computed by the formula
$$f[i_0][q_{_{\nwarrow}}'][q_{_{\swarrow}}'][q_{_{\nearrow}}'][q_{_{\searrow}}']=\min_{q_{_{\nwarrow}},q_{_{\swarrow}},q_{_{\nearrow}},q_{_{\searrow}}} \Big(f[i_0-1][q_{_{\nwarrow}}][q_{_{\swarrow}}][q_{_{\nearrow}}][q_{_{\searrow}}]+w(\mathcal{Q}_{\mathrm{appx}}(t'))$$
$$+\sum_{d\in \{\nwarrow,\swarrow,\nearrow,\searrow\}} I[q_d'\neq q_d]\cdot w(q_d')\Big).$$
The weight of an $O(1)$-approximate set cover solution $\mathcal{Q}_{\mathrm{appx}}(t')$ for the rectangle $t'$ can be obtained by querying the column substructure $\mathcal{D}_{\bullet,i_0}$. The final solution is the one with minimum weight among $f[r][q_{_{\nwarrow}}][q_{_{\swarrow}}][q_{_{\nearrow}}][q_{_{\searrow}}]$ for all possible long quadrants $q_{_{\nwarrow}},q_{_{\swarrow}},q_{_{\nearrow}},q_{_{\searrow}}$.

We maintain pointers during the dynamic programming process about how the minimum weight is obtained, so that the actual solution can be easily recovered.

To compute the solution for each grid-aligned rectangle $\Box_{i,j,k,l}$ we need to perform $(r\log U)^{O(1)}$ queries, since there are only $O(r\log U)$ choices for each of the long quadrants $q_{_{\nwarrow}}',q_{_{\swarrow}}',q_{_{\nearrow}}',q_{_{\searrow}}'$. There are $O(r^4)$ such grid-aligned rectangles, so the total running time is $(r\log U)^{O(1)}\cdot Q(O(\frac{n}{r}))$, which is $O(n^{O(\delta)})$ as calculated later.

%\quad\newline

\subsubsection*{Construction time.} Let $T(n)$ denote the construction time for the data structure $\mathcal{D}$ when the instance size is $n$. The construction time satisfies the recurrence
$$T(n)=\sum_{i=1}^{2r} T(n_i)+\tilde{O}(n)+O(n^{O(\delta)}),$$
where $n_i=O(\frac{n}{r})$ is the instance size of a row/column, and we have $\sum_{i=1}^{2r} n_i\leq 2n$. Set $r=N^{\delta}$ where $N$ is the global upper bound on the instance size and $\delta>0$ is an arbitrarily small constant. The recurrence solves to $T(n)=\tilde{O}(n)$.

\subsubsection*{Query.} Given a query rectangle $t$, we first \emph{guess} that the maximum weight among all trivial quadrants with respect to $[0,1]^2$ used in the optimal solution $\mathcal{Q}_{\mathrm{opt}}(t)$ is within $[2^{k-1},2^{k})$, and include $\mathcal{M}_{[0,1]^2,k}$ in the approximate solution $\mathcal{Q}_{\mathrm{appx}}(t)$. Let $t'$ denote the complement region of the union of quadrants in $\mathcal{M}_{[0,1]^2,k}$, it suffices to query for the rectangular region $t_0=t\cap t'$. There are $O(\log U)$ possible choices for $k=1,\dots,\log U$, so we need to perform $O(\log U)$ queries and then take the minimum among the results.

The query rectangle $t_0$ can be decomposed into a grid-aligned rectangle $\Box_{i,j,k,l}$ and at most four rectangles $L_\leftarrow,L_\rightarrow,L_\uparrow,L_\downarrow$ which are contained within a row/column, as shown in Figure~\ref{fig-annuluslong}. To compute the approximate set cover $\mathcal{Q}_{\mathrm{appx}}(t_0)$, it suffices to take the union of the approximate solutions within $\Box_{i,j,k,l}$ and $L_\leftarrow,L_\rightarrow,L_\uparrow,L_\downarrow$, i.e., let $\mathcal{Q}_{\mathrm{appx}}(t_0)=\mathcal{Q}_{\mathrm{appx}}(\Box_{i,j,k,l})\cup \mathcal{Q}_{\mathrm{appx}}(L_\leftarrow)\cup \mathcal{Q}_{\mathrm{appx}}(L_\rightarrow)\cup \mathcal{Q}_{\mathrm{appx}}(L_\uparrow)\cup \mathcal{Q}_{\mathrm{appx}}(L_\downarrow)$. The approximation factor will only grow by a factor of $5$.

$\mathcal{Q}_{\mathrm{appx}}(\Box_{i,j,k,l})$ has already been maintained, so we can retrieve its weight in $O(1)$ time. To compute $\mathcal{Q}_{\mathrm{appx}}(L_\uparrow)$ (and similarly for $L_\leftarrow,L_\rightarrow,L_\downarrow$), we perform a query on the rectangle $L_\uparrow$ using the substructure $\mathcal{D}_{i-1,\bullet}$ for row $i-1$, since the rectangle $L_\uparrow$ is contained in row $i-1$.

Let $Q(n)$ denote the query time for instance size $n$. It satisfies the recurrence
\begin{equation}
    Q(n)=O(\log U)\cdot \left(4\cdot Q\left(O\left(\frac{n}{r}\right)\right)+O(1)\right).
\label{eqn:query}    
\end{equation}
The recursion depth is $O(\log_r n)=O(\frac{1}{\delta})=O(1)$, so we have $Q(n)=O(\log U)^{O(\frac{1}{\delta})}=\log^{O(1)}n$. The approximation factor grows by a constant factor at each level, so the whole approximation factor is $O(1)^{O(\frac{1}{\delta})}=O(1)$.
%=O(n^{O(\delta)})

%report the solution
%It is not hard to retrieve the actual solution in $O(opt)$ time.

\subsubsection*{Update.} When we insert/delete a quadrant $q$, recursively update the substructures $\mathcal{D}_{i^*,\bullet}$ and $\mathcal{D}_{\bullet,j^*}$ for the $i^*$-th row and $j^*$-th column that contain the vertex of $q$. Update the sets of maximal quadrants $\mathcal{M}_{R_i,k}$ and $\mathcal{M}_{C_j,k}$ in the rows and columns, for $i,j\in [r]$ and $k=1,\dots,\log U$. Recompute the $O(1)$-approximate set cover solutions $\mathcal{Q}_{\mathrm{appx}}(\Box_{i,j,k,l})$ for each of the $O(r^4)$ grid-aligned rectangles $\Box_{i,j,k,l}$, using dynamic programming in $(r\log U)^{O(1)}\cdot Q(O(\frac{n}{r}))$ time.

When we insert/delete a point $p$, recursively update the substructures $\mathcal{D}_{i^*,\bullet}$ and $\mathcal{D}_{\bullet,j^*}$ for the $i^*$-th row and $j^*$-th column that contain $p$, and also recompute  $\mathcal{Q}_{\mathrm{appx}}(\Box_{i,j,k,l})$ for each of the $O(r^4)$ grid-aligned rectangles $\Box_{i,j,k,l}$.

We reconstruct the entire data structure after every $\frac{n}{r}$ updates, so that the row and column sizes are always bounded by $O(\frac{n}{r})$. The amortized cost per update is $\frac{T(n)}{n/r}=\tilde{O}(r)$.

Let $\mathcal{U}(n)$ denote the update time for instance size $n$. It satisfies the recurrence
$$\mathcal{U}(n)=2\mathcal{U}\left(O\left(\frac{n}{r}\right)\right)+(r\log U)^{O(1)}\cdot \mathcal{Q}\left(O\left(\frac{n}{r}\right)\right)+\tilde{O}(r),$$
which solves to $\mathcal{U}(n)=O(n^{O(\delta)})$.

%\paragraph{Running time analysis.} 

\begin{theorem}
There exists a dynamic data structure for $O(1)$-approximate weighted unit-square set cover with $O(n^{\delta})$ amortized update time and $\tilde{O}(n)$ construction time, for any constant $\delta > 0$ (assuming polynomially bounded integer weights).
\end{theorem}

\subsubsection*{Remark.}
It is possible to remove the assumption of polynomially bounded weights with
more work.  One way is to directly modify the recursive query algorithm, as we now briefly sketch:  

First,
we solve the approximate decision problem, of deciding whether the optimal value
is approximately less than a given value $W_0$.  To this end, it suffices to consider $O(\log n)$ choices for $k$ instead of $O(\log U)$ (namely, 
$k=\lceil\log\frac{W_0}{cn}\rceil,\ldots,\lceil\log W_0\rceil$ for a large constant~$c$), since replacing a quadrant with weight less than $\frac{W_0}{cn}$ with another one 
weight less than $\frac{W_0}{cn}$ causes only additive error $O(\frac{W_0}{cn})$, which is tolerable even when summing over all $O(n)$ quadrants.
(We don't explicitly store the maximal quadrants in
$\mathcal{M}_{\Gamma,i}$ 
for all $i=1,\ldots,\log U$, but can generate them on demand by orthogonal range searching.) 

Having solved the approximate decision problem, we can next obtain an $O(n)$-approximation of the optimal value, by binary search on the quadrant weights (since the total weight in the optimal solution is 
within an $O(n)$ factor of the maximum
quadrant weight in the optimal solution); this requires $O(\log n)$ calls to the decision oracle.  Knowing an $O(n)$-approximation, 
we can finally obtain an $O(1)$-approximation, by another binary search; this requires
$O(\log\log n)$ additional calls to the decision oracle.
Thus, we get the same recurrence as Equation~\ref{eqn:query}, but with $\log U$
replaced by $\log^{O(1)}n$.  We still obtain $O(n^\delta)$ update time in the end.

%\TIMOTHY{please check @Qizheng}\Qizheng{ok, I'll check it. Yesterday Jie told me an alternative way is to first get a $poly(n)$-approx using the unweighted 2D algorithm (like what he did in 1D).}

%It should be possible to remove the assumption of  polynomially bounded weights with more work.

%\paragraph{Remark.} For arbitrary weights, .
%de-amortize.

\ignore{
\subsubsection{3-sided dominance range with one direction}

using dynamic programming.

\subsubsection*{Data structures.}

\subsubsection*{Running time analysis.}

\begin{theorem}
There exists a dynamic data structure for $O(\log n)$-approximation weighted $3$-sided dominance range set cover with $O(\mathrm{polylog}~n)$ (amortized) update time and $\widetilde{O}(n)$ construction time.
\end{theorem}
}

\section{Conclusion and Future Work} \label{sec-conclusion}
%future works:
%conditional lower bound
%reporting query for 3D halfspaces

We have described improved dynamic data structures for various
versions of the geometric set cover problem, and in particular, achieving very low (polylogarithmic or $n^{o(1)}$) update time for 1D intervals and 2D unit squares, in both the unweighted and weighted settings.  Besides obtaining further improvements of
our update time bounds, there are a number of interesting
directions to explore for future work:

\begin{itemize}
\item We have given sublinear results for unweighted
2D halfplanes with regards to reporting queries, 
but could similar results be obtained
for unweighted 3D halfspaces
and 2D disks?  As mentioned, previous work by Chan and He~\cite{chan2021dynamic} 
can only handle size queries.
\item Are there data structures with sublinear update time
for the dynamic hitting set problem for
ranges such as 2D arbitrary squares?  We can use duality to reduce 
hitting set to set cover in the unit square case, but not in the
arbitrary square case (not even for ``nearly unit'' squares with side lengths in $[1,1+\eps]$).
\item Are there data structures with sublinear update time
for 2D arbitrary rectangles with polylogarithmic approximation factor?
(Demanding constant approximation factor would be unreasonable,
because of the lack of known efficient static $O(1)$-approximation algorithms,
but there are static $O(\log n)$-approximation algorithms
with near-linear running time for 2D rectangles~\cite{agarwal2014near}.)
\item In view of the recent developments in fine-grained complexity and reductions~\cite{virgisurvey},
could one prove $n^{\Omega(1)}$ conditional
lower bounds on the update time for dynamic approximate geometric set cover, e.g., for arbitrary squares or other ranges, based on 
the conjectured hardness of standard problems such as 3SUM, all-pairs shortest paths, or orthogonal vectors?  
\end{itemize}

\bibliographystyle{plain}
\bibliography{my_bib}

\begin{thebibliography}{10}

\bibitem{AbboudA0PS19}
Amir Abboud, Raghavendra Addanki, Fabrizio Grandoni, Debmalya Panigrahi, and
  Barna Saha.
\newblock Dynamic set cover: improved algorithms and lower bounds.
\newblock In {\em Proceedings of the 51st Annual ACM Symposium on Theory of
  Computing (STOC)}, pages 114--125, 2019.
\newblock URL: \url{https://doi.org/10.1145/3313276.3316376}, \href
  {http://dx.doi.org/10.1145/3313276.3316376}
  {\path{doi:10.1145/3313276.3316376}}.

\bibitem{agarwal2020dynamic}
Pankaj~K. Agarwal, Hsien{-}Chih Chang, Subhash Suri, Allen Xiao, and Jie Xue.
\newblock Dynamic geometric set cover and hitting set.
\newblock In {\em Proceedings of the 36th Symposium on Computational Geometry
  (SoCG)}, pages 2:1--2:15, 2020.
\newblock \href {http://dx.doi.org/10.4230/LIPIcs.SoCG.2020.2}
  {\path{doi:10.4230/LIPIcs.SoCG.2020.2}}.

\bibitem{AgaEriSURV}
Pankaj~K. Agarwal and Jeff Erickson.
\newblock Geometric range searching and its relatives.
\newblock In B.~Chazelle, J.~E. Goodman, and R.~Pollack, editors, {\em Advances
  in Discrete and Computational Geometry}, pages 1--56. AMS Press, 1999.

\bibitem{agarwal2012near}
Pankaj~K. Agarwal, Esther Ezra, and Micha Sharir.
\newblock Near-linear approximation algorithms for geometric hitting sets.
\newblock {\em Algorithmica}, 63(1-2):1--25, 2012.

\bibitem{agarwal2014near}
Pankaj~K. Agarwal and Jiangwei Pan.
\newblock Near-linear algorithms for geometric hitting sets and set covers.
\newblock {\em Discrete \& Computational Geometry}, 63(2):460--482, 2020.
\newblock Preliminary version in SoCG'14.

\bibitem{bentley1980decomposable}
Jon~Louis Bentley and James~B. Saxe.
\newblock Decomposable searching problems {I}: {S}tatic-to-dynamic
  transformation.
\newblock {\em Journal of Algorithms}, 1(4):301--358, 1980.

\bibitem{BhattacharyaHNW21}
Sayan Bhattacharya, Monika Henzinger, Danupon Nanongkai, and Xiaowei Wu.
\newblock Dynamic set cover: Improved amortized and worst-case update time.
\newblock In {\em Proceedings of the 32nd {ACM-SIAM} Symposium on Discrete
  Algorithms (SODA)}, pages 2537--2549, 2021.
\newblock URL: \url{https://doi.org/10.1137/1.9781611976465.150}, \href
  {http://dx.doi.org/10.1137/1.9781611976465.150}
  {\path{doi:10.1137/1.9781611976465.150}}.

\bibitem{BhoreCIK20}
Sujoy Bhore, Jean Cardinal, John Iacono, and Grigorios Koumoutsos.
\newblock Dynamic geometric independent set.
\newblock {\em CoRR}, abs/2007.08643, 2020.
\newblock To appear in ESA 2021.
\newblock \href {http://arxiv.org/abs/2007.08643} {\path{arXiv:2007.08643}}.

\bibitem{bronnimann1995almost}
Herv{\'e} Br{\"o}nnimann and Michael~T. Goodrich.
\newblock Almost optimal set covers in finite {VC}-dimension.
\newblock {\em Discrete \& Computational Geometry}, 14(4):463--479, 1995.

\bibitem{bus2018practical}
Norbert Bus, Nabil~H. Mustafa, and Saurabh Ray.
\newblock Practical and efficient algorithms for the geometric hitting set
  problem.
\newblock {\em Discrete Applied Mathematics}, 240:25--32, 2018.

\bibitem{Chan12DCG}
Timothy~M. Chan.
\newblock Optimal partition trees.
\newblock {\em Discrete \& Computational Geometry}, 47(4):661--690, 2012.
\newblock \href {http://dx.doi.org/10.1007/s00454-012-9410-z}
  {\path{doi:10.1007/s00454-012-9410-z}}.

\bibitem{ChanG14}
Timothy~M. Chan and Elyot Grant.
\newblock Exact algorithms and {APX}-hardness results for geometric packing and
  covering problems.
\newblock {\em Comput. Geom.}, 47(2):112--124, 2014.
\newblock URL: \url{https://doi.org/10.1016/j.comgeo.2012.04.001}, \href
  {http://dx.doi.org/10.1016/j.comgeo.2012.04.001}
  {\path{doi:10.1016/j.comgeo.2012.04.001}}.

\bibitem{chan2012weighted}
Timothy~M. Chan, Elyot Grant, Jochen K{\"o}nemann, and Malcolm Sharpe.
\newblock Weighted capacitated, priority, and geometric set cover via improved
  quasi-uniform sampling.
\newblock In {\em Proceedings of the 23rd Annual ACM-SIAM Symposium on Discrete
  Algorithms (SODA)}, pages 1576--1585, 2012.

\bibitem{ChanH20}
Timothy~M. Chan and Qizheng He.
\newblock Faster approximation algorithms for geometric set cover.
\newblock In {\em Proceedings of the 36th Symposium on Computational Geometry
  (SoCG)}, volume 164, pages 27:1--27:14, 2020.
\newblock \href {http://dx.doi.org/10.4230/LIPIcs.SoCG.2020.27}
  {\path{doi:10.4230/LIPIcs.SoCG.2020.27}}.

\bibitem{chan2021dynamic}
Timothy~M. Chan and Qizheng He.
\newblock More dynamic data structures for geometric set cover with sublinear
  update time.
\newblock In {\em Proceedings of the 37th International Symposium on
  Computational Geometry (SoCG)}, pages 25:1--25:14, 2021.
\newblock URL: \url{https://doi.org/10.4230/LIPIcs.SoCG.2021.25}, \href
  {http://dx.doi.org/10.4230/LIPIcs.SoCG.2021.25}
  {\path{doi:10.4230/LIPIcs.SoCG.2021.25}}.

\bibitem{ChanH15}
Timothy~M. Chan and Nan Hu.
\newblock Geometric red-blue set cover for unit squares and related problems.
\newblock {\em Comput. Geom.}, 48(5):380--385, 2015.
\newblock URL: \url{https://doi.org/10.1016/j.comgeo.2014.12.005}, \href
  {http://dx.doi.org/10.1016/j.comgeo.2014.12.005}
  {\path{doi:10.1016/j.comgeo.2014.12.005}}.

\bibitem{clarkson1993algorithms}
Kenneth~L. Clarkson.
\newblock Algorithms for polytope covering and approximation.
\newblock In {\em Proceedings of the 3rd Workshop on Algorithms and Data
  Structures (WADS)}, pages 246--252, 1993.

\bibitem{clarkson1989applications}
Kenneth~L. Clarkson and Peter~W. Shor.
\newblock Applications of random sampling in computational geometry, {II}.
\newblock {\em Discrete \& Computational Geometry}, 4(5):387--421, 1989.

\bibitem{clarkson2007improved}
Kenneth~L. Clarkson and Kasturi Varadarajan.
\newblock Improved approximation algorithms for geometric set cover.
\newblock {\em Discrete \& Computational Geometry}, 37(1):43--58, 2007.

\bibitem{BerBOOK}
Mark de~Berg, Otfried Cheong, Marc~J. van Kreveld, and Mark~H. Overmars.
\newblock {\em Computational Geometry: Algorithms and Applications}.
\newblock Springer, 3rd edition, 2008.

\bibitem{ErlebachL10}
Thomas Erlebach and Erik~Jan van Leeuwen.
\newblock {PTAS} for weighted set cover on unit squares.
\newblock In {\em Proceedings of 13th International Workshop on Approximation,
  Randomization, and Combinatorial Optimization (APPROX)}, pages 166--177,
  2010.
\newblock \href {http://dx.doi.org/10.1007/978-3-642-15369-3_13}
  {\path{doi:10.1007/978-3-642-15369-3_13}}.

\bibitem{GuptaK0P17}
Anupam Gupta, Ravishankar Krishnaswamy, Amit Kumar, and Debmalya Panigrahi.
\newblock Online and dynamic algorithms for set cover.
\newblock In {\em Proceedings of the 49th Annual {ACM} Symposium on Theory of
  Computing (STOC)}, pages 537--550, 2017.
\newblock URL: \url{https://doi.org/10.1145/3055399.3055493}, \href
  {http://dx.doi.org/10.1145/3055399.3055493}
  {\path{doi:10.1145/3055399.3055493}}.

\bibitem{Har-PeledL12}
Sariel Har{-}Peled and Mira Lee.
\newblock Weighted geometric set cover problems revisited.
\newblock {\em J. Comput. Geom.}, 3(1):65--85, 2012.
\newblock URL: \url{https://doi.org/10.20382/jocg.v3i1a4}, \href
  {http://dx.doi.org/10.20382/jocg.v3i1a4} {\path{doi:10.20382/jocg.v3i1a4}}.

\bibitem{Henzinger0W20}
Monika Henzinger, Stefan Neumann, and Andreas Wiese.
\newblock Dynamic approximate maximum independent set of intervals, hypercubes
  and hyperrectangles.
\newblock In {\em Proceedings of the 36th Symposium on Computational Geometry
  (SoCG)}, pages 51:1--51:14, 2020.
\newblock URL: \url{https://doi.org/10.4230/LIPIcs.SoCG.2020.51}, \href
  {http://dx.doi.org/10.4230/LIPIcs.SoCG.2020.51}
  {\path{doi:10.4230/LIPIcs.SoCG.2020.51}}.

\bibitem{matouvsek1992efficient}
Ji{\v{r}}{\'\i} Matou{\v{s}}ek.
\newblock Efficient partition trees.
\newblock {\em Discrete \& Computational Geometry}, 8(3):315--334, 1992.

\bibitem{matouvsek1990net}
Ji{\v{r}}{\'\i} Matou{\v{s}}ek, Raimund Seidel, and Emo Welzl.
\newblock How to net a lot with little: Small $\varepsilon$-nets for disks and
  halfspaces.
\newblock In {\em Proceedings of the 6th Symposium on Computational Geometry
  (SoCG)}, pages 16--22, 1990.

\bibitem{MustafaRR15}
Nabil~H. Mustafa, Rajiv Raman, and Saurabh Ray.
\newblock Quasi-polynomial time approximation scheme for weighted geometric set
  cover on pseudodisks and halfspaces.
\newblock {\em {SIAM} Journal on Computing}, 44(6):1650--1669, 2015.
\newblock URL: \url{https://doi.org/10.1137/14099317X}, \href
  {http://dx.doi.org/10.1137/14099317X} {\path{doi:10.1137/14099317X}}.

\bibitem{mustafa2009ptas}
Nabil~H. Mustafa and Saurabh Ray.
\newblock Improved results on geometric hitting set problems.
\newblock {\em Discrete \& Computational Geometry}, 44(4):883--895, 2010.
\newblock \href {http://dx.doi.org/10.1007/s00454-010-9285-9}
  {\path{doi:10.1007/s00454-010-9285-9}}.

\bibitem{varadarajan2010weighted}
Kasturi Varadarajan.
\newblock Weighted geometric set cover via quasi-uniform sampling.
\newblock In {\em Proceedings of the 42nd ACM Symposium on Theory of Computing
  (STOC)}, pages 641--648, 2010.

\bibitem{virgisurvey}
Virginia {Vassilevska Williams}.
\newblock On some fine-grained questions in algorithms and complexity.
\newblock In {\em Proceedings of the ICM}, volume~3, pages 3431--3472. World
  Scientific, 2018.

\end{thebibliography}

%{\small
%\bibliography{my_bib.bib}
%}

\end{document}